\documentclass[manuscript,times]{aastex631}

\usepackage{amsmath}

\begin{document}

\title{Non-equilibrium Ionization Effects on Synthetic Spectra in the AWSoM Solar Corona\footnote{Released on \today}}

\correspondingauthor{Judit Szente}
\email{judithsz@umich.edu}

\author[0000-0002-9465-7470]{J.\ Szente}
\affiliation{Climate and Space Sciences and Engineering Department\\
  University of Michigan\\
  Ann Arbor, MI 48109}

\author[0000-0002-9325-9884]{E.\ Landi}
\affiliation{Climate and Space Sciences and Engineering Department\\
  University of Michigan\\
  Ann Arbor, MI 48109}
  
\author[0000-0001-5260-3944]{B.\ van der Holst}
\affiliation{Climate and Space Sciences and Engineering Department\\
  University of Michigan\\
  Ann Arbor, MI 48109}

\begin{abstract}
In this work we combined AWSoM's non-equilibrium ionization [NEI] calculations from \citet{Szente:2022} with the synthetic spectral computations of SPECTRUM \citet{Szente:2019}, to predict non-equilibrium line intensities across the entire domain of the AWSoM 3D global model. We find that the resulting spectra are strongly affected by non-equilibrium effects in the fast wind regions and streamer edges and that these effects propagate to narrowband images from SoHO/EIT, SECCHI/EUVI and SDO/AIA. The dependence shows a different nature for each line observed resulting in significant changes in line intensity, which need to be accounted for during plasma diagnostics. However, we also find that these effects depend on the local plasma properties, and that no single correction can be developed to account for non-equilibrium effects in observed spectra and images. Comparing to observational data we saw that the changes due to NEI, while significant, are not sufficient to account for the differences between Hinode/EIS spectra and AWSoM/SPECTRUM predictions.
\end{abstract}

\keywords{Solar corona (1483) --- Spectroscopy (1558) --- Solar coronal lines (2038) --- Ionization (2068) --- Magnetohydrodynamical simulations (1966)}

\section{Introduction} \label{sec:intro}

The solar particle and radiation output from the Sun plays a fundamental role in the solar system. Through the solar wind plasma and magnetic field, the Sun forms the Heliosphere, interacts with the planets, and influences their magnetospheres; solar radiation interacts with planetary atmospheres and, in the case of the Earth, drives climate. The Sun is also responsible for Space Weather via solar flares and Coronal Mass Ejections (CMEs), which can cause sudden disturbances potentially very dangerous to human assets in space and on the ground in a variety of ways. In order to predict the effects that the Sun has on the Earth, its environment and its near space, as well as the conditions in the heliosphere, it is crucial to be able to predict the solar radiation and particle output at all phases of solar activity. Hence, we need to understand how the solar atmosphere and wind are created, heated, and evolve in time.

Remote sensing provides the most direct information about the Sun, and solar spectra provide a wealth of information about the solar atmosphere and the source regions of the solar wind through X-ray, EUV and UV narrow-band imaging and by measuring the intensities, widths and Doppler shifts of emission lines in high resolution spectra. From these spectral properties one can determine densities, temperatures, wave amplitudes and periods, particle abundances and other plasma parameters, which are needed to understand coronal heating and solar wind acceleration. 

The diagnostic techniques that provide most plasma properties rely on the analysis of spectral line intensities emitted by the ions present in the plasma. Such intensities critically depend on the density of these ions, which is determined by the interplay of ionization and recombination processes occurring locally at every place along the line of sight. Usually, it is assumed that the plasma evolution is slow enough to allow each element to settle into equilibrium ionization [EI]. However, there are many instances in the solar atmosphere where plasma evolution is faster than ionization and recombination times, so that this assumption can lead to inaccurate results. For example, during the flare impulsive phase the variation in plasma temperature is much faster than the local ionization and recombination times so that plasma ionization lags behind the electron temperature; however, the high electron density allows the plasma to reach equilibrium again soon in the decay phase \citep{Landi:2003}. Nanoflare-induced departures from ionization equilibrium can also take place, although they are more difficult to identify. Another example is the acceleration of the solar wind from the chromosphere to the corona, because as solar wind plasma crosses the transition region it changes its temperature by 1-2 orders of magnitude in a very short time, while the decrease in electron density slows the plasma ionization and recombination processes. As a result, the solar wind ionization lags behind the plasma local temperature, until the density becomes so low that recombination and ionization stop being effective. At this height, called {\it freeze-in point}, the wind ionization stops evolving at a certain height above the photosphere. The height of this point depends on wind speed and on each ion's recombination and ionization rates, and can range between 2 and 5 solar radii in heliocentric coordinates \citep[and references therein.]{Landi:2012}. For example, Ko et al. (1997) studied the evolution of the charge states of C, O, Mg, Si, and Fe in the fast solar wind and indicated that freeze-in occurs within two solar radii for all elements except Fe, which stopped evolving at around 4–5 Rsun, while \citet{Gilly:2020} suggests values as low as 1.1 or 1.2 for some heavier ions. Similarly, flows in closed coronal loops can also induce departures from ionization equilibrium, as loop plasma is again accelerated from the chromosphere into the coronal section of the loop \citep{Bradshaw:2004}.

Departures from equilibrium ionization significantly impact the charge state composition, affecting line intensities. Such effects have direct consequences for diagnostic techniques making use of absolute line intensities or intensity ratios of lines from different ions, such as elemental abundances, plasma thermal distribution, and electron temperature, and can also affect the plasma radiative losses \citep{Landi:2012b}. It is therefore important to use non-equilibrium ionization (NEI) assumption in coronal plasma diagnostics, see e.g. \cite{Shi:2019}, \cite{Lee:2019}, and \citet{Gilly:2020}. Reviews about NEI effects on spectra in the solar atmosphere can be found in e.g. \cite{Bradshaw:2013} and \cite{Dudik:2017}.

In order to predict the ionization evolution of a plasma, several ingredients are necessary. First, the required ionization and recombination rate coefficients are necessary for every ion of a given element, usually as a function of temperature. Then, it is necessary to know the plasma electron temperature ($T_e$) and density ($n_e$) as a function of time, to calculate the ionization and recombination rates to be included in the system of equations governing the plasma charge state distribution. In the case of the solar wind, the evolution of $T_e$ and $n_e$ is calculated by considering the variation of these quantities along the wind trajectory, and the speed of the solar wind, which determines how much time a wind plasma parcels spends at every single location. The density, temperature and velocity profiles are either taken from theoretical models or from observations (although sometimes interpolation/extrapolation is needed).

Several models have been developed that can predict charge state composition. \cite{Shen:2015} used an eigenvalue method to solve the time-dependent ionization equations; additionally, an adaptive time stepping was introduced for computational efficiency. The Michigan Ionization Code (MIC, \cite{Landi:2012}) uses the electron density, electron temperature and bulk speed along the 1D wind trajectory combined with ionization and recombination rates from the CHIANTI database \citep{DelZanna:2021} to determine the charge state evolution. Their method uses a fourth-order Runge–Kutta method with adaptive step size. MIC was combined with 3D magnetohydrodynamic models of the solar wind as a post-processing tool to validate against Ulysses/Solar Wind Ion Composition Experiment (SWICS; \cite{Landi:2014}). \cite{Lionello2019} developed a fractional charge state composition module integrated in the Magnetohydrodynamic Algorithm outside a Sphere (MAS) model with Alfv\'en wave turbulence, suitable for both steady state and time-dependent simulations. More recently, \cite{Szente:2022} integrated a full 3D charge state composition module in the Alfv\'en Wave Solar Atmosphere Model (AWSoM, \cite{vanderHolst:2014}), making it possible to calculate the charge state composition of solar plasmas anywhere in the solar corona and solar wind self-consistently. 

Recently, there is a growing interest in synthetic spectral calculations for first-principles solar corona model validation. \cite{Gibson:2016} developed the FORWARD model to calculate high-resolution spectra with a special emphasis on coronal magnetometry, to facilitate the science return of UCoMP and DKIST. While FORWARD models emission too, it is not considering NEI effects on the line intensities. RADYN \citep{Carlsson:1992ApJ...397L..59C} is a 1D radiative hydrodynamic code used for energetic event simulations of the chromosphere, considering the multi-ion structure of the non-LTE plasma and its varying opacity. The Global Heliospheric Optically thin Spectral Transport Simulation (GHOSTS, \citet{Gilly:2020}) uses inputs from other models to simulate charge states and NEI spectral lines.

\cite{Szente:2019} developed SPECTRUM, a post-processing module within the Space Weather Modeling Framework \citep[SWMF,][]{Toth:2012} that calculates synthetic spectra for the optically thin solar corona by combining the results of the AWSoM model with CHIANTI-tabulated emission rates. SPECTRUM allows detailed studies of line formation everywhere along the user-defined line-of-sight (LOS), by including the thermal broadening, non-thermal broadening due to Alfv\'en wave turbulence, Doppler shifts due to solar wind bulk speed projected along the LOS, and instrumental broadening. SPECTRUM is publicly available though SWMF at the \url{http://csem.engin.umich.edu/Tools/SWMF} website. In \cite{Szente:2019}, SPECTRUM was used to validate AWSoM by comparing with the Extreme-ultraviolet Imaging Spectrometer (EIS, \cite{Culhane:2007}) observations for Carrington rotations 2063 and 2082.

In this paper we investigate the effects of NEI on the emission of the solar atmosphere predicted by the SPECTRUM module, focusing on both high resolution spectra and narrow band EUV imaging instruments (SoHO/EIT, SDO/AIA, SECCHI/EUVI). We first introduce the components of the non-equilibrium spectral calculation in Section~\ref{sec:methods} such as the MHD coronal model with non-equilibrium charge state calculation (Section~\ref{subsec:chargestates}), and then we describe the spectral calculation in Section~\ref{subsec:spectra}. We then discuss the changes in the solar emission induced by non-equilibrium in Section~\ref{sec:results} considering several instruments, and summarize our findings in Section~\ref{sec:summary}.

\section{NEI and Line Intensity Calculation}\label{sec:methods} 

\subsection{Solar Corona with Charge States}\label{subsec:chargestates}
The Alfv\'en Wave Solar atmosphere Model \citep[AWSoM,][]{vanderHolst:2022} is the solar model of the Solar Corona and Inner Heliosphere components of the Space Weather Modeling Framework \citep[SWMF,][]{Toth:2012}, publicly available at \url{https://github.com/MSTEM-QUDA/SWMF}. AWSoM utilizes magnetograms for the inner boundary condition of the radial component of the magnetic field and produces a realistic solar corona model from the lower transition region all the way to $24~R_{\odot}$. At the inner boundary, at $1~R_{\odot}$ the electron temperature is set to 50,000~K. The Solar Corona component is then coupled to the Inner Heliosphere module, which propagates the solution out to 1AU. The solar corona is heated via Alfv\'en wave dissipation and the solar wind is accelerated via Alfv\'en wave pressure gradient; the model takes into account 3 temperatures: isotropic electron temperature, as well as parallel and perpendicular proton temperatures (relative to the local magnetic field). The model incorporates radiative losses, electron heat-conduction, Coulomb collisions, gravity, and improved energy partitioning based on stochastic heating \citet{Chandran:2011}. For the governing equations and details see \citet{vanderHolst:2022}. In this paper we carried out a global simulation using a magnetogram by the Global Oscillation Network Group \citep[GONG,][]{Harvey:1996} with the radial magnetic field of Carrington Rotation (CR) 2063 (between 2007-11-07 and 2007-12-04).  

\citet{Szente:2022} recently implemented in AWSoM the capability of calculating NEI for multiple ions self-consistently as part of the main simulation. The charge states are calculated throughout the three-dimensional domain of the solar corona and inner heliosphere by solving the set of continuity equations for ions $X^{+m}$ with number density $N(X^{+m})$:

\begin{eqnarray}
\label{eq:chargestate}
\begin{aligned}
     &\frac{\partial N(X^{+m})}{\partial t} + \nabla\cdot\left[N(X^{+m}){\bf u}\right]= \\
     &N_e \{ N(X^{+,{m-1}})C_{m-1}\left(T_e\right) - N(X^{+m}) \left[ C_{m}\left(T_e\right) + R_{m}\left(T_e\right)\right] + N(X^{+,{m+1}}) R_{m+1}\left(T_e\right) \}.
     \end{aligned}
\end{eqnarray}
Here, $T_e$ is the electron temperature, $N_e$ is the electron density and ${\bf u}$ is the solar wind bulk speed. The charge states are calculated using tables for the total ionization and recombination rates, $C_m(T_e)$ and $R_m(T_e)$, obtained from CHIANTI~10.0 \citep{Dere:1997, DelZanna:2021}. The details of the implementation of charge state calculations can be found in \citet{Szente:2022}. In this paper we calculated charge states of ions of C, N O, Ne, Mg, Si, S and Fe, as these are the ions providing the bulk of the emission observed by current remote-sensing instrumentation working on the UV, EUV and X-ray wavelength ranges; AWSoM can calculate charge states for all elements included in the CHIANTI database, namely from H to Zn.

\subsection{Non-equilibrium Ionization SPECTRUM Calculation}\label{subsec:spectra}

In SWMF the spectral calculations used to be performed as a post-simulation step either on a Cartesian box \citep{Szente:2019} or the original grid \citep{Shi:2022}. SPECTRUM has now been fully implemented into the Block Adaptive Tree Solarwind Roe Upwind Scheme \citep[BATS-R-US,][]{Powell:1999}, so that the line-of-sight (LOS) integration of the spectral emission is calculated on the original,block adaptive grid, along with the simulation, in the same way as in \citet{Downs:2010}. The spectral calculation is performed on a voxel-by-voxel basis in the same way as in \citet{Szente:2019}, but replacing equilibrium charge state distribution values with non-equilibrium values in the following manner.

The flux $F$ reaching the observer at distance $d$ coming from volume element $V$ emitted by an ion $X_{j}^{+m}$ in the excited state $j$ can be written as
\begin{eqnarray}
  \label{eq:flux}
  F = \frac{1}{4\pi d^2} \int_V N\left(X_{j}^{+m} \right) A_{ji}h\nu_{ij}dV,
\end{eqnarray}
where $N\left(X_{j}^{+m}\right)$ is the density of the emitting ion in level $j$, 
$A_{ji}$ is the Einstein coefficient for the transition $j\rightarrow i$, 
$h$ is the Planck constant, 
$\nu_{ij}$ is emission frequency. 
Assuming that level populations are in statistical equilibrium the density of emitting ions can be rewritten as the product of several individual components:
\begin{eqnarray}
  \label{eq:fluxwithG}
  F = \frac{1}{4\pi d^2} \int_V \frac{N\left(X_{j}^{+m}\right)}{N\left(X^{+m}\right)}
  \frac{N\left(X^{+m}\right)}{N\left(X\right)}
  \frac{N\left(X\right)}{N\left(H\right)} \frac{N\left(H\right)}{N_e}
  \frac{A_{ij}}{N_e}h\nu_{ij} N_e^2 dV ,
\end{eqnarray}
where $N_e$ is the electron density. 
$\frac{N\left(X_{j}^{+m}\right)}{N\left(X^{+m}\right)}$ is the level $j$ relative population calculated assuming balance between excitation and de-excitation processes, 
$\frac{N\left(X^{+m}\right)}{N\left(X\right)}$ is the abundance of ion $X^{+m}$ relative to that total element abundance (the {\em ion fraction}),
$\frac{N\left(X\right)}{N\left(H\right)}$ is the abundance relative to hydrogen, which we assumed to be the {\it feldman\_1992\_extended} coronal abundances from CHIANTI 10.0, while
$\frac{N\left(H\right)}{N_e}=0.833$ assuming fully ionized plasma.

Implementing the non-equilibrium ion fractions into line intensity calculation only involves replacing the equilibrium values of the ion fractions used in the previous version of SPECTRUM \citep{Szente:2019} with the non-equilibrium values calculated voxel-by-voxel by AWSoM \citep{Szente:2022}. Defining the Contribution Function of any given voxel in ionization equilibrium as
\begin{eqnarray}
  \label{eq:contributionfunction}
  G_{equil}\left(T_e, N_e \right) = \frac{N\left(X_{j}^{+m}\right)}{N\left(X^{+m}\right)}
  \left[\frac{N\left(X^{+m}\right)}{N\left(X\right)}\right]_{equil}
  \frac{N\left(X\right)}{N\left(H\right)} \frac{N\left(H\right)}{N_e}
  \frac{A_{ij}}{N_e}h\nu_{ij} ,
\end{eqnarray}
the non-equilibrium value is given by:
\begin{eqnarray}
  \label{eq:magic_ratio}
  G_{non-equil}\left(T_e, N_e \right) = G_{equil}\left(T_e, N_e \right) \frac{\left[\frac{N\left(X^{+m}\right)}{N\left(X\right)}\right]_{non-equil}}{\left[\frac{N\left(X^{+m}\right)}{N\left(X\right)}\right]_{equil}}
 \end{eqnarray} 
In this way, the observed non-equilibrium line flux can be written as
\begin{eqnarray}
  \label{eq:newfluxwithG}
  F = \frac{1}{4\pi d^2} \int_V G_{non-equil}\left(T_e,N_e \right) N_e^2 dV .
\end{eqnarray}
This flux can then be used for comparison with observations or to calculate the response function of narrow-band imagers releasing the ionization equilibrium assumption.

\section{Results}\label{sec:results}

\subsection{High resolution spectra}

We first studied how individual line intensities change when the assumption of ionization equilibrium is released, utilizing the same AWSoM model results used by \citet{Szente:2022}, which included the calculation of non-equilibrium charge states. This simulation used the GONG magnetogram of Carrington Rotation 2063 as boundary condition (corresponding to the minimum of solar cycle 24 between November 4 and December 1, 2007), and provided, after evolving for 200,000 time steps, a realistic steady state plasma solution in the 3D solar corona from 50,000~K (at 1 solar radii) out to 24 solar radii. To resolve the transition region's steep density and temperature change, the solution is artificially stretched in the radial direction. In order to avoid the large, unphysical line intensity enhancement caused by this artificial numerical solution, we excluded all emission coming from plasma below the temperature of 230,000K ($\log{T} = 5.36~K$). This temperature cut is corresponding to what is traditionally used for constructing synthetic narrowband images with AWSoM.

In the simulation we calculated non-equilibrium charge states of ions of the following elements: C, N, O, Ne, Mg, Si, S, and Fe. These elements provide the bulk of the emission observed both by high-resolution and narrow-band instruments working in the wavelength range between X-rays and UV. The spectra used for the present work were calculated for the Earth-Sun line of sight of 4 November 2007, 19:12:27, corresponding to the EIS full spectrum observed outside the solar west limb in a 14"$\times$512" raster centered at around (-37",991").

\subsubsection{Coronal ions}

\begin{figure}[htb!]
\includegraphics[trim={5cm 0.cm 3cm 4cm},clip,width=6cm]{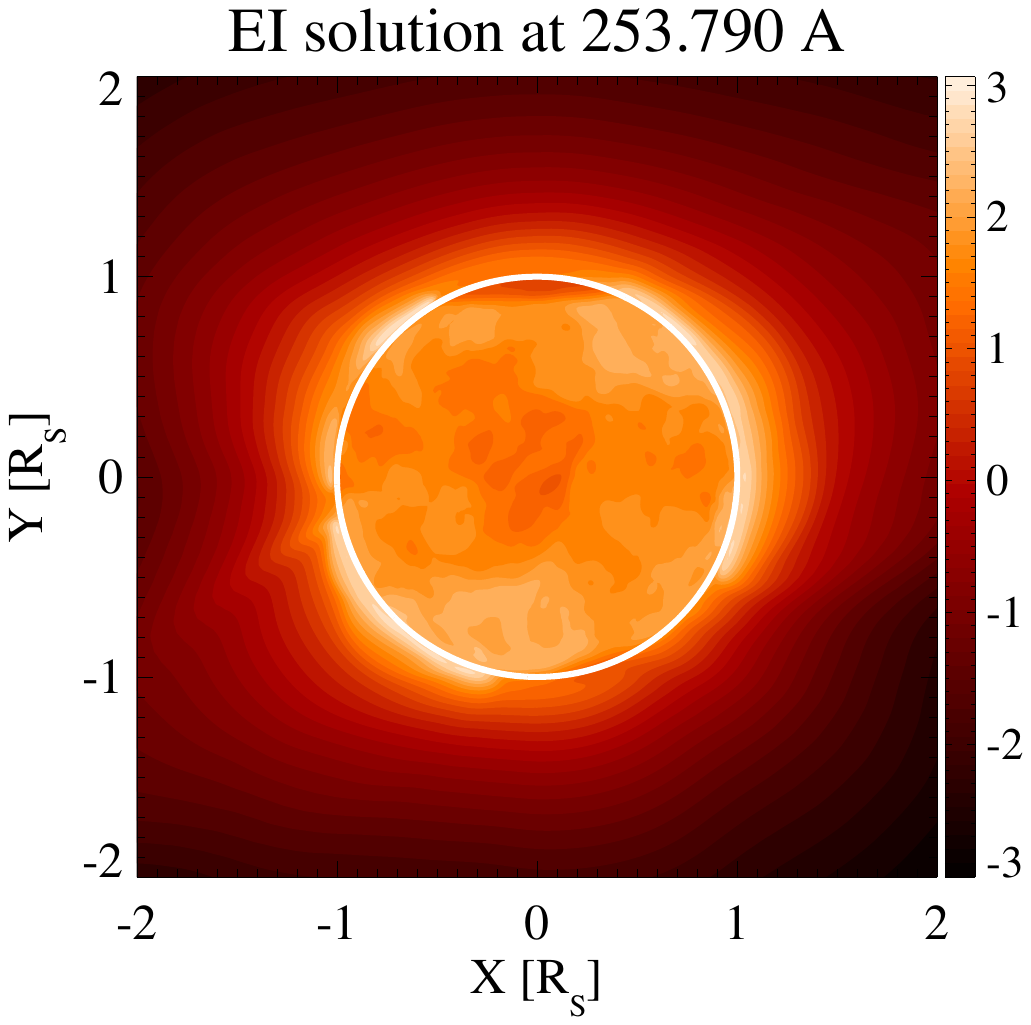}
\includegraphics[trim={5cm 0.cm 3cm 4cm},clip,width=6cm]{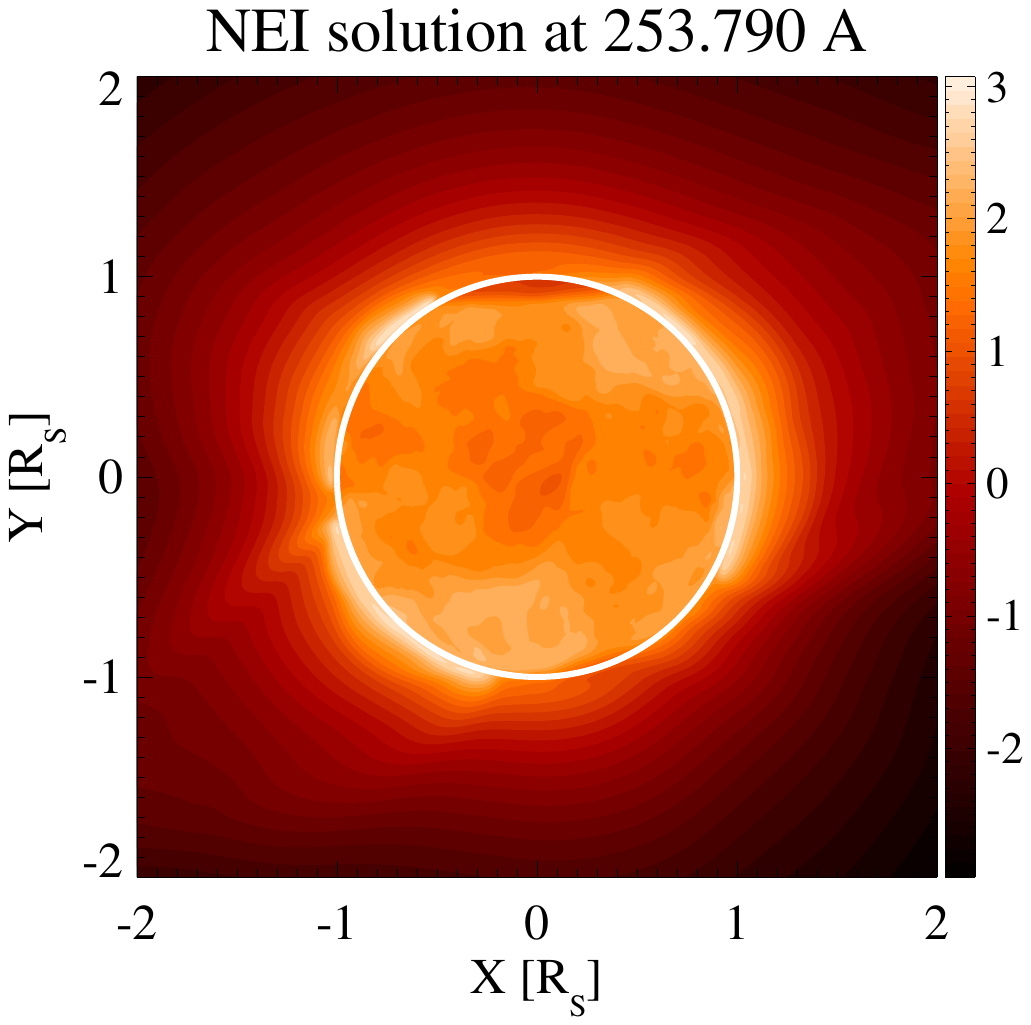}
\includegraphics[trim={5cm 0.cm 3cm 4cm},clip,width=6cm]{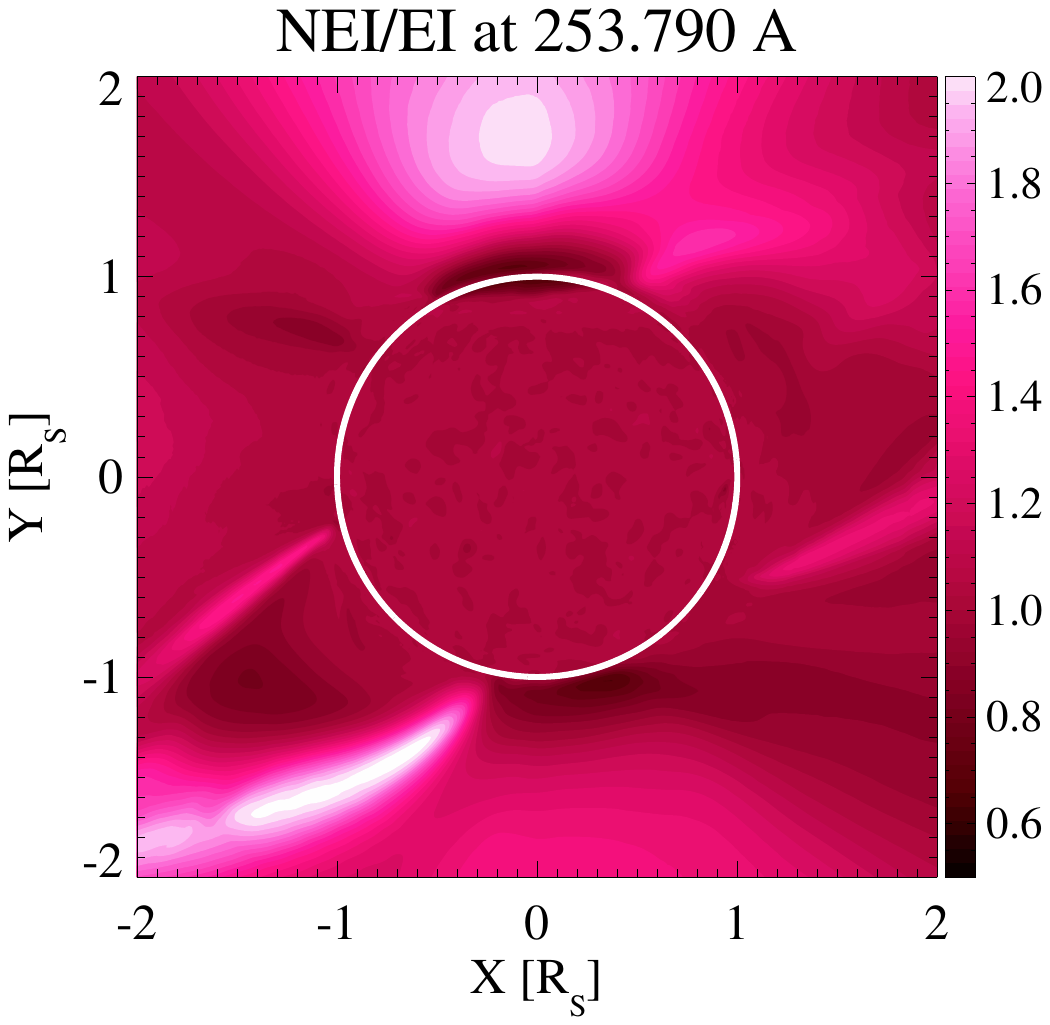}\\
\includegraphics[trim={5cm 0.cm 3cm 4cm},clip,width=6cm]{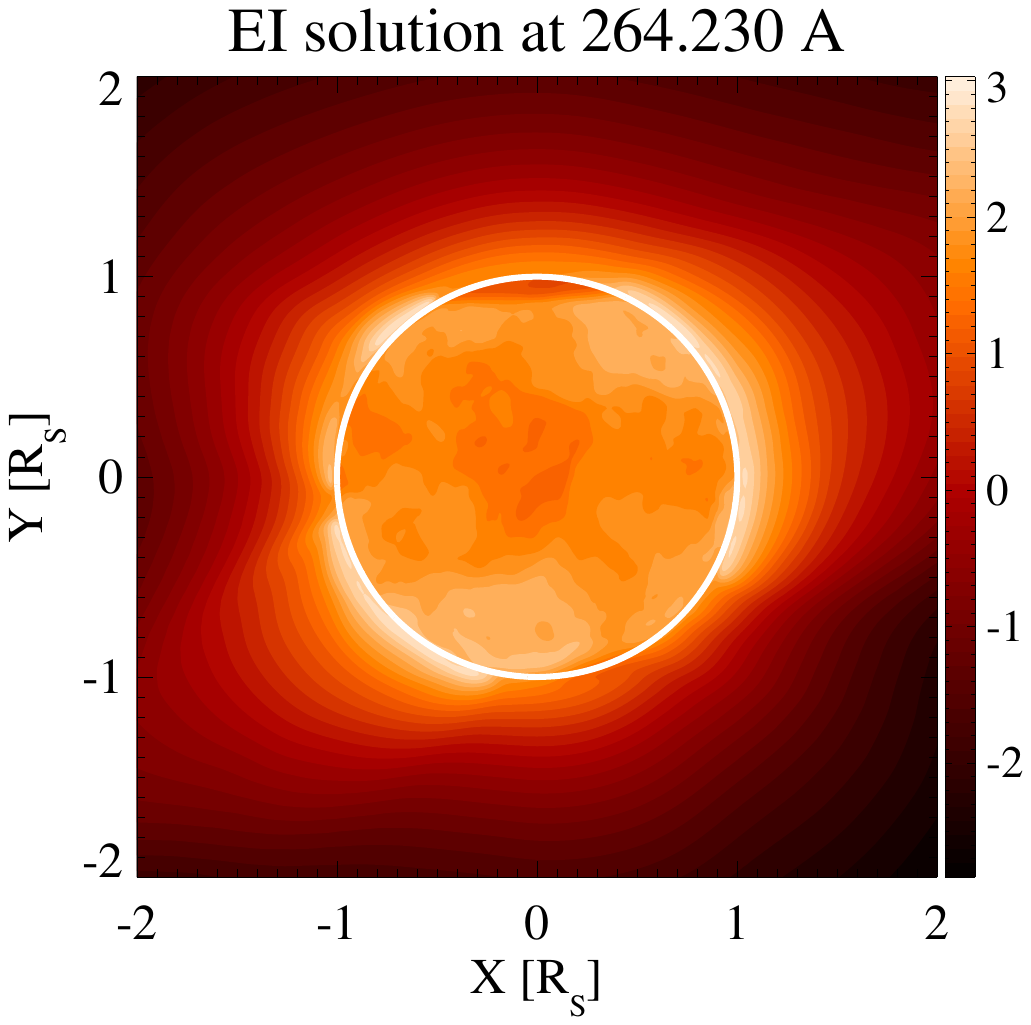}
\includegraphics[trim={5cm 0.cm 3cm 4cm},clip,width=6cm]{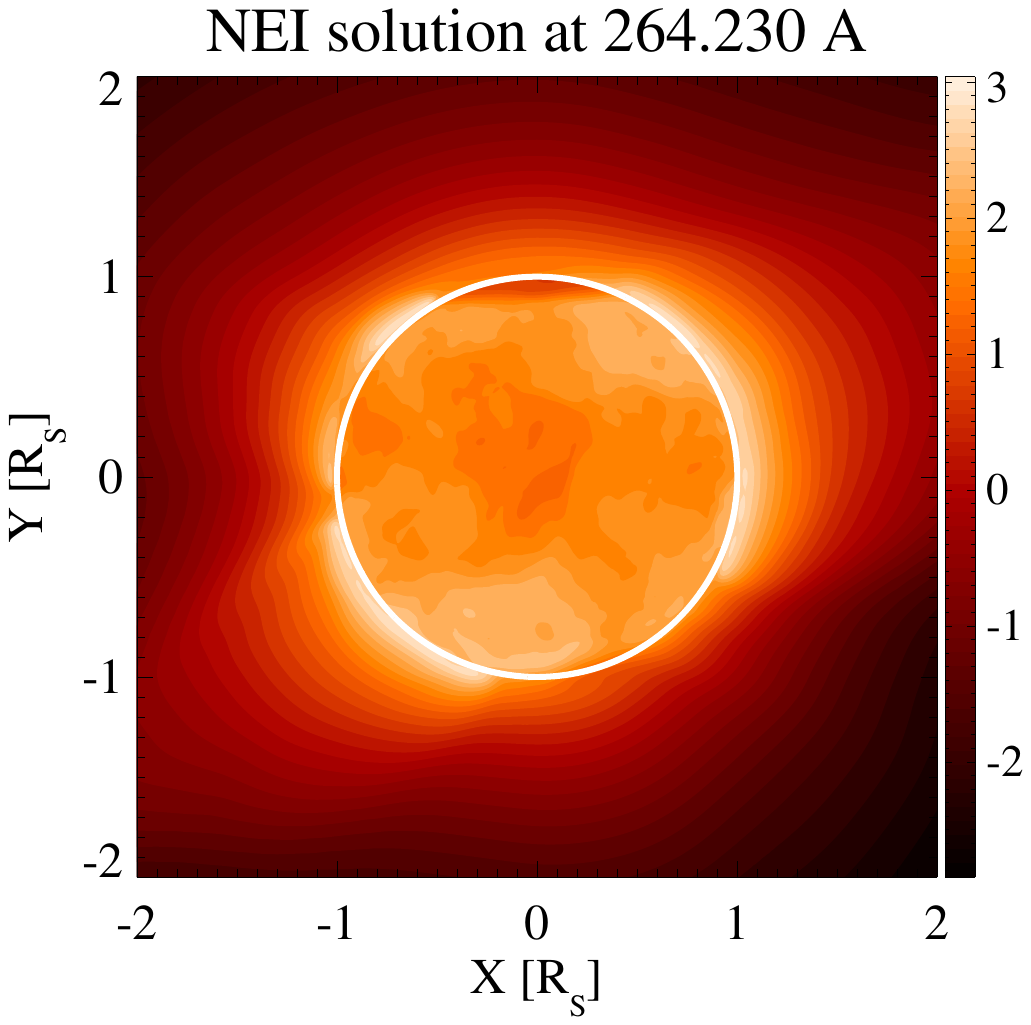}
\includegraphics[trim={5cm 0.cm 3cm 4cm},clip,width=6cm]{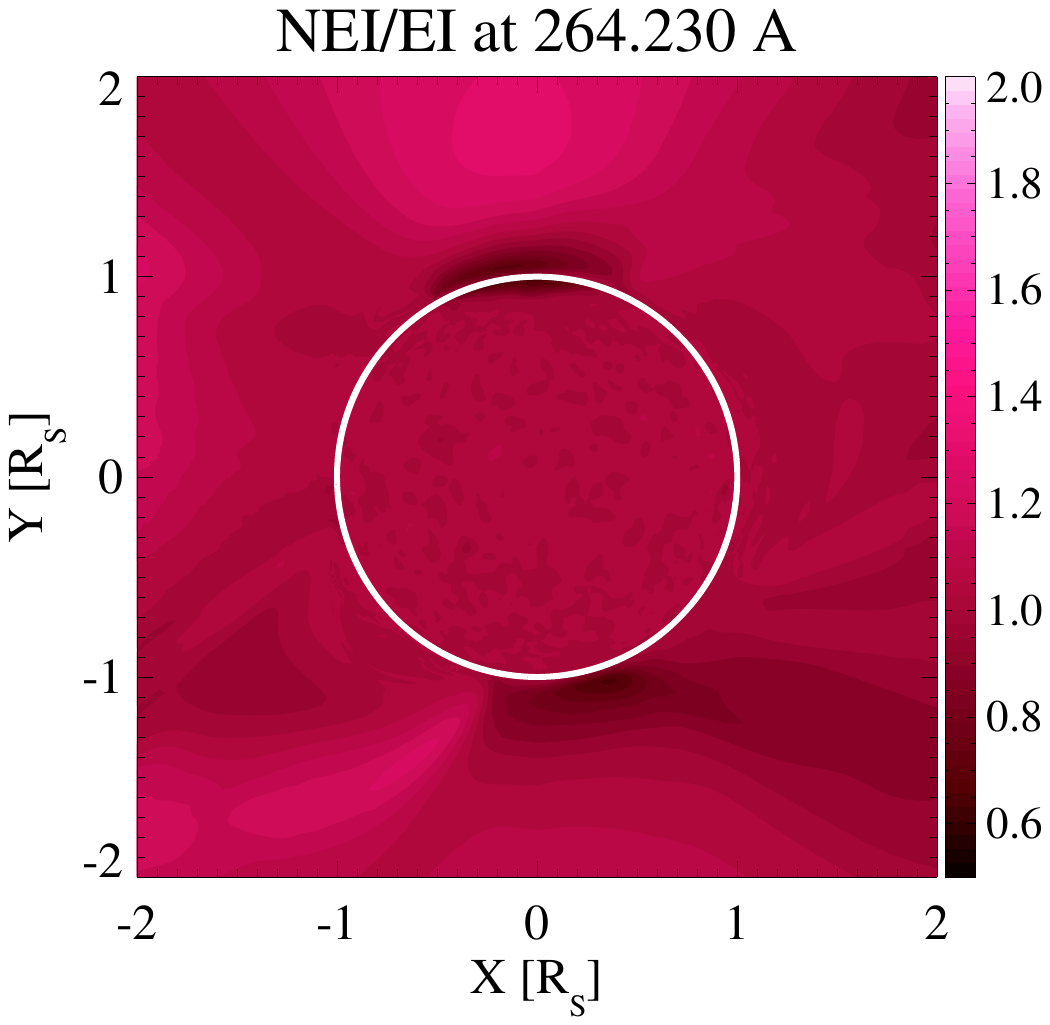}\\
\includegraphics[trim={5cm 0.cm 3cm 4cm},clip,width=6cm]{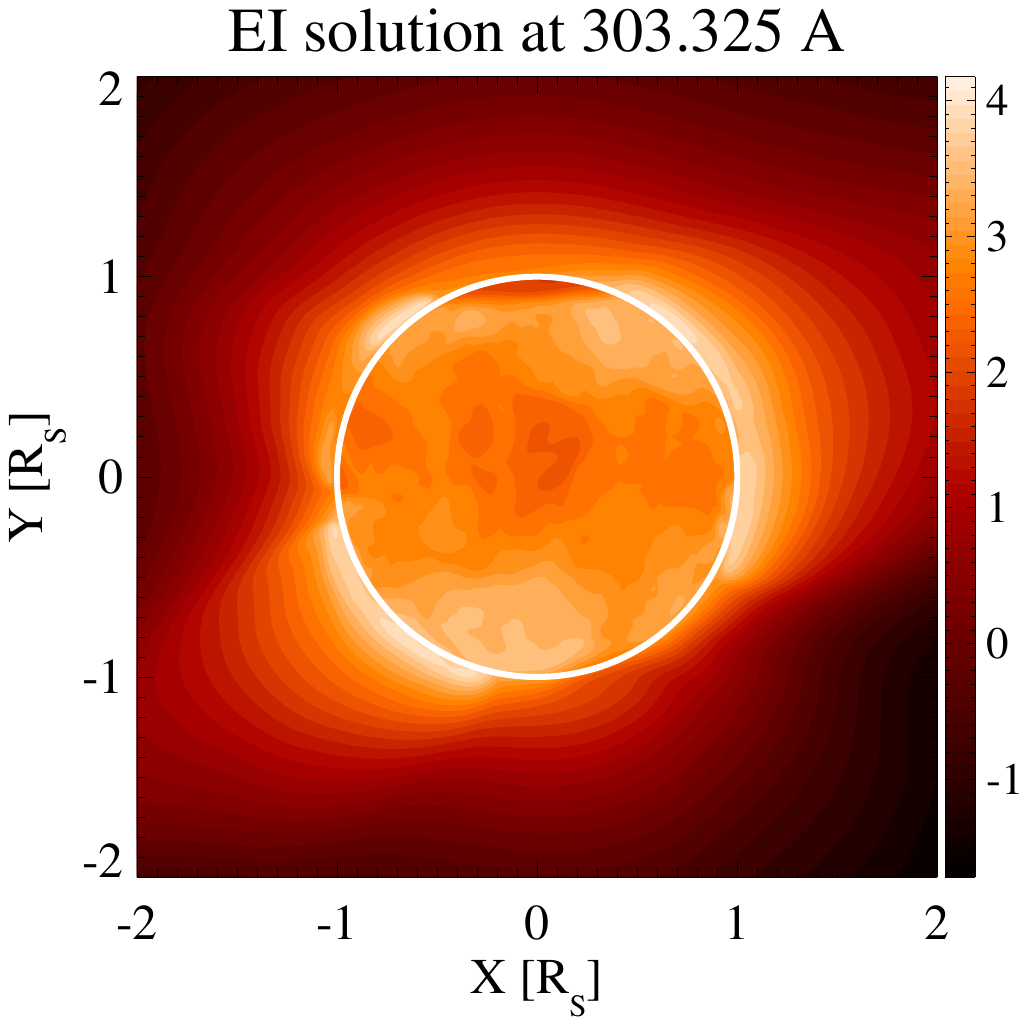}
\includegraphics[trim={5cm 0.cm 3cm 4cm},clip,width=6cm]{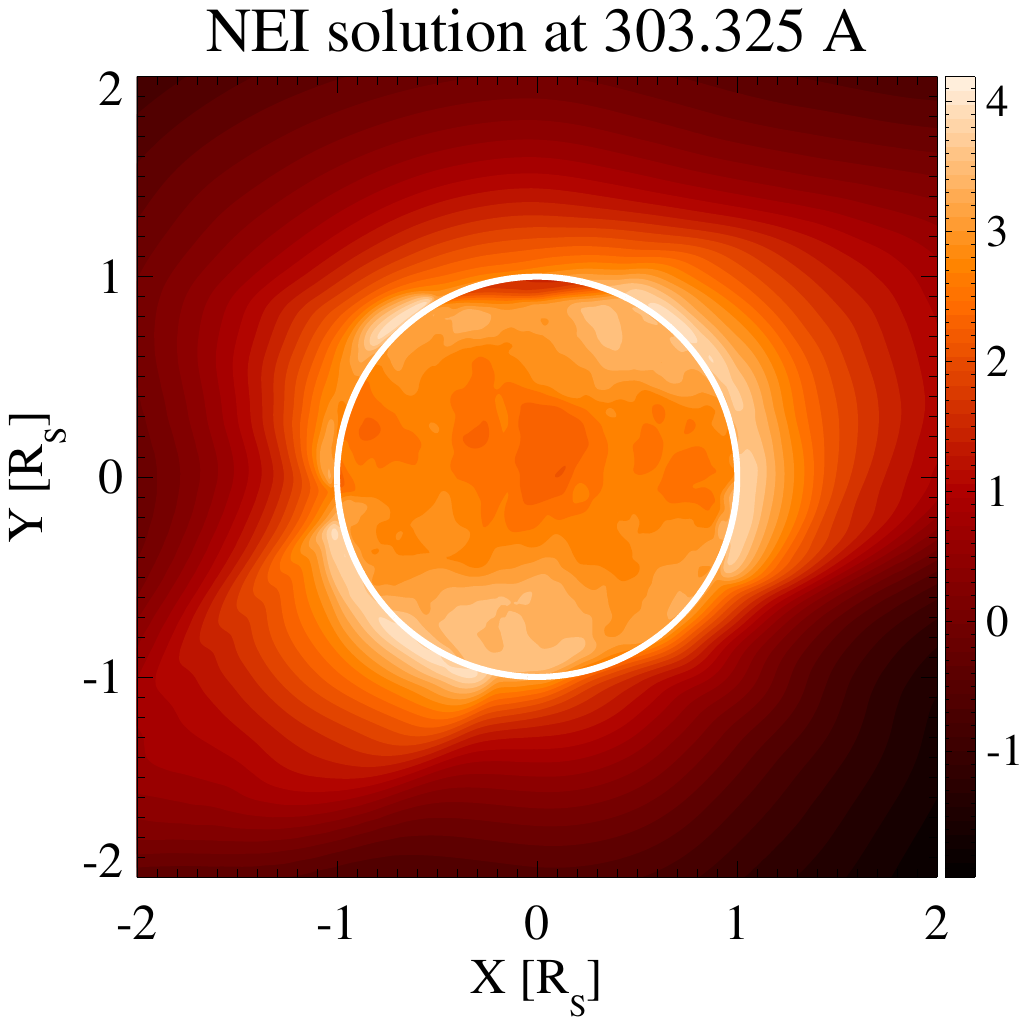}
\includegraphics[trim={5cm 0.cm 3cm 4cm},clip,width=6cm]{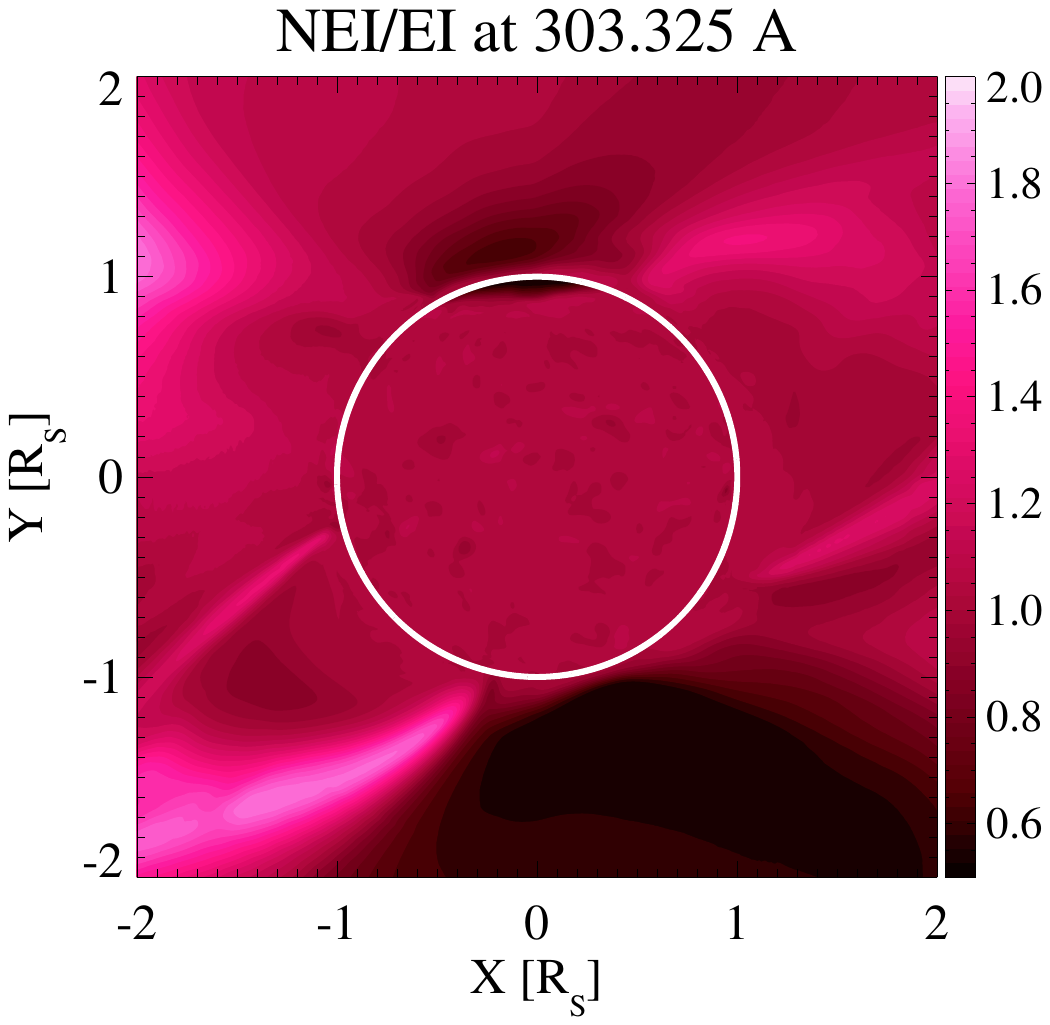}\\
\includegraphics[trim={5cm 0.cm 3cm 4cm},clip,width=6cm]{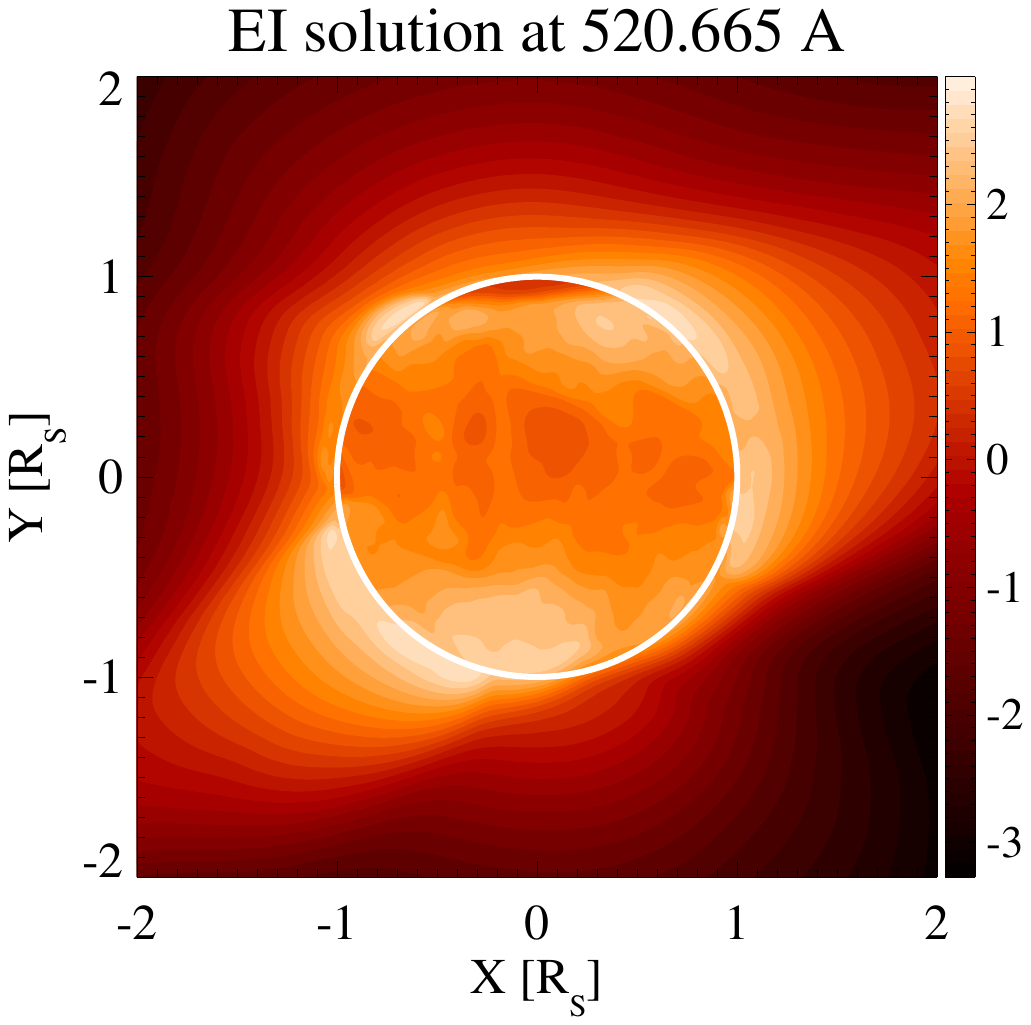}
\includegraphics[trim={5cm 0.cm 3cm 4cm},clip,width=6cm]{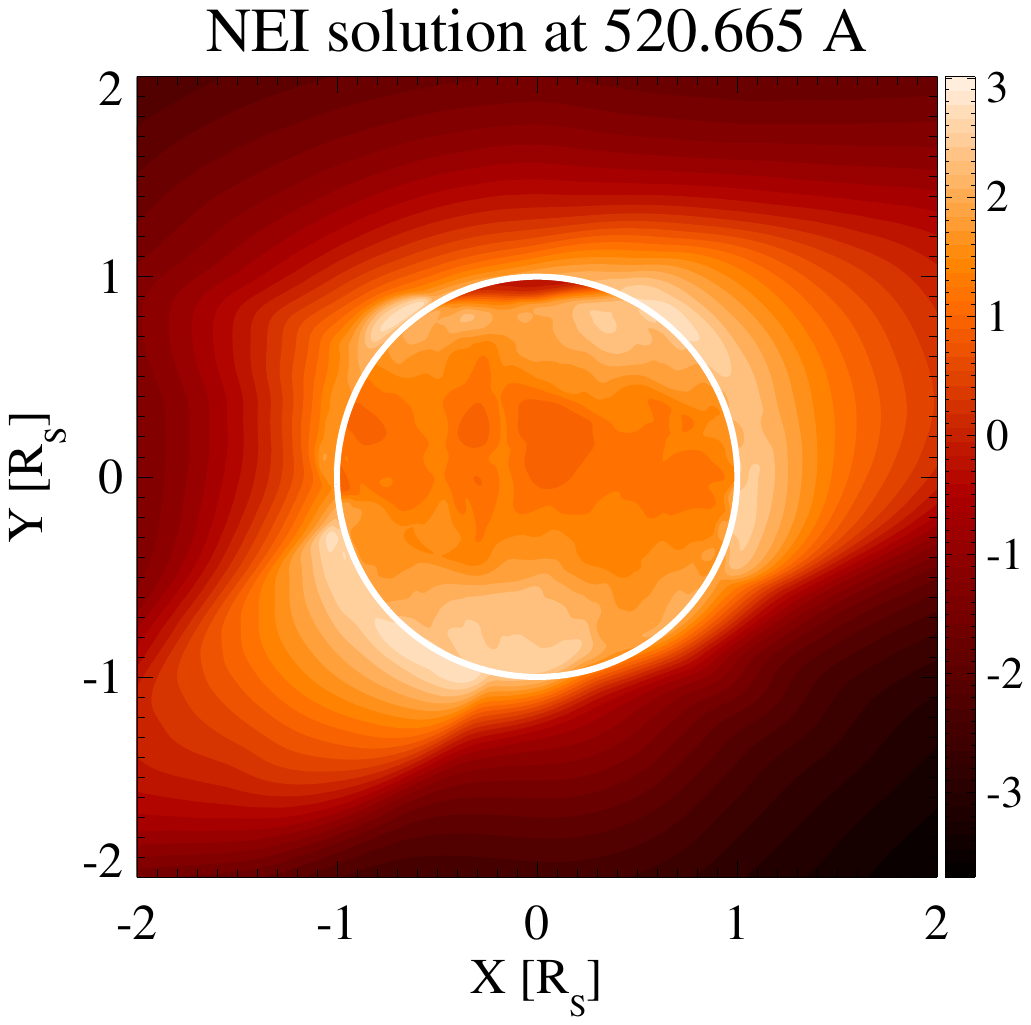}
\includegraphics[trim={5cm 0.cm 3cm 4cm},clip,width=6cm]{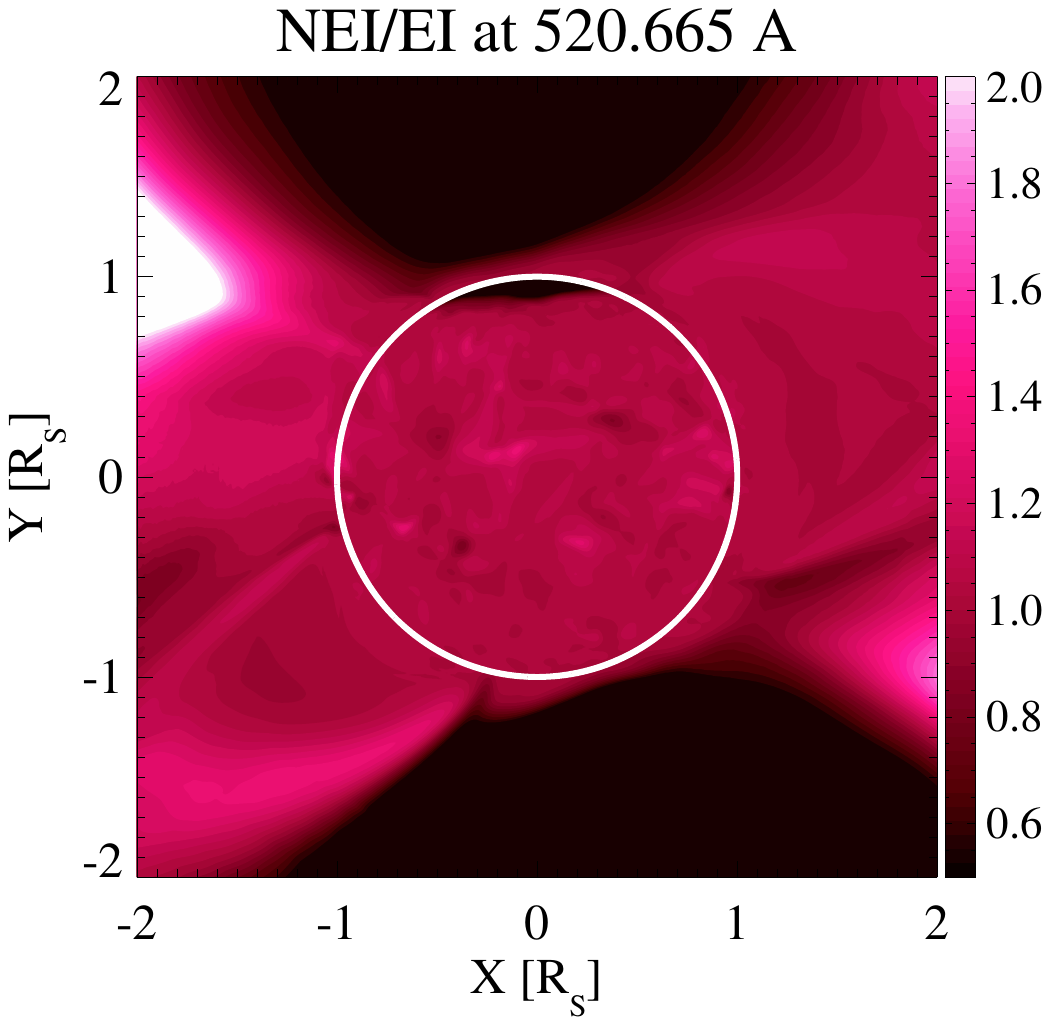}
\caption{Plasma emission calculated assuming equilibrium ionization (\emph{left}), NEI (\emph{center}) and showing the non-equilibrium/equilibrium ratio (\emph{right}) for ions: Si~X (253.790 \AA, $\log{T} = 6.15$), S~X (264.230 \AA, $\log{T} = 6.17$), Si~XI (303.325 \AA, $\log{T} = 6.22$), Si~XII (520.665 \AA, $\log{T} = 6.29$). }  
\label{fig:eqneq1}
\end{figure}

Figure~\ref{fig:eqneq1} shows emission from select Si and S ions calculated with equilibrium assumption (left column), non-equilibrium charge state ratios (middle) and their ratios. The images corresponding to emission are on the same logarithmic scale, while the ratio images are on a linear scale. These were chosen because their maximum formation temperature (taken arbitrarily as the temperature of maximum line emissivity under the assumption of equilibrium, listed in Table~\ref{tbl:lines}) is approximately similar at $\log T_e \simeq 6.15-6.29$ (and in particular Si~X and S~X are formed almost at the same temperature), to show that even in the case of ions formed at similar temperatures, non-equilibrium effects are highly variable due to ion electronic configuration (from Li-like Si~XII to N-like S~X), local plasma conditions in the solar atmosphere (coronal holes, quiet Sun and streamers) as well as locations in the solar image (e.g. disk vs. limb locations). Also, the comparison of three consecutive stages of ionization of Silicon (X to XII) allows to understand a typical trend of non-equilibrium effects.

While absolute line intensities provide visually nearly identical images, ratio images show significant differences due to non-equilibrium. Within the same element, different stages of ionization respond differently to non-equilibrium conditions. This is due to two separate effects: first, each stage of ionization is heavily affected by the behavior of the previous ones, especially when the plasma is fast ionizing or recombining; second, each ion's recombination and ionization rates are different, and depend on its electronic configuration, so that they react differently to the same plasma conditions. 

\noindent
\begin{table}[htb!]
\centering
\begin{tabular}{|l| l| c||l| l| c|}
    \hline
    \textbf{Ion} & \textbf{Line [A]} & \textbf{Temperature [logK]} & \textbf{Ion} & \textbf{Line [A]} & \textbf{Temperature [logK]} \\
    \hline
Mg~VIII & 772.2600 & 5.90 & Fe~VIII & 185.213 & 5.66 \\
\hline
Mg~VII & 276.1540 & 5.78 & Fe~IX & 171.073 & 5.88 \\ 
\hline
Mg~IX & 706.0600 & 5.99 & Fe~X & 174.531 & 6.02 \\
\hline
Mg~X & 624.9410 & 6.07 & Fe~XI & 188.299 & 6.11 \\
\hline		             						
Si~VII & 275.3610 & 5.78 & Fe~XII & 195.119 & 6.18 \\
\hline
Si~VIII & 316.2180 & 5.93 & Fe~XIII & 202.044 & 6.24\\
\hline
Si~IX & 290.6900 & 6.05 & Fe~XIV & 264.788 & 6.29 \\ 
\hline
Si~X & 253.7900 & 6.15 & Fe~XV & 284.163 & 6.35  \\
\hline
Si~XI & 303.3250 & 6.22 & Fe~XVI & 262.976 & 6.43 \\
\hline
Si~XII & 520.6650 & 6.29 &
S~X & 264.2300 & 6.17 \\
\hline
O~VI & 1031.9120 & 5.47 & 
Ne~VIII & 770.4280 & 5.80 \\
\hline
\end{tabular}
    \label{tbl:lines}
    \caption{Ions with corresponding emission wavelengths and formation temperature of lines discussed in this paper.}
\end{table}

\begin{figure}[htb!]
\includegraphics[trim={0.2cm 0.5cm 0.3cm 0.4cm},clip,width=9cm]{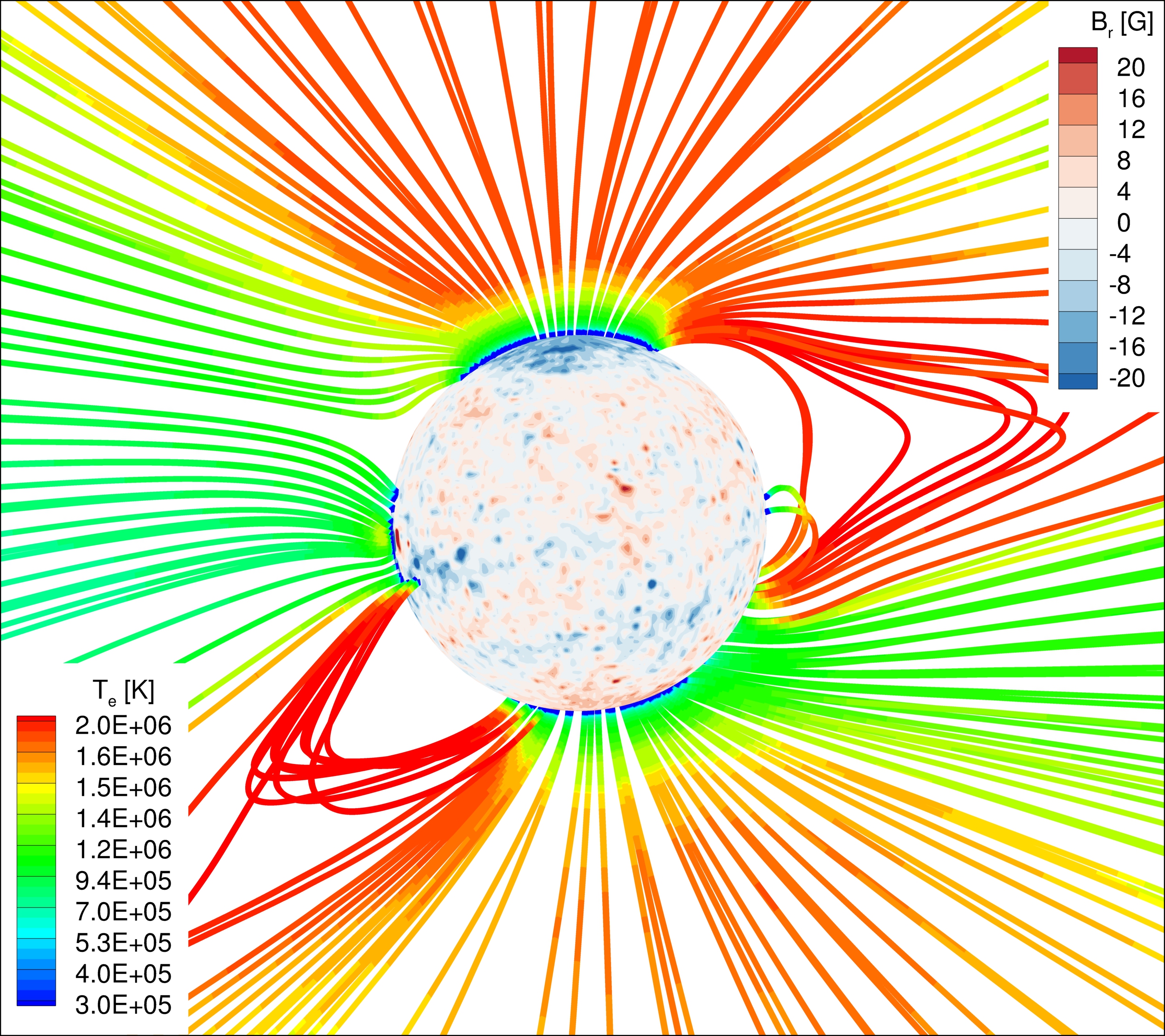}
\includegraphics[trim={0.2cm 0.5cm 0.3cm 0.4cm},clip,width=9cm]{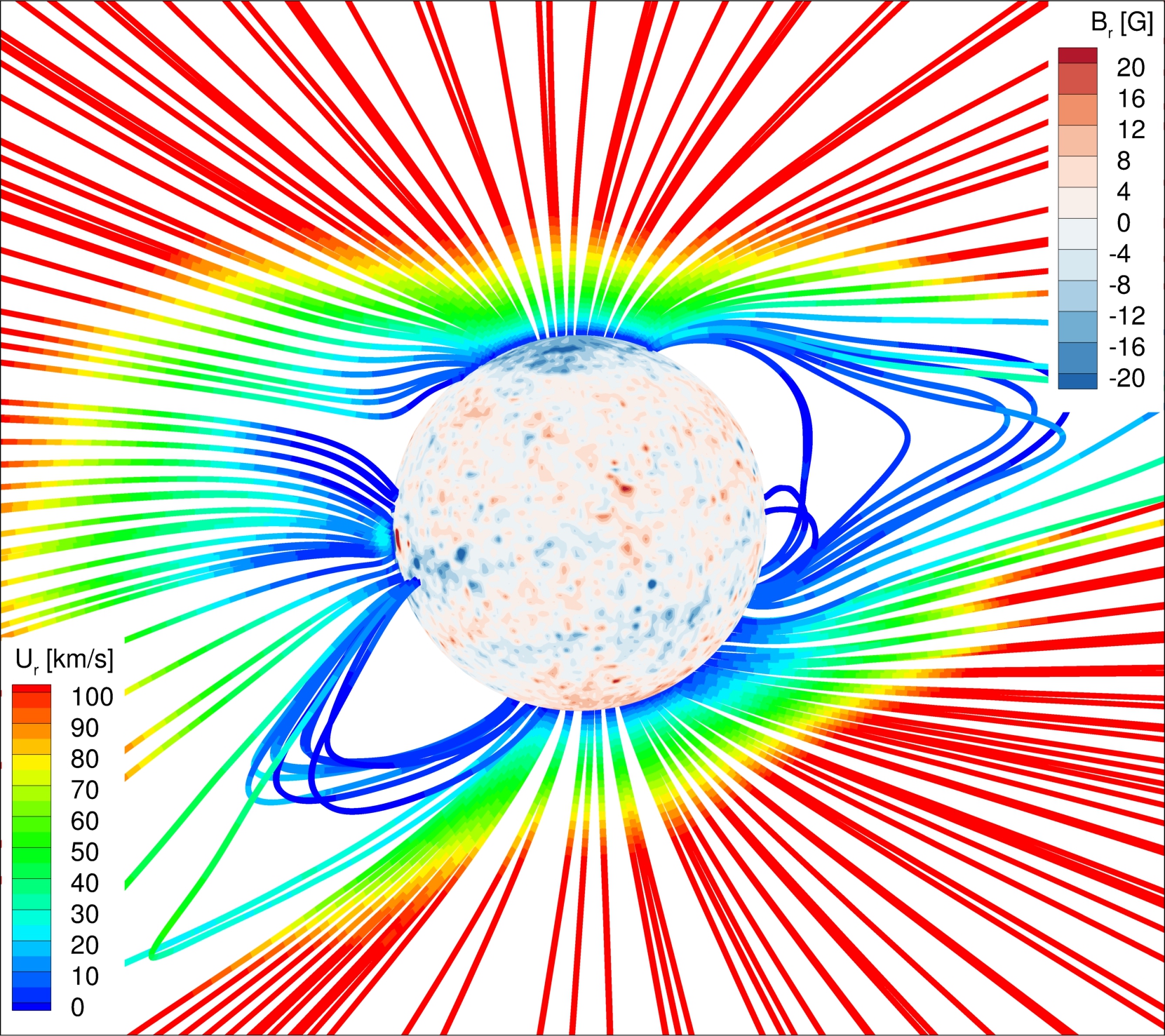}
\caption{Electron temperature (\emph{left}) and radial speed (\emph{right}) on magnetic fieldlines. The solar surface is colored with the contour of the magnetic field strength.  
\label{fig:cooling}}
\end{figure}

The first effect is clearly visible in the three Si ions shown in Figure~\ref{fig:eqneq1}. Non-equilibrium induced changes are strongly dependent on the solar structure they take place in. In both the south and north coronal holes, the ion abundance ratio changes from larger than 1 (that is, the ion is more abundant than in equilibrium) to much smaller than 1. In this case, the so-called "Delay effect" described by \citet{Landi:2012} is at play: the accelerating solar wind travels through the transition region and the corona faster than the plasma is able to adapt while the electron density is decreasing quickly, so that this element can not ionize fast enough. As a result, lower ionization stages do not lose electrons and are overabundant, while higher ones are produced less and less causing under-abundance. A similar effect is present in streamer legs (relatively fast flows and high temperatures are shown in Figure~\ref{fig:cooling} at the boundary of the south-west streamer and south coronal hole), where the overall higher temperature shifts the decrease of abundance towards species even more ionized than Si~XII; still, the increase relative to equilibrium is smaller for the more ionized species (Si~XII) than for the less ionized ones (Si~X). Inside streamer cores instead conditions are close to equilibrium, though some decrease over equilibrium values is more evident in Si~X than in the other two ions. In the north-east region, instead, we have the opposite situation, where lower ionization stages are depleted and higher ones are enhanced over equilibrium. This is due to the cooling at both in the North-East and South-West regions, more pronounced in the North-East, shown in Figure~\ref{fig:cooling}. In fact, at those heights non-equilibrium charge states are almost frozen in and thus are not responsive to changes in the local electron temperature; on the contrary, equilibrium values tend to decrease. However, the decrease in temperature from $\approx 2$~MK to $\approx 1.5$~MK causes the equilibrium abundance of Si~XII to decrease by a large amounts, while the Si~X value, closer to the peak abundance, changes much less.

The second effect is exemplified by the comparison of S~X (N-like ion) with Si~X and Si~XI: even if their temperature of formation is really close, S~X behavior is intermediate. In fact, in the northern coronal hole S~X shows  an initial decrease of abundance followed by an increase, whose values are more limited than those shown by Si~X (whose increase exceeds a factor 2 already at $\approx 1.4$ solar radii) and Si~XII, which on the contrary is always under-abundant; at the streamers' edges, S~X non-equilibrium abundance is much closer to equilibrium than those of whether Si~X and XI, which show significant enhancements. Still, some general trends are common to each of these three ions, such as the smaller variations from equilibrium in the southern coronal hole than in the northern one.

Most differences correspond to the coronal holes at the poles and to the edges of streamers. At the Northern coronal hole the non-equilibrium assumption decreases the relative charge states at the immediate vicinity of the limb in all three lines presented. Then as the radial distance from the limb increases, the ratio of emissions follows, and in case of the Si X and S X lines the non-equilibrium emission the charge states are overly abundant by 1.5-2 solar radii. The southern coronal hole shows similar behaviour with a more extended region of decreased emission due to the non-equilibrium assumption. At the boundaries of streamers we see an occasional enhancement due to non-equilibrium, especially pronounced on the east streamers' southern side. While at the center of the streamers the ratio is close to unity, the boundaries are visible on the ratio images in case of each line, from the footpoint out and beyond 2 solar radii. The behavior of the four different lines corresponding to 2 different elements show the complexity the NEI brings into diagnostics: all four line formation are close to $\log{T} = 6.2~K$ yet the non-equilibrium effects are vastly different.

Departures from equilibrium are most evident in limb spectra, thanks to the longer line of sight, the largest range of heights from the photosphere which contribute to the observed emission, and the lower electron density. On the disk, where the emission is dominated by plasmas located at the lowest heights, differences are much more limited with the only exception of the north coronal hole, where the intensity is lower than equilibrium in all ions. The reason for this behavior is the larger electron density of the emitting plasma, which reduces the departures from equilibrium. Si~X, XI and S~X behave in a very similar way, with localized, structure-dependent differences ranging within 10\% of the equilibrium values. The only exception is Si~XII, whose disk emission is enhanced in more locations, though differences are never larger than 20-30\%. 


It is also important to stress that the north and south coronal holes, though undergoing qualitatively the same process, show different degrees of departures from equilibrium, the south pole being closer to equilibrium than the north one. This is a clear reminder than even if the process behind these departure is the same for both regions, the individual plasma properties in each hole yield very different results: no one-box-fits-all set of empirical corrections can be adopted to account for non-equilibrium effects, but a complete 3D solution of the charge state evolution is necessary.

\subsubsection{Upper transition region to quiet Sun coronal ions}

\begin{figure}[htb!]
\includegraphics[trim={4cm 0.cm 1cm 4cm},clip,width=9cm]{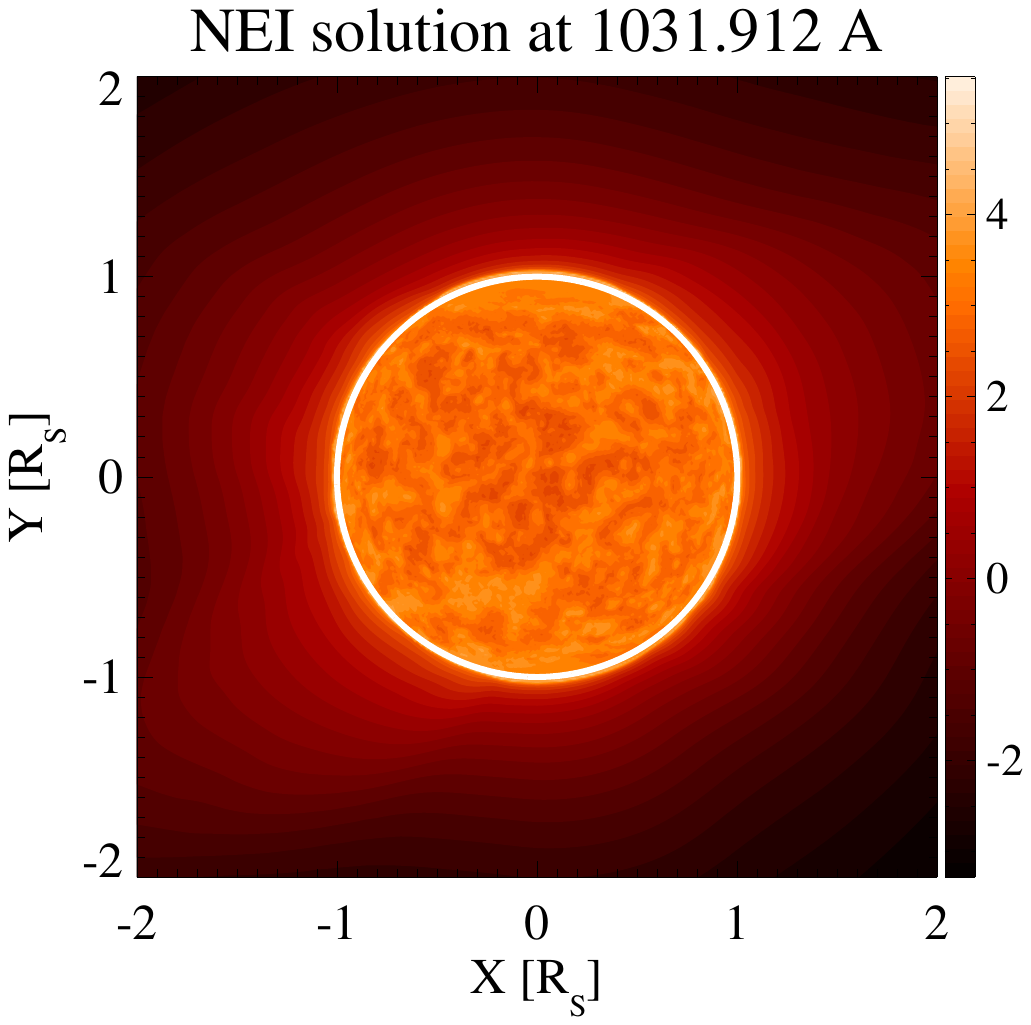}
\includegraphics[trim={4cm 0.cm 1cm 4cm},clip,width=9cm]{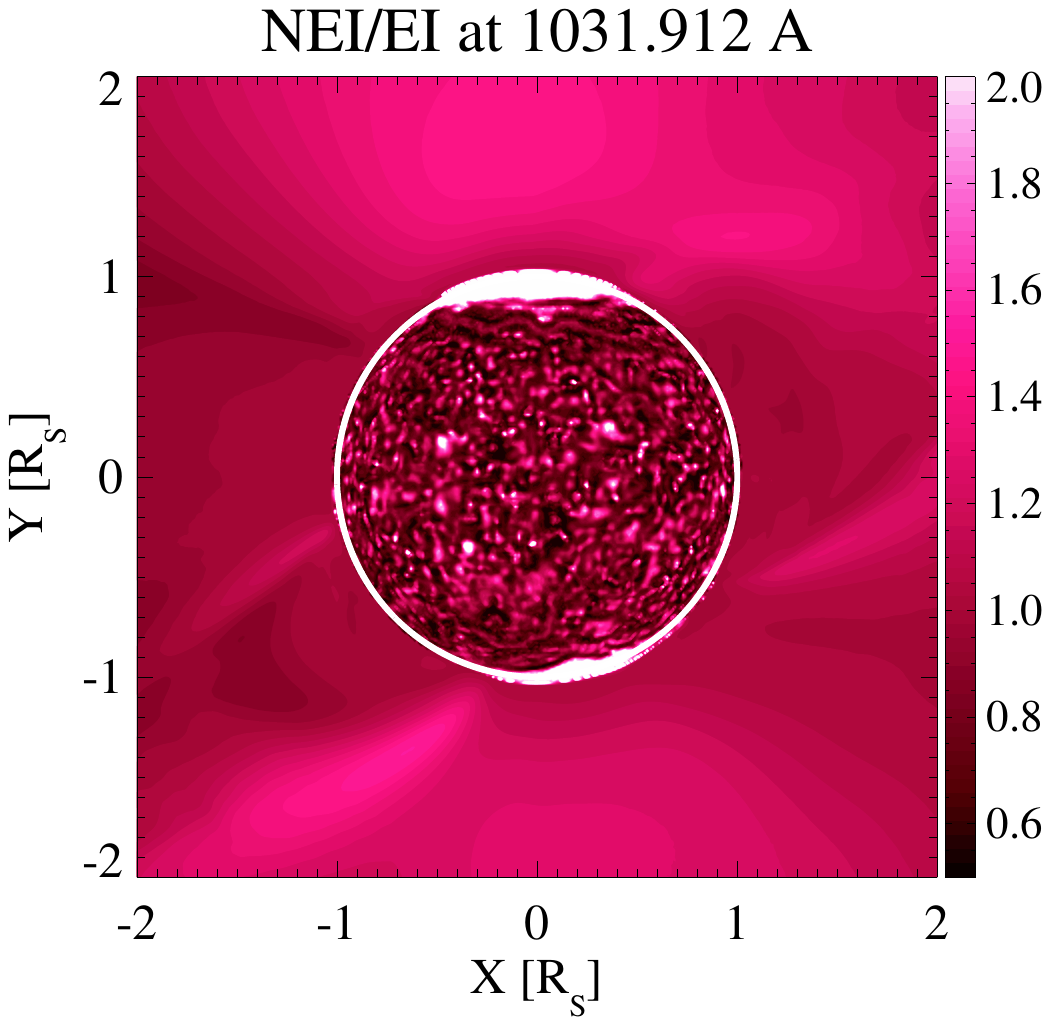}\\
\includegraphics[trim={4cm 0.cm 1cm 4cm},clip,width=9cm]{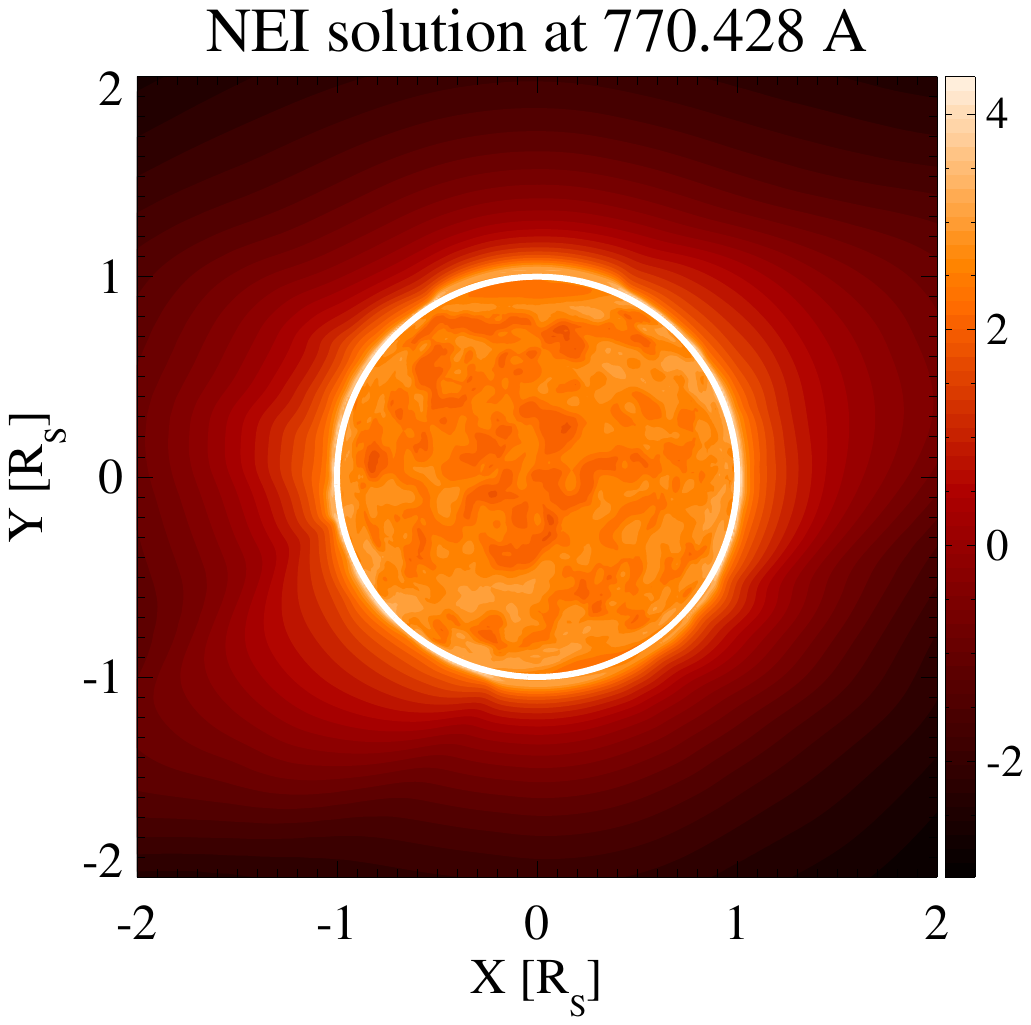}
\includegraphics[trim={4cm 0.cm 1cm 4cm},clip,width=9cm]{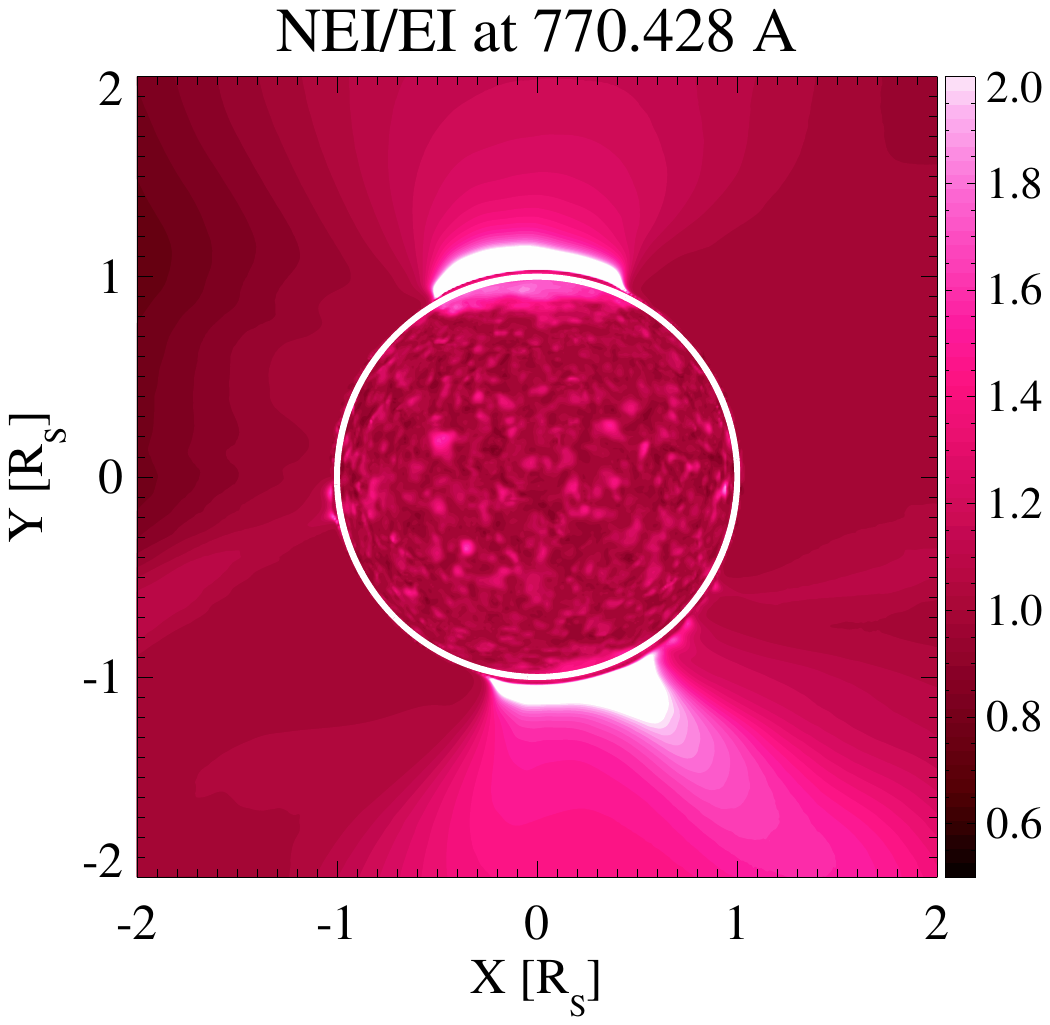}
\caption{Plasma emission in terms of the non-equilibrium/equilibrium ratio for ions (\emph{top to bottom}): 
O~VI (1031.912~\AA, $\log{T} = 5.5~K$)
and Ne~VIII (770.428~\AA, $\log{T} = 5.8~K$).
\label{fig:eqneq2}}
\end{figure}

Figure~\ref{fig:eqneq2} shows the non-equilibrium to equilibrium ratio in emission of lines corresponding to relatively low formation temperatures: O~VI (1031.912~\AA) at $\log{T} = 5.5~K$ and Ne~VIII (770.428~\AA) at $\log{T} = 5.8~K$. The low formation temperature reflects changes due to NEI in the low-corona and on-disk. Both these ions belong to the Li-like isoelectronic sequence and are characterized on one side by large ionization rates, and the other to be the ions immediately before He-like ions, which on the contrary have very low ionization rate coefficients, while maintaining high recombination rate coefficients. This creates the situation called "Cold effect" described by \citet{Landi:2012}, whereby a relatively light element (such as O and Ne, with few ionization stages available) is accelerated into the solar corona either along an open field line in a coronal hole, or a closed loop through a siphon flow. This element rapidly ionizes until the He-like stage (O~VII and Ne~IX in this case), whose low ionization rates cause it to increase its abundance almost to unity. In increasingly less dense plasmas, such ions almost stop ionizing, but still are capable of recombining; their very large relative abundance cause the recombined ions (O~VI and Ne~VIII) to increase their abundance relative to their equilibrium values. Such behavior is more extreme in coronal holes, where the emission of both Li-like ions is increased by large amounts; such increase is larger at lower heights (on the disk, and in the inner corona) for O~VI, due to the lower temperature of formation, while it lasts for a larger range of heights for Ne~VIII; such a difference is due to the fact that in order to reach the He-like stage Neon requires to be ionized through more stages than Oxygen. Such a difference is also evident on the disk, where the more localized intensity enhancements of O~VI are due to their occurring at lower heights than Ne~VIII, where the magnetic field of coronal structures has not yet expanded in the corona. It is worth noting that even on the disk, where the enhancements are smaller, the intensity of these important ions can easily change by 20-30\%.

\begin{figure}[htb!]
\includegraphics[trim={5cm 0.cm 3cm 4cm},clip,width=8cm]{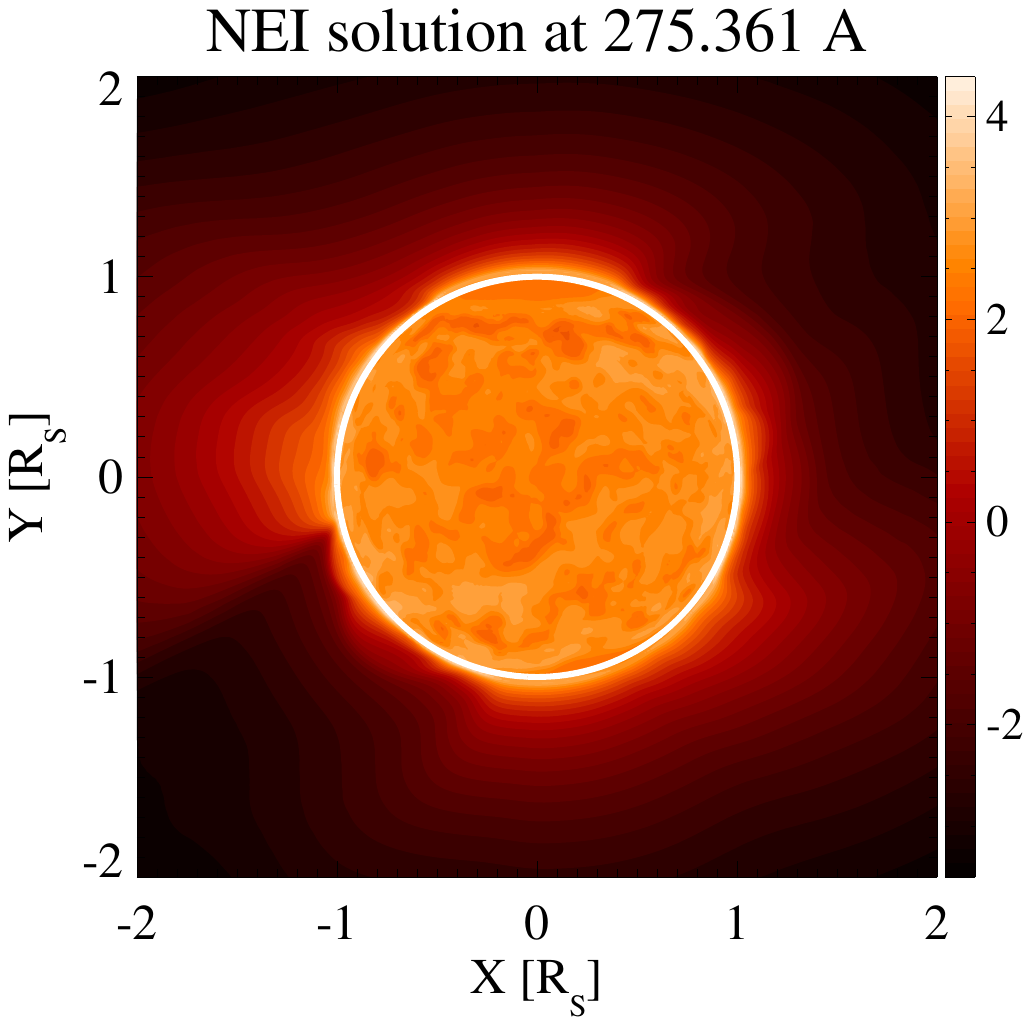}
\includegraphics[trim={5cm 0.cm 3cm 4cm},clip,width=8cm]{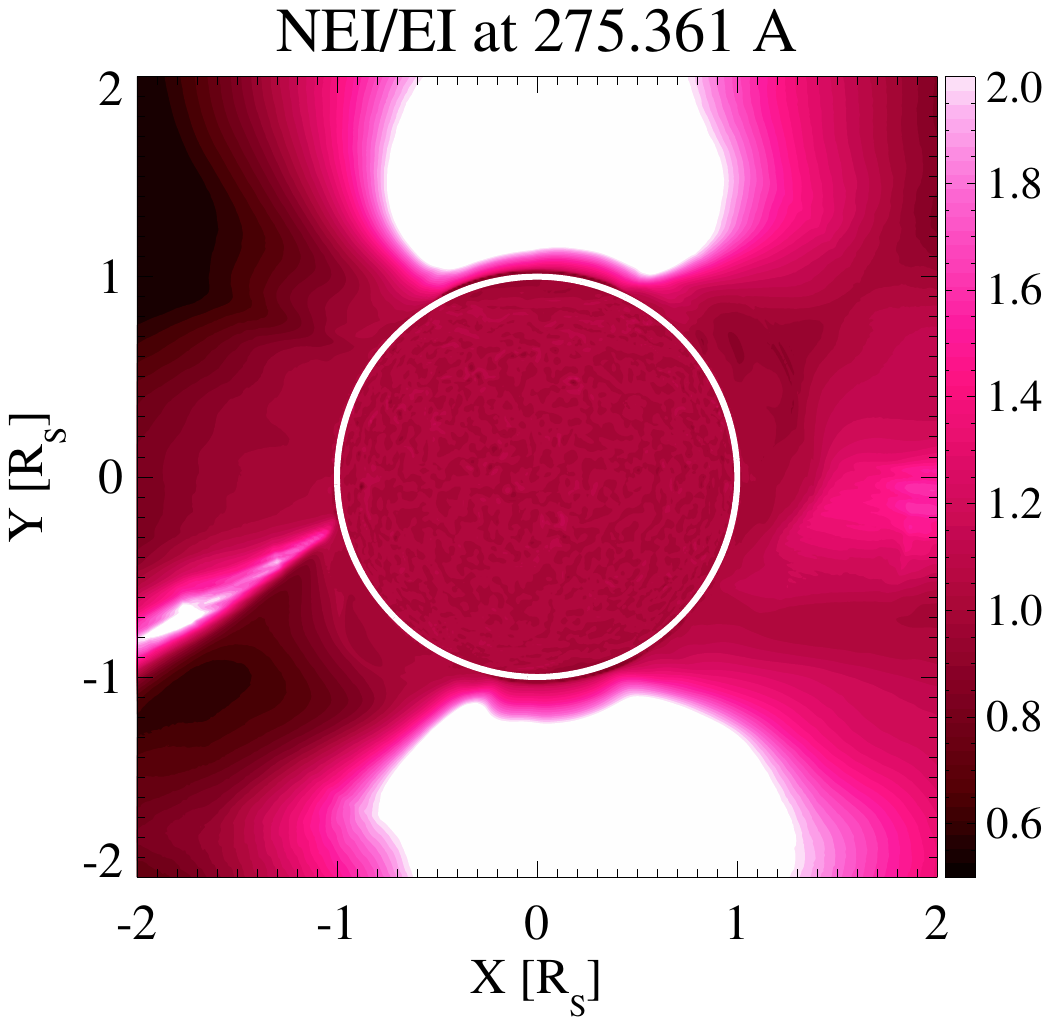}\\
\includegraphics[trim={5cm 0.cm 3cm 4cm},clip,width=8cm]{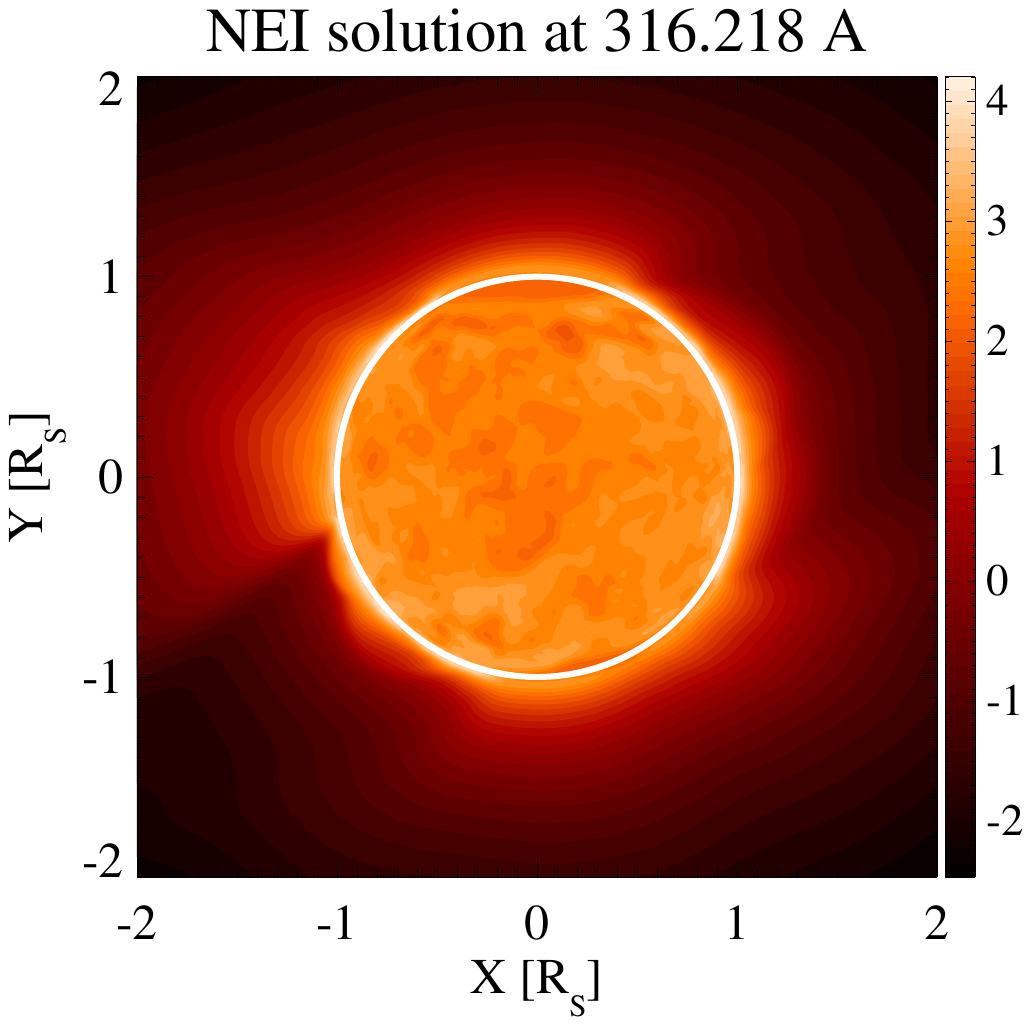}
\includegraphics[trim={5cm 0.cm 3cm 4cm},clip,width=8cm]{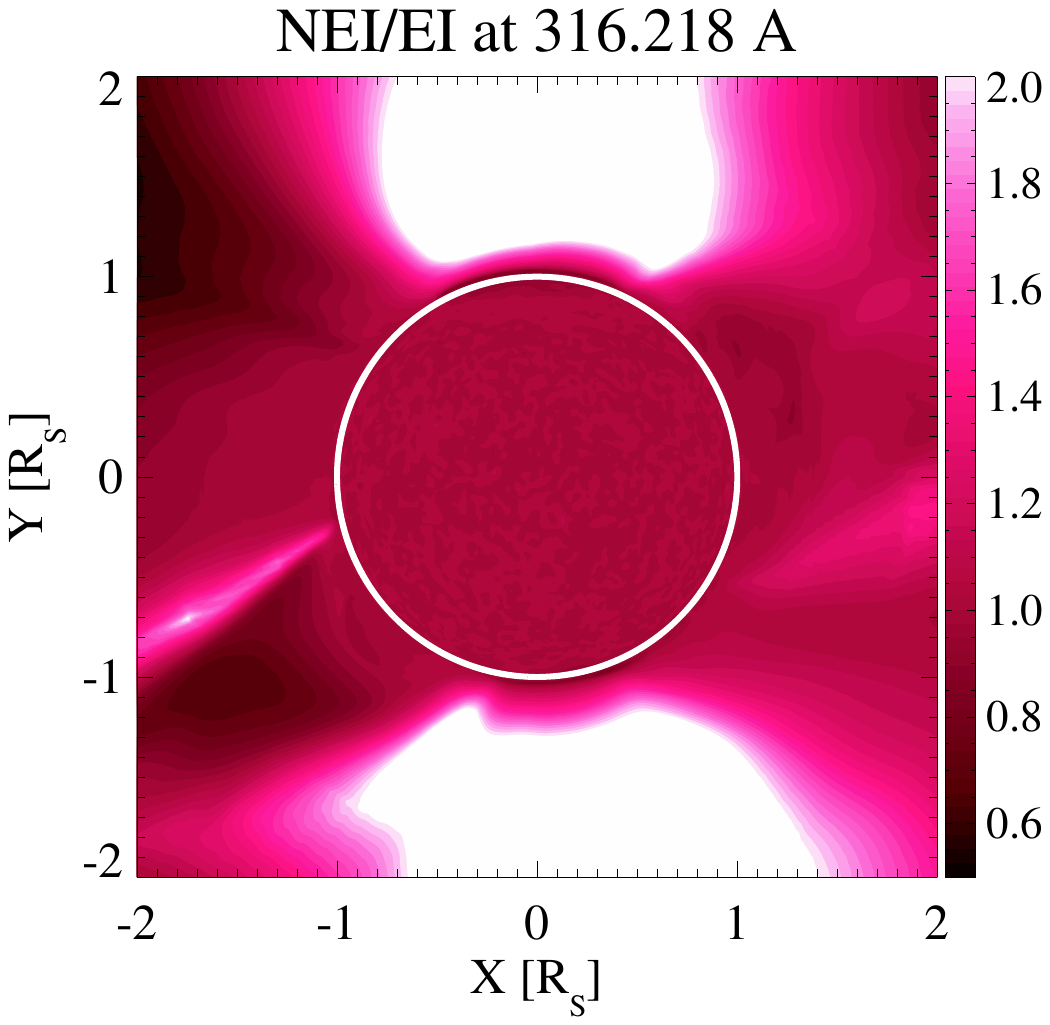}\\
\includegraphics[trim={5cm 0.cm 3cm 4cm},clip,width=8cm]{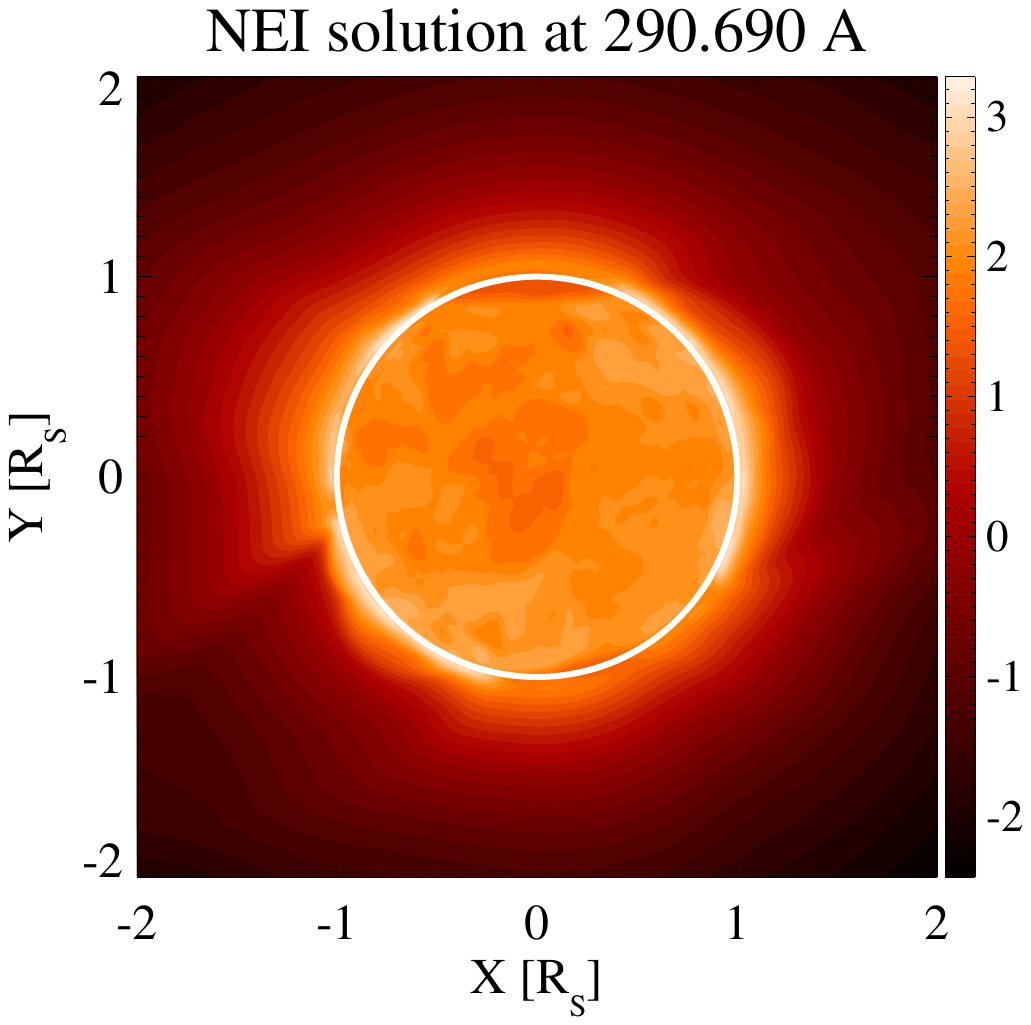}
\includegraphics[trim={5cm 0.cm 3cm 4cm},clip,width=8cm]{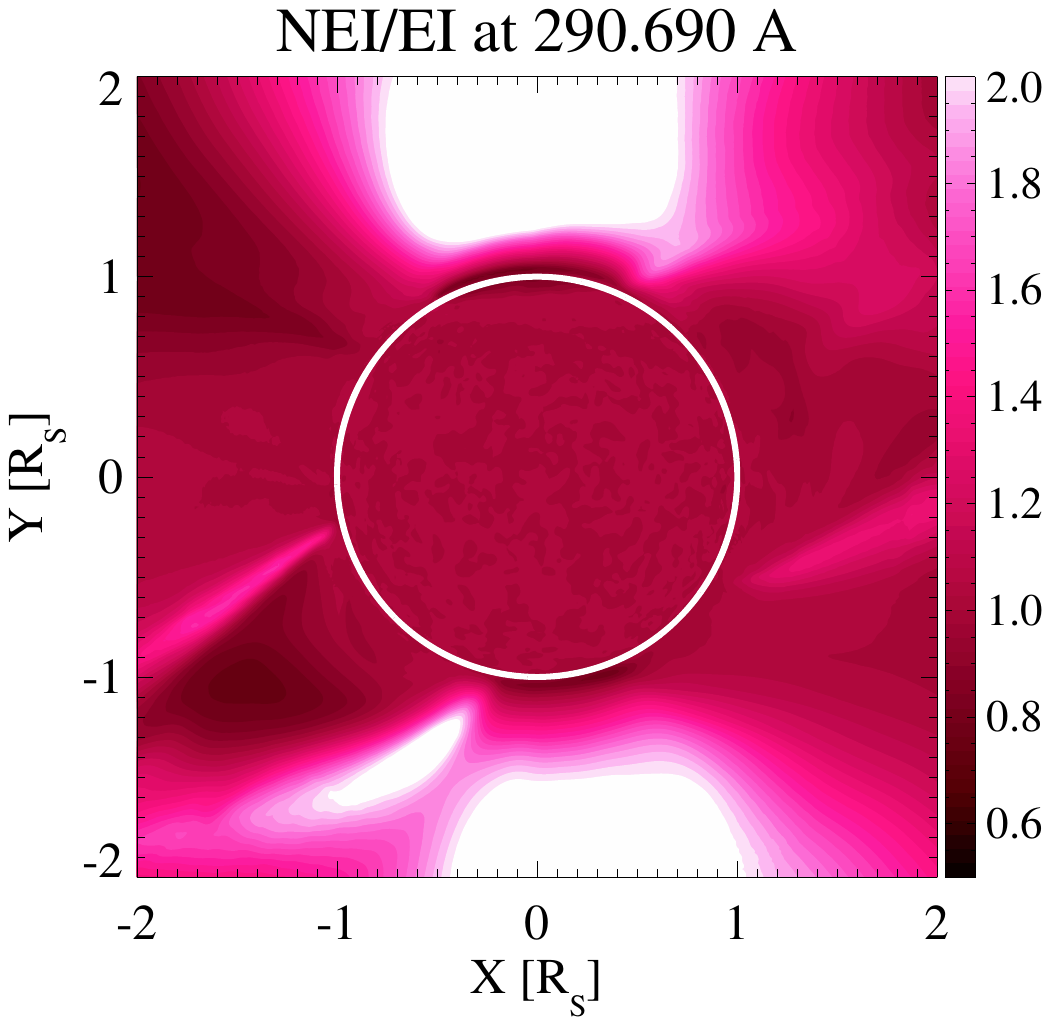}
\caption{Plasma emission in terms of the non-equilibrium/equilibrium ratio for ions (\emph{left to right}): 
Si~VII (275.361~\AA, $\log{T} = 5.8~K$), 
Si~VIII (316.218~\AA,  $\log{T} = 5.9~K$), 
Si~IX (290.690~\AA, $\log{T} = 6.1~K$), 
\label{fig:si}}
\end{figure}

\begin{figure}[htb!]
\includegraphics[trim={5cm 0.cm 3cm 4cm},clip,width=9cm]{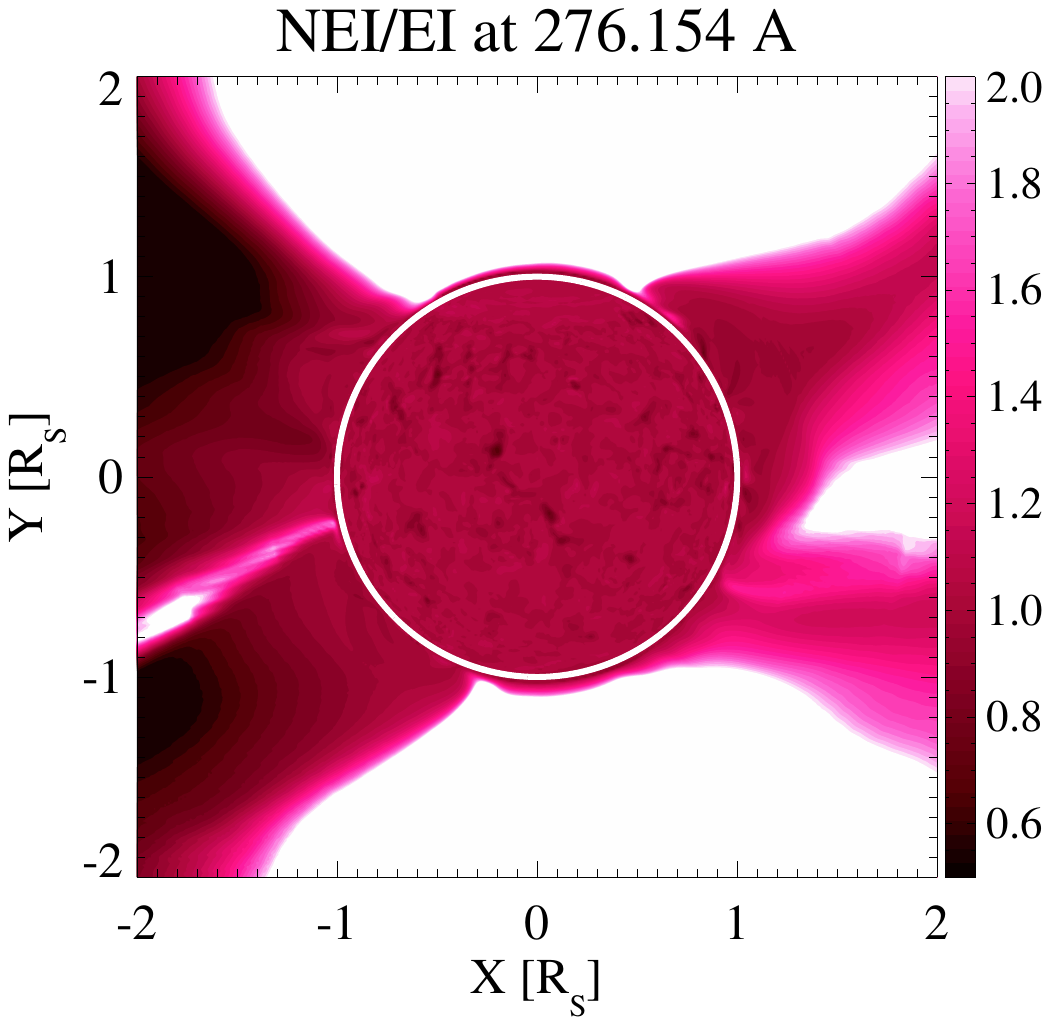}
\includegraphics[trim={5cm 0.cm 3cm 4cm},clip,width=9cm]{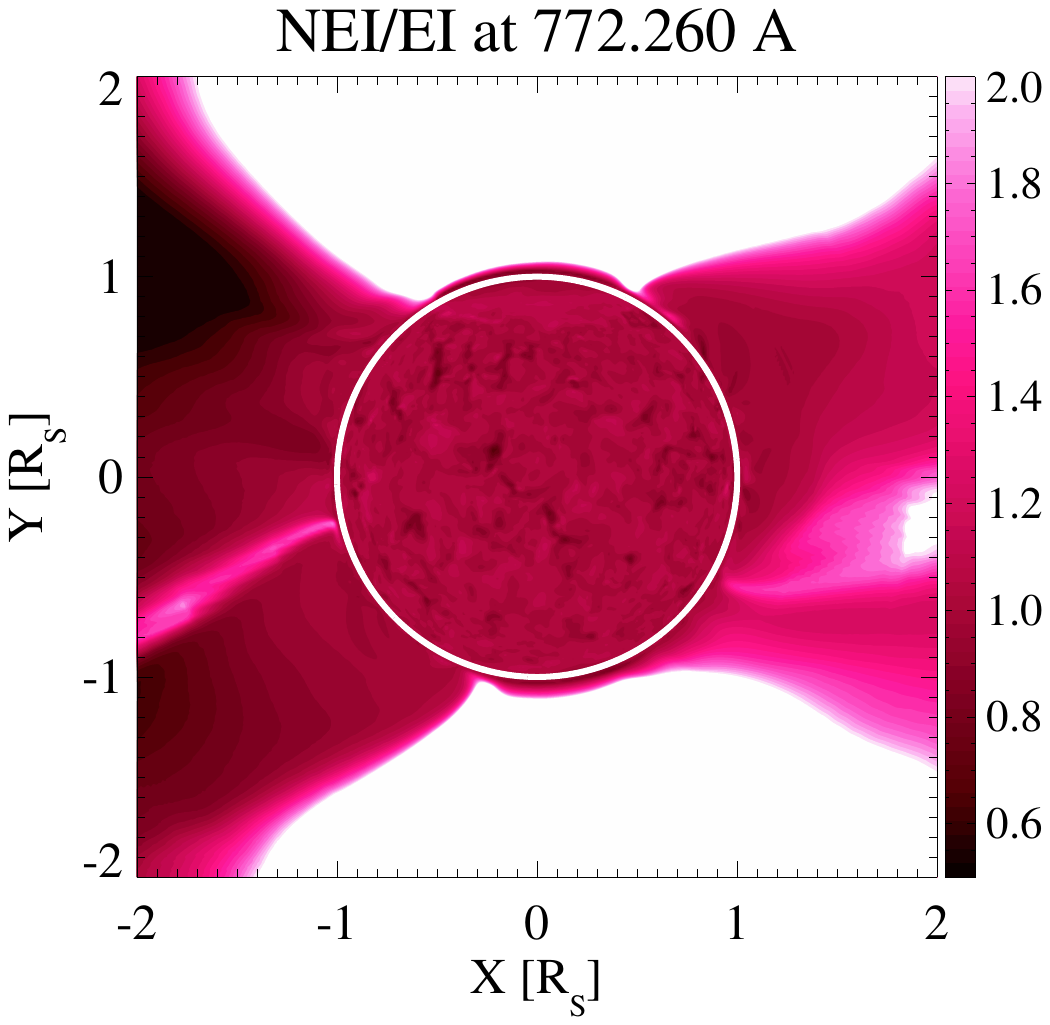}\\
\includegraphics[trim={5cm 0.cm 3cm 4cm},clip,width=9cm]{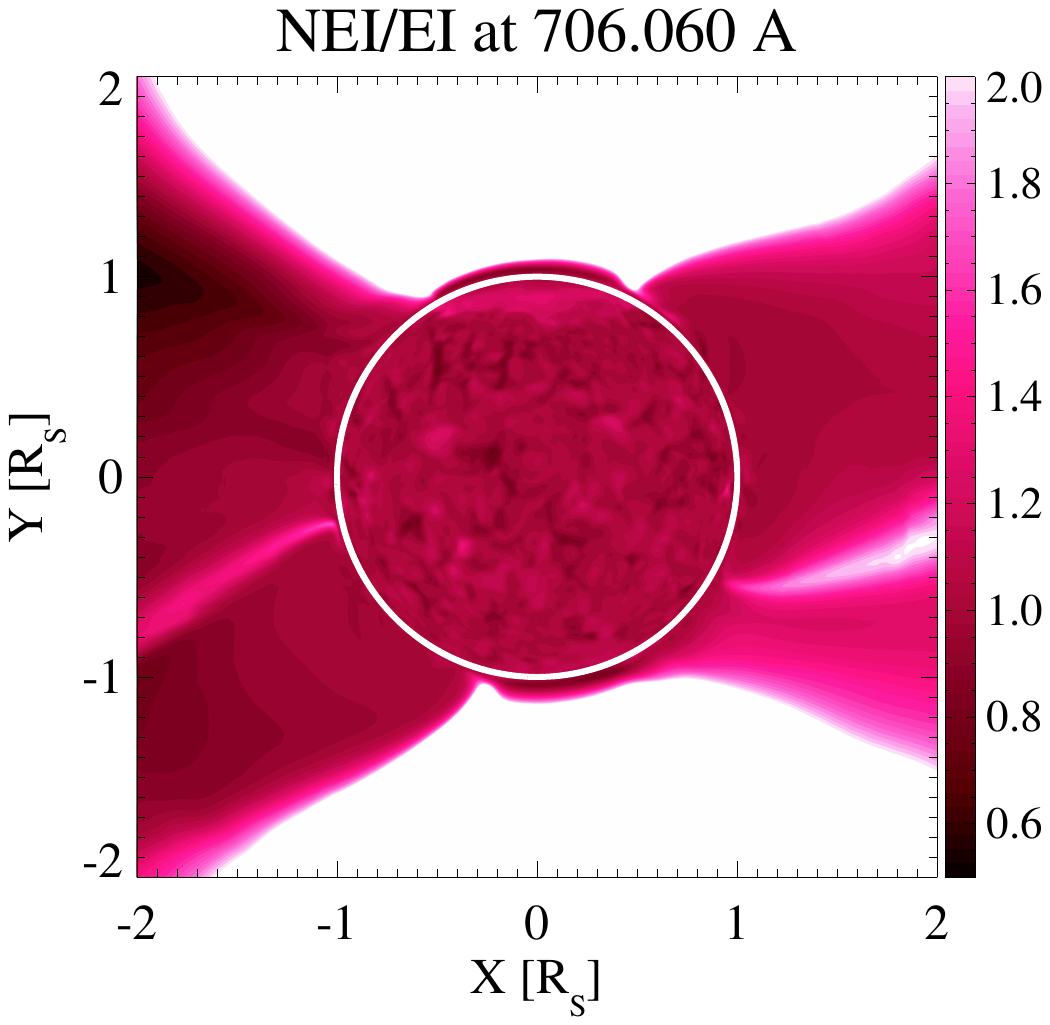}
\includegraphics[trim={5cm 0.cm 3cm 4cm},clip,width=9cm]{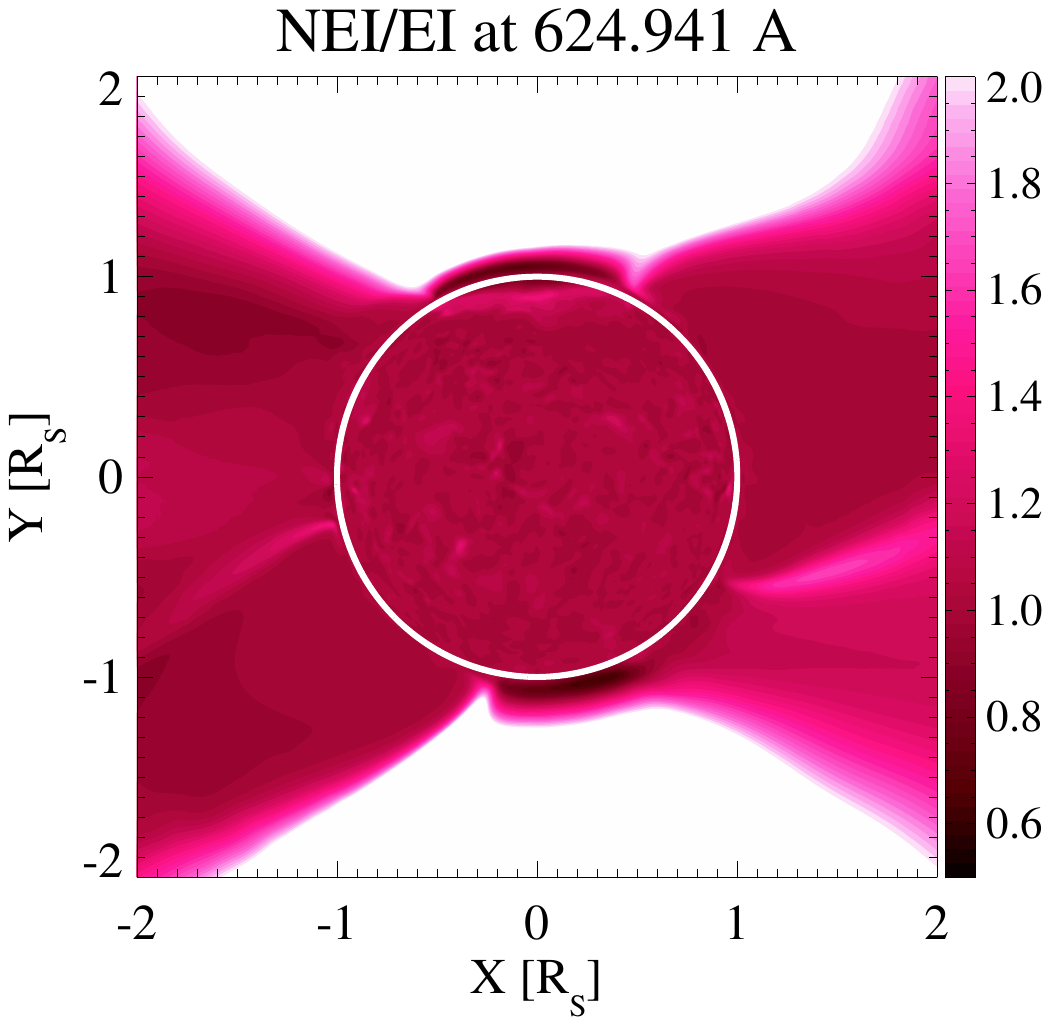}
\caption{Plasma emission in terms of the non-equilibrium/equilibrium ratio for ions (\emph{left to right}): 
Mg~VII (276.154~\AA, $\log{T} = 5.8~K$), 
Mg~VIII (772.260~\AA, $\log{T} = 5.9~K$), 
Mg~IX (706.060~\AA,  $\log{T} = 6.0~K$), 
Mg~X (624.941~\AA, $\log{T} = 6.1~K$). 
\label{fig:mg}}
\end{figure}

Figures~\ref{fig:si} and \ref{fig:mg} show the Si and Mg sequences of low-corona ions Si~VII-IX and Mg~VII-X. As expected, differences are largest in coronal holes because of the low density and the plasma speed; also, they are larger than for O~VI and Ne~VIII, and show the typical "Delay effect" behavior with an initial ion abundance underestimation and a subsequent overestimation over the equilibrium values. As the ion charge increase, these effects take place at an increasingly large distance; the ratio values indicate that departures from equilibrium are significant in height ranges typically covered by past, current and planned EUV high resolution spectrometers. Again, each coronal hole shows a different spatial distribution of the amount of change, indicating that every structure needs to be modeled independently. Differences are smaller in closed-field structures, but still significant at streamer legs (where lower ionization stages show the largest differences) and in the NE structure.

\subsubsection{Iron ions} 

Ions belonging to Fe emit a large number of lines in the EUV and soft X-ray wavelength range, which constitute the bulk of the radiative losses in the solar corona, and provide a very large number of diagnostic tools to study the solar corona at all temperatures and activity state. Also, they provide the bulk of the emission observed by narrow-band EUV imaging instruments, so that their departures from equilibrium are likely to affect the analysis of the images provided by these instruments.

\begin{figure}[htb!]
\includegraphics[trim={4.5cm 0.cm 1.5cm 4cm},clip,width=7.8cm]{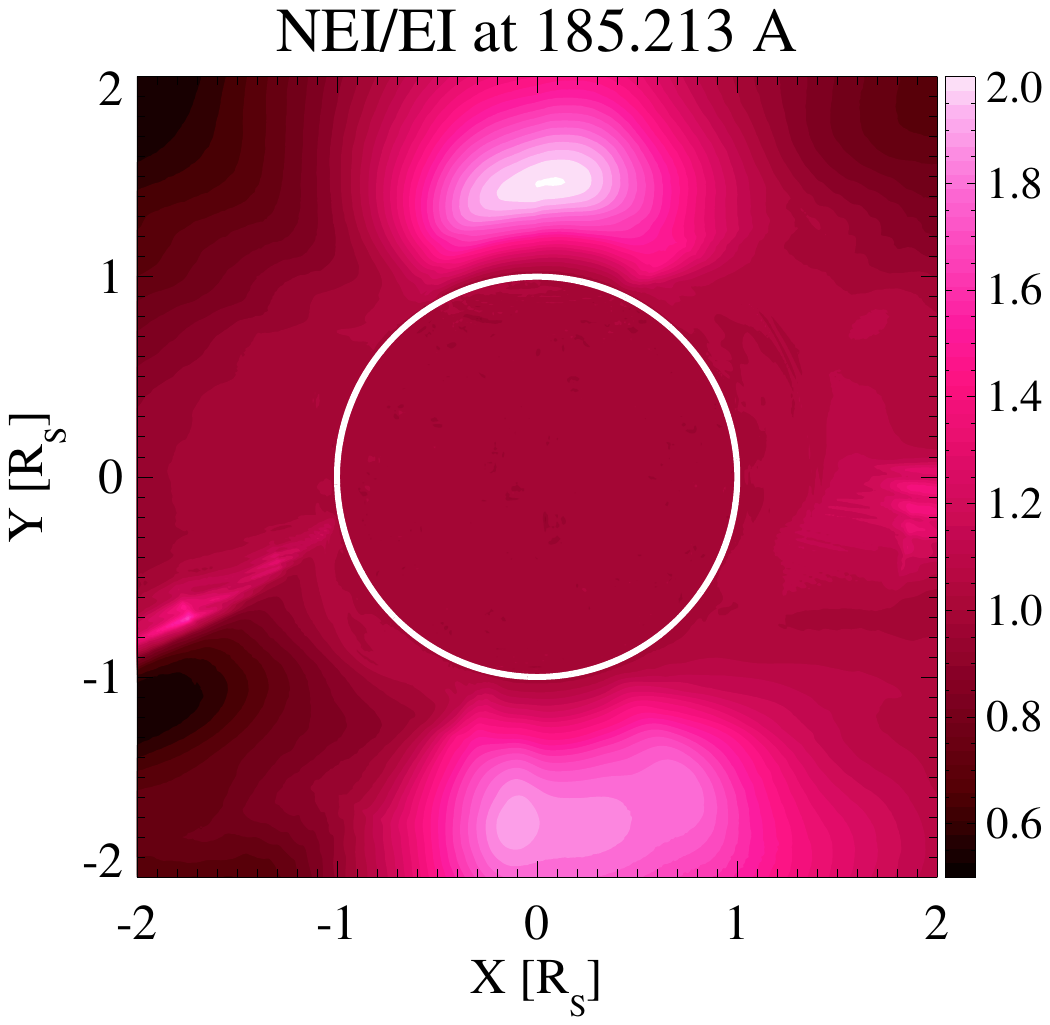}
\includegraphics[trim={4.5cm 0.cm 1.5cm 4cm},clip,width=7.8cm]{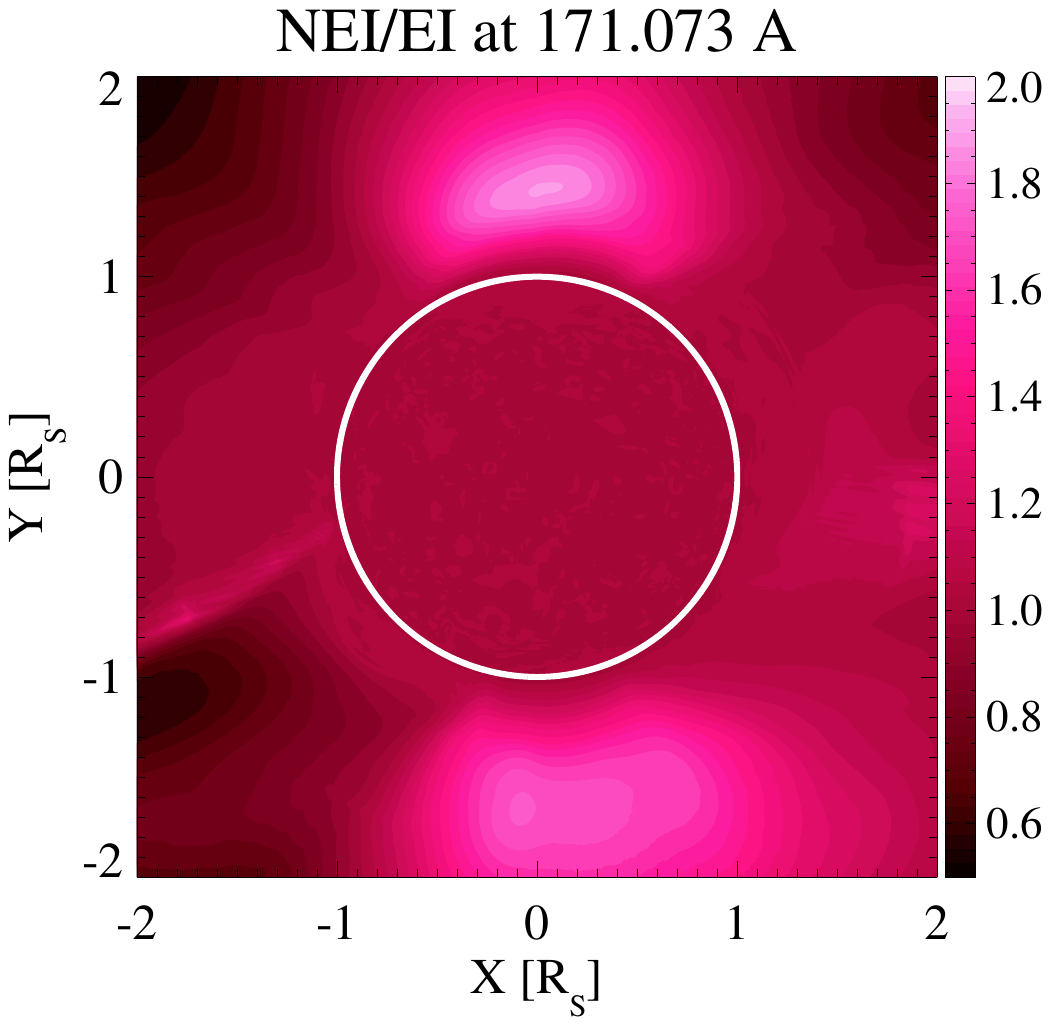}\\
\includegraphics[trim={4.5cm 0.cm 1.5cm 4cm},clip,width=7.8cm]{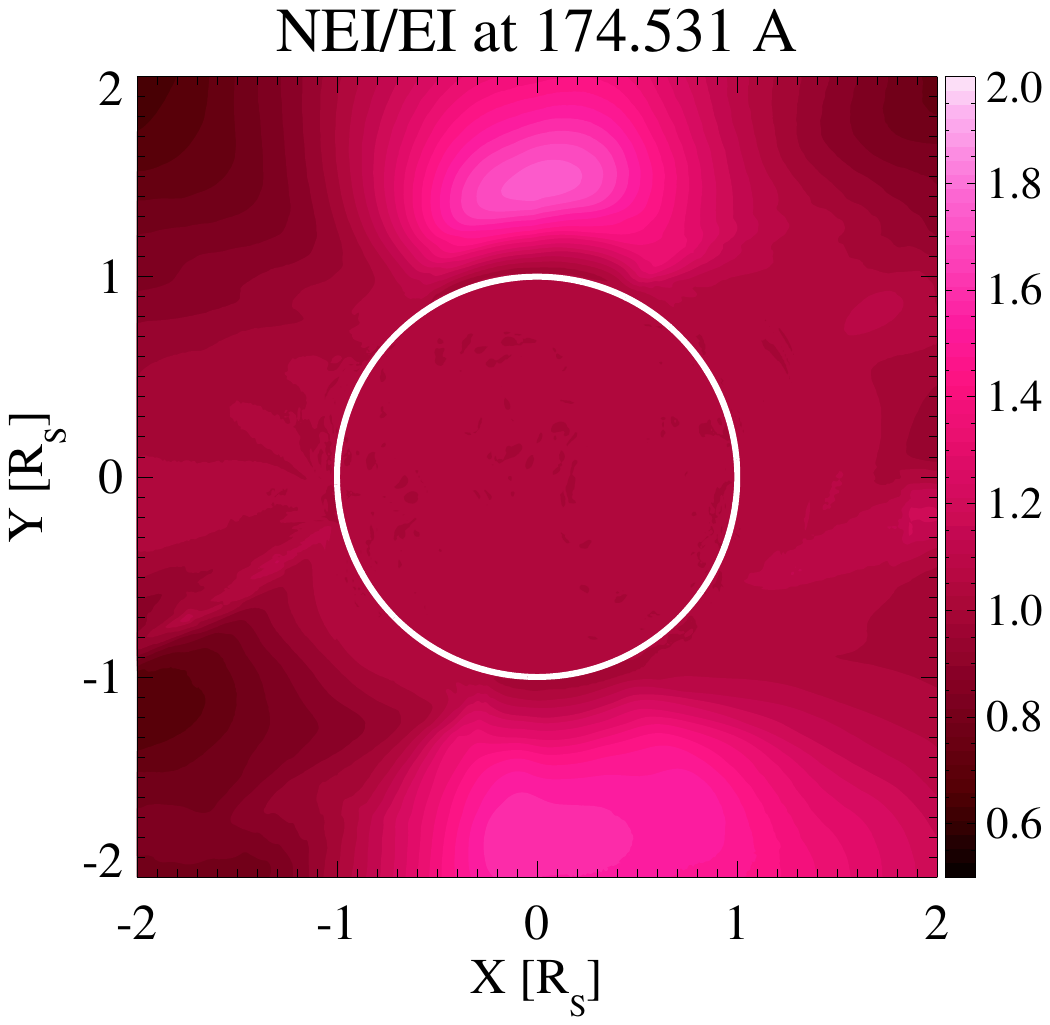}
\includegraphics[trim={4.5cm 0.cm 1.5cm 4cm},clip,width=7.8cm]{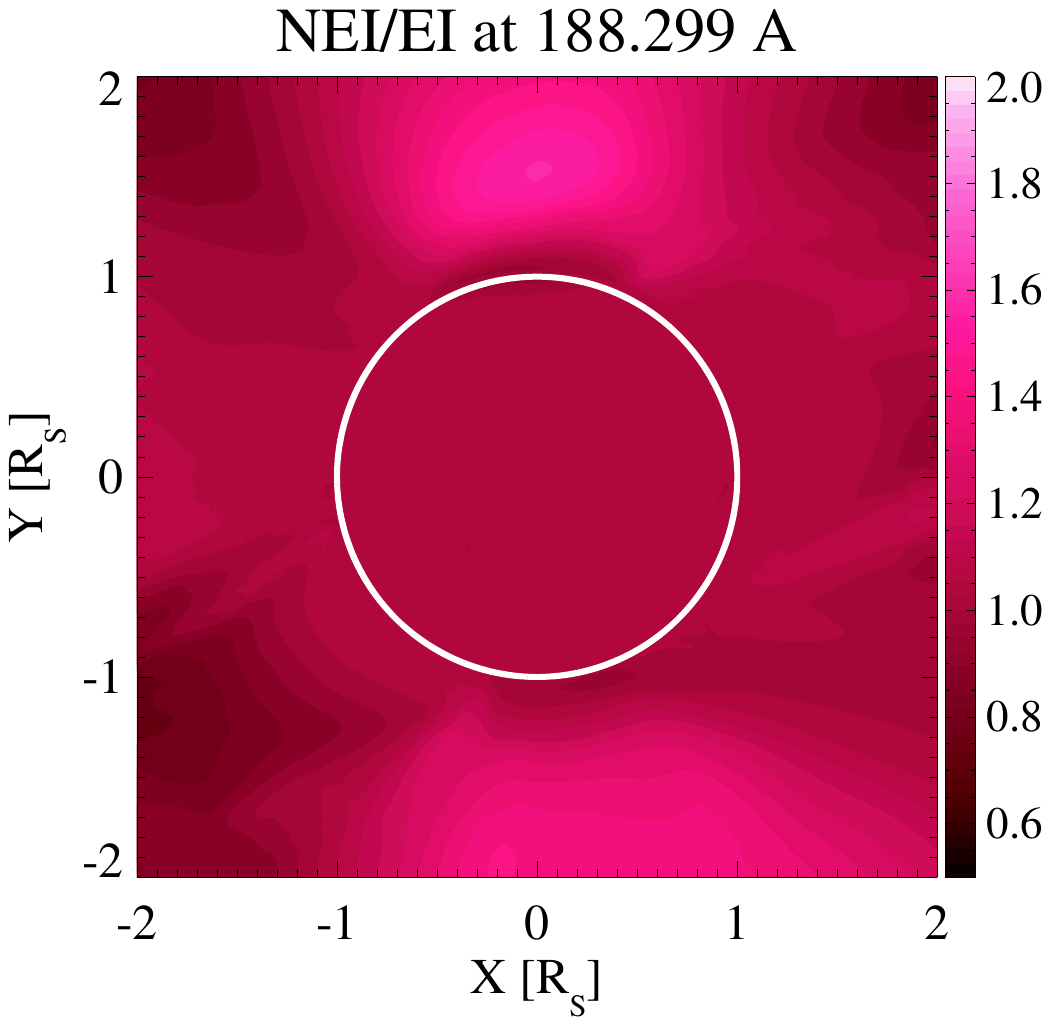}\\
\includegraphics[trim={4.5cm 0.cm 1.5cm 4cm},clip,width=7.8cm]{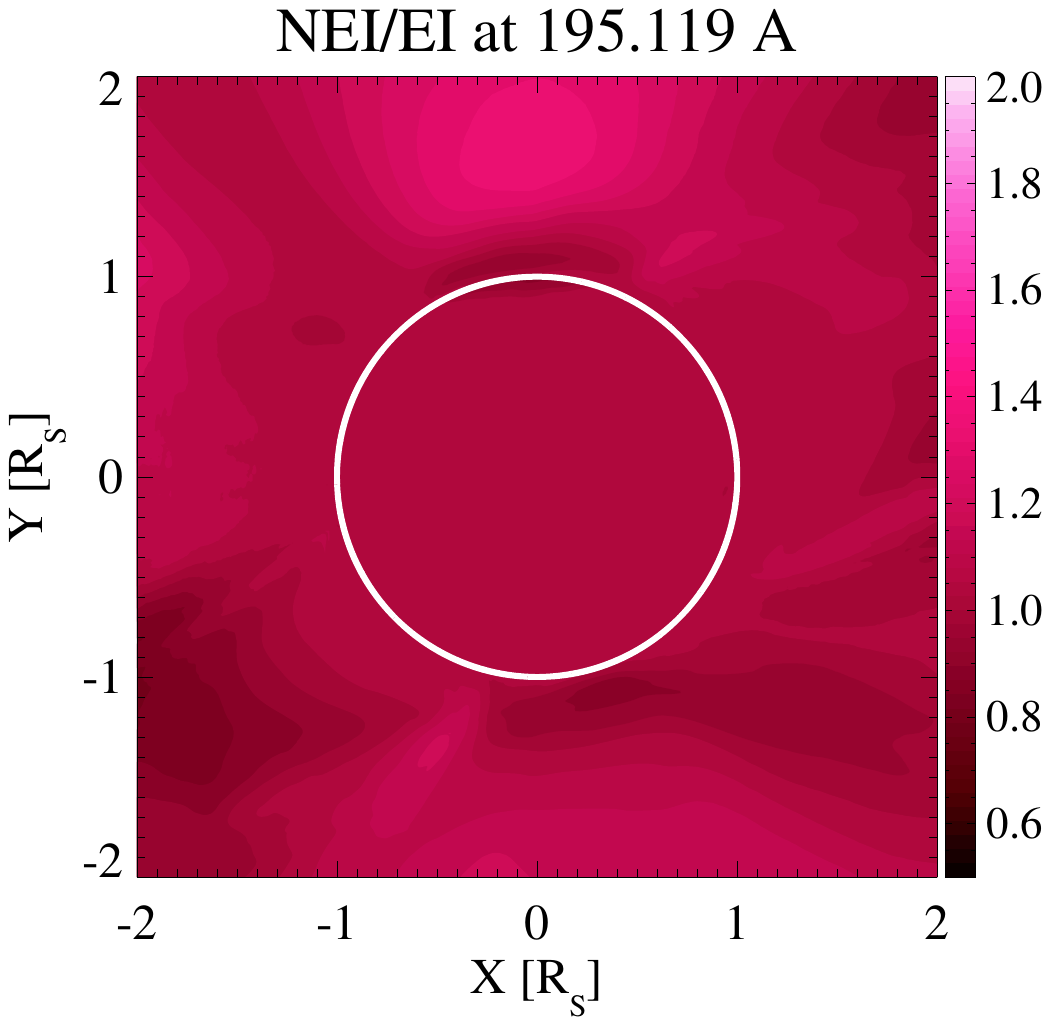}
\caption{Plasma emission ratios of plasma assuming NEI versus equilibrium ionization for iron ions Fe~VIII to Fe~XII. The peak formation temperatures for the lines presented in this figure can be found in table~\ref{tbl:lines}. 
\label{fig:Fes1}}
\end{figure}

\begin{figure}[htb!]
\includegraphics[trim={4.5cm 0.cm 1.5cm 4cm},clip,width=9cm]{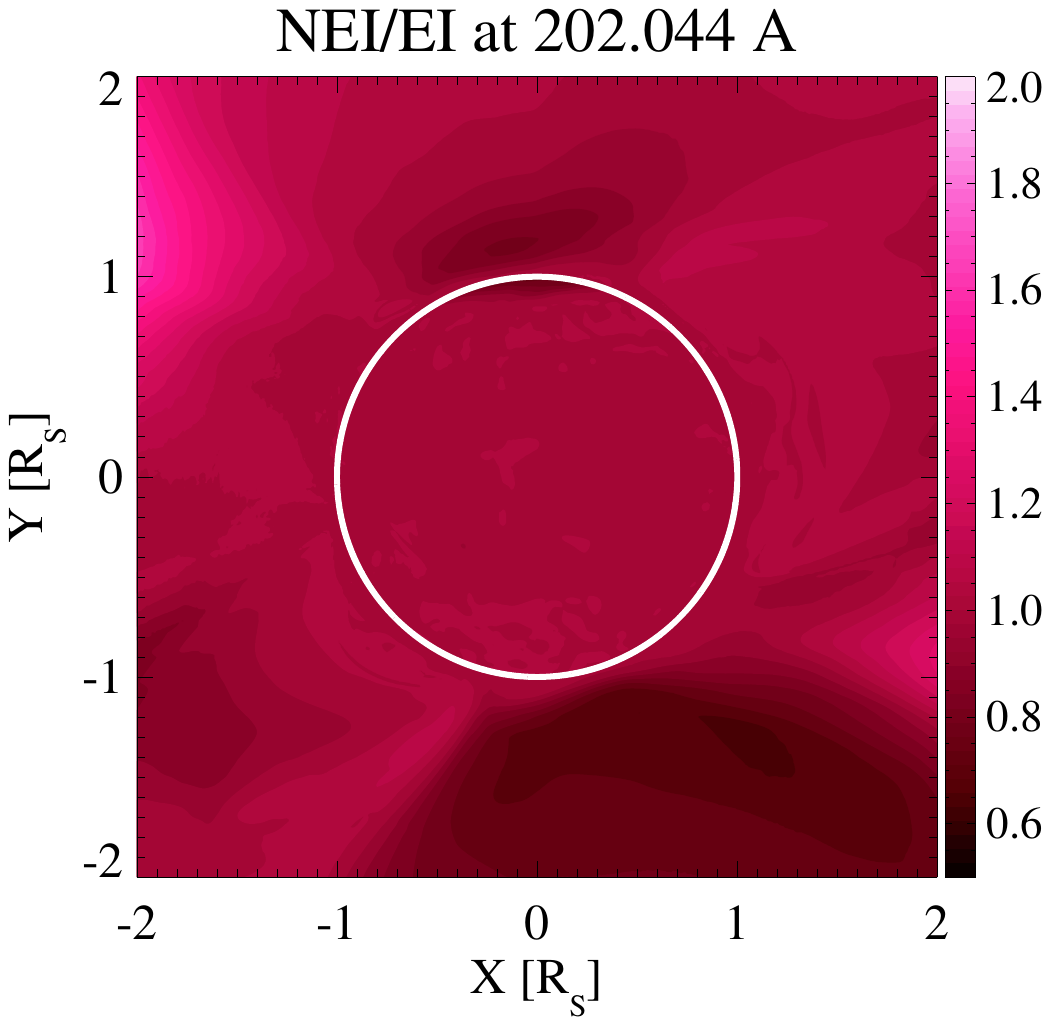}
\includegraphics[trim={4.5cm 0.cm 1.5cm 4cm},clip,width=9cm]{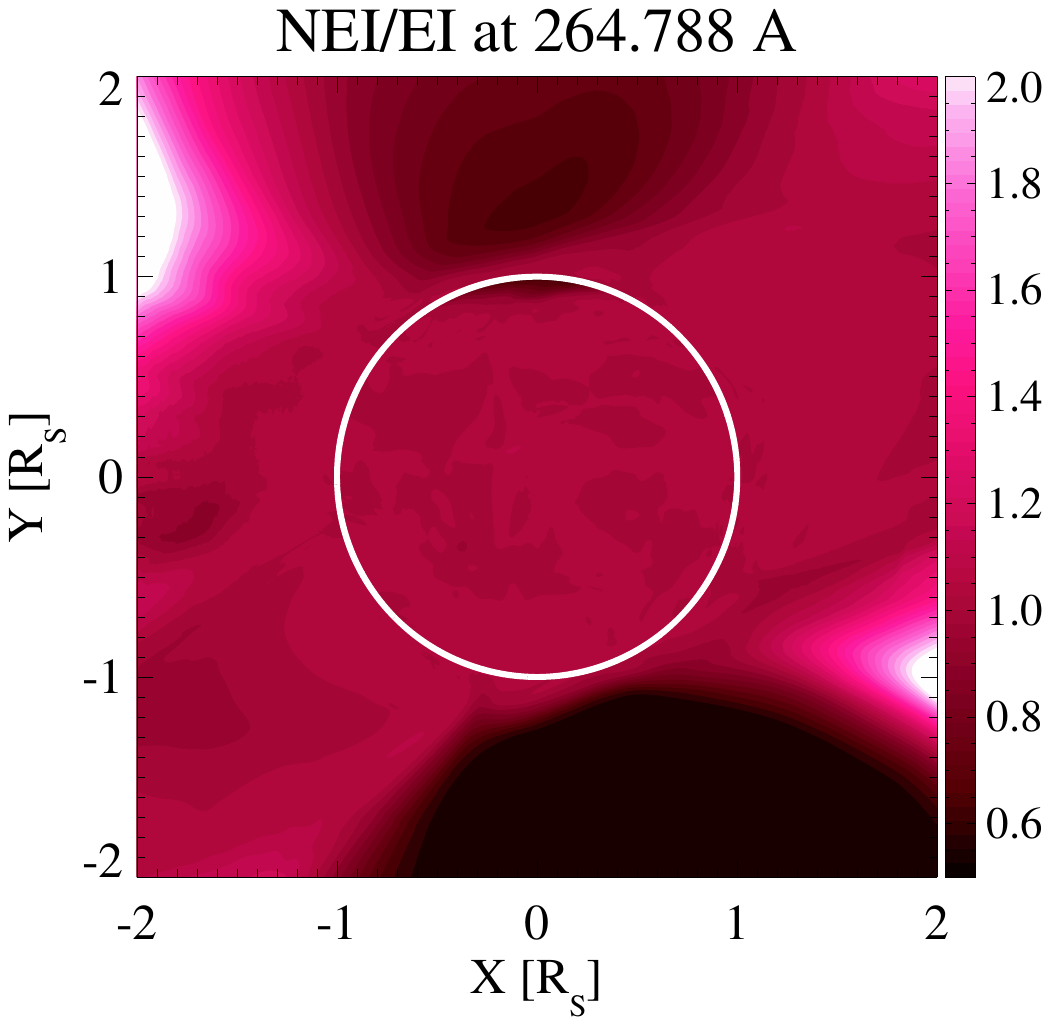}\\
\includegraphics[trim={4.5cm 0.cm 1.5cm 4cm},clip,width=9cm]{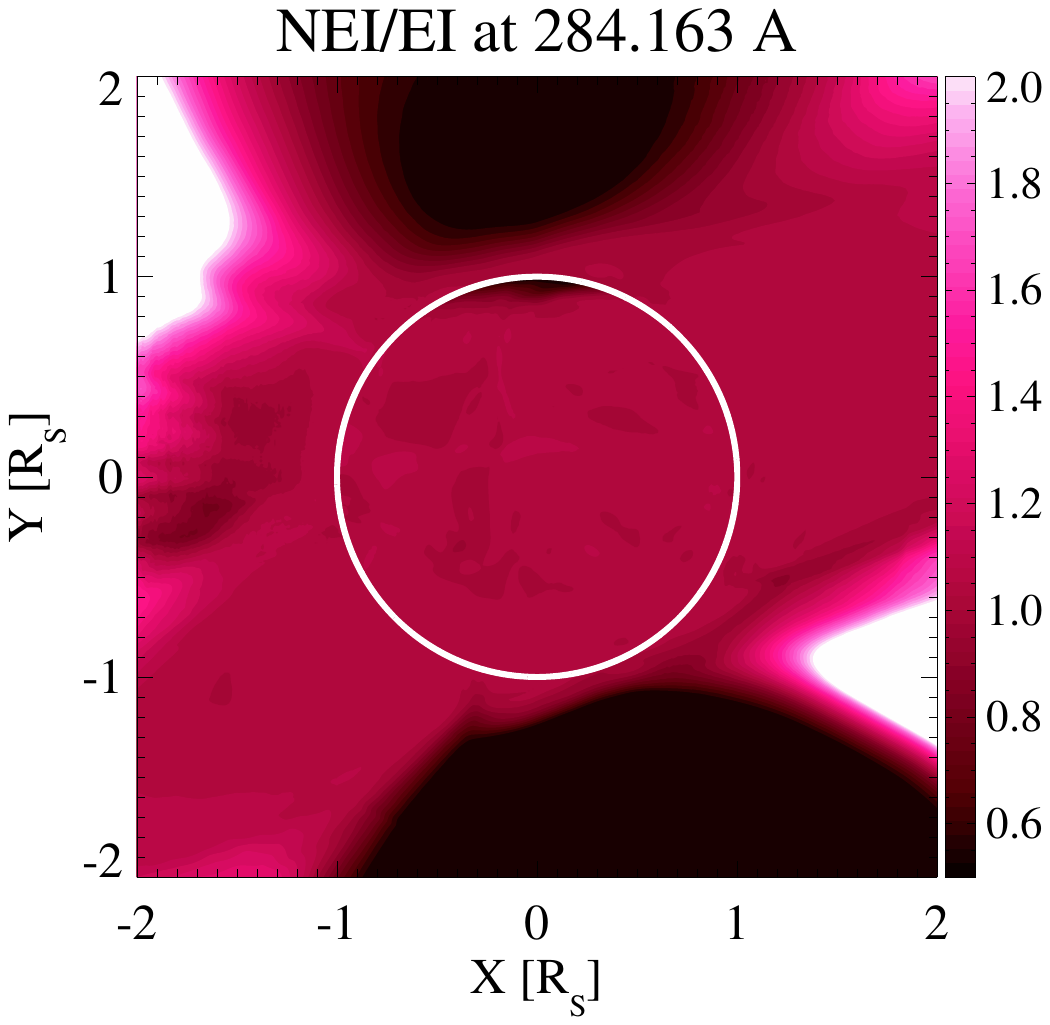}
\includegraphics[trim={4.5cm 0.cm 1.5cm 4cm},clip,width=9cm]{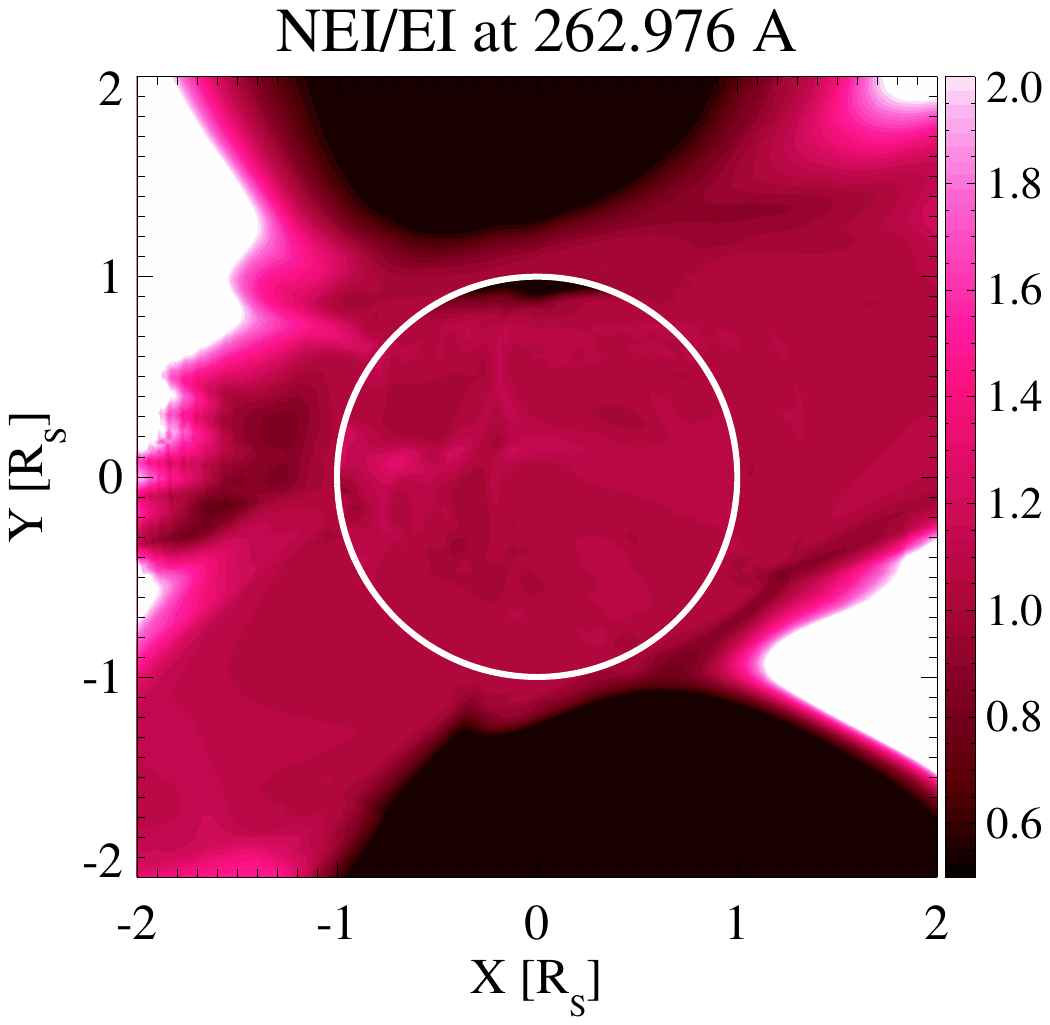}
\caption{Plasma emission ratios of plasma assuming NEI versus equilibrium ionization for iron ions Fe~XIII to Fe~XVI. The peak forming temperatures for the lines presented in this figure can be found in Table~\ref{tbl:lines}. 
\label{fig:Fes2}}
\end{figure}

Figures~\ref{fig:Fes1} and \ref{fig:Fes2} show the sequence of iron ions from Fe~VIII to Fe~XVI in two groups and how the different iron ions' emission changes due to NEI in the corona. Ions beyond Fe~XVI, though present in the simulation, are too weak to provide significant emission in the quiescent corona simulated in this work, so we have not displayed them and will not discuss them any further.

Again, non-equilibrium effects are largest in the two polar coronal holes, due to the mix of low density and wind speed. The largest increases in the emission of the low-ionization species such as Fe~VIII to Fe~X) takes place above 1.4 solar radii, reaching and exceeding a factor two; at lower heights, however, they still provide significant intensity enhancement capable of affecting line intensity analyses. As the ions' charge increases, the polar holes indicate increasingly smaller intensity ratios, indicating severe depletion in the abundance of these ions in line with the "Delay Effect"; for these ions, available observations of intensities of ions like Fe~XIII to XV, routinely reported by EUV spectrometers, can indicate the presence of two possible effects: 1) a larger presence of instrument-scattered light, or 2) inaccuracies in the modeled speed and electron temperature and density of the nascent solar wind. In case instrument-scattered light is properly corrected, the residual emission of these ions can provide and extremely important  tools to infer these wind properties, and improve solar wind models.

Differences in streamer emission are smaller at the heights normally covered by available EUV spectrometers. This is due to the streamers' higher density, and to higher ionization rates of Fe ions when compared to other elements; in this case, EUV spectral diagnostics seems to be more robust against non-equilibrium effects. At distances larger than 2 solar radii, however, the Delay Effect seems to be more important.

The structure at the NE limb can be studied more effectively with Fe than with Mg and Si thanks to the availability of more ionization stages sampling a larger temperature range. While with Mg and Si the NE structure only showed a decrease in emission, the higher stages of ionization from Fe allow to see that this region increases the intensity for the higher ions (Fe~XIII and beyond) by values which {\em increase} with height, providing an opposite effect for the Delay Effect, due to the temperature drop in that region.
A second, similar feature is visible thanks to the higher Fe ionization stages in the SW quadrant of the solar corona, whose presence was not visible with the other elements. 

\begin{figure}[htb!]
\includegraphics[trim={4cm 0.cm 3.8cm 4.cm},clip,width=6cm]{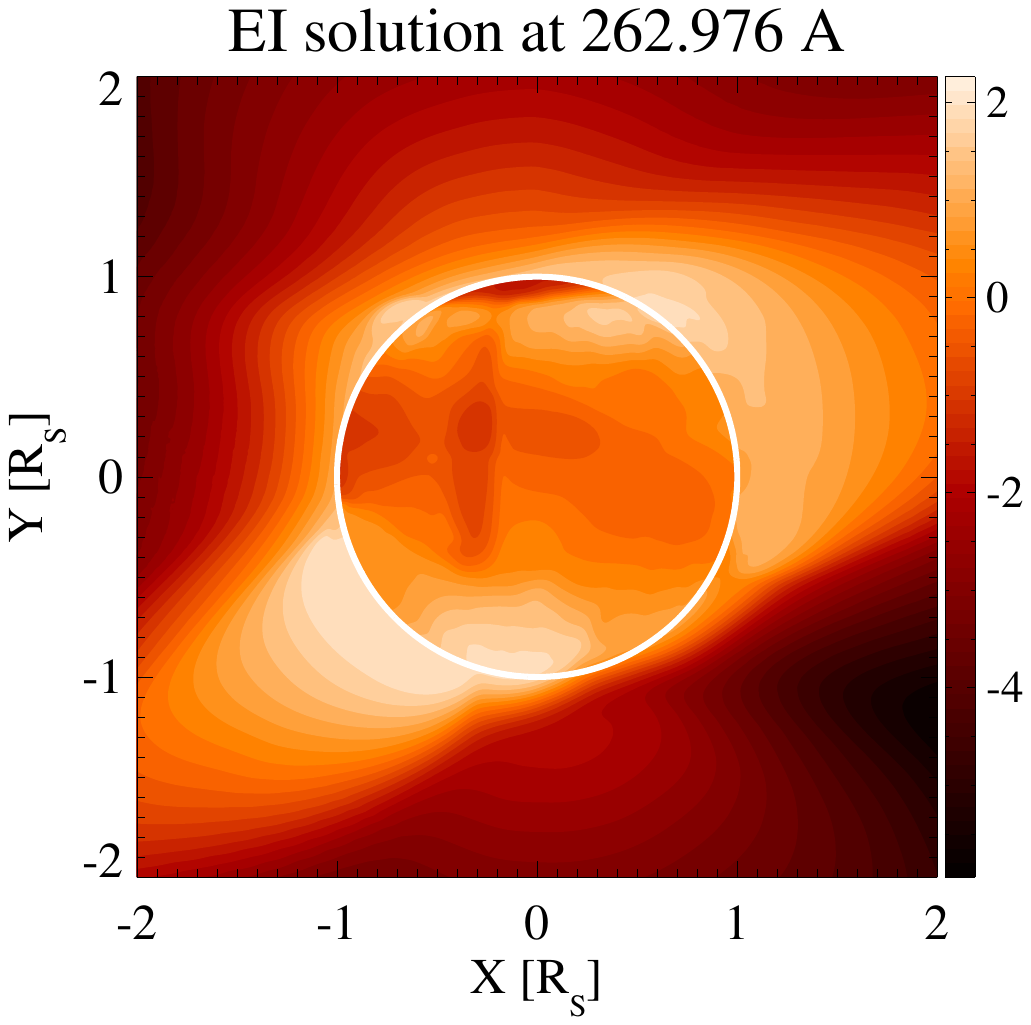}
\includegraphics[trim={4cm 0.cm 3.8cm 4.cm},clip,width=6cm]{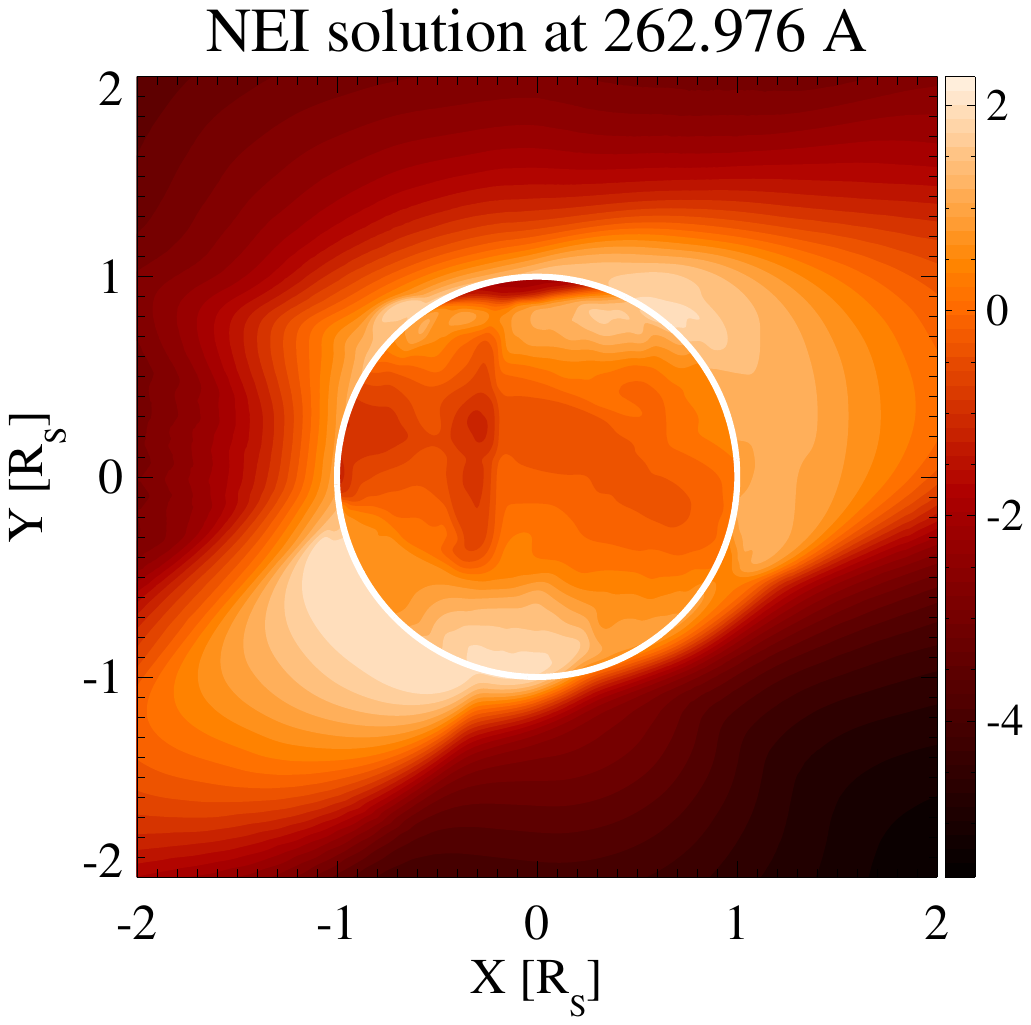}
\includegraphics[trim={4cm 0.cm 3.8cm 4.cm},clip,width=6cm]{figures_new/262.976_rat.pdf}
\caption{Plasma emission ratios of plasma assuming NEI versus equilibrium ionization for iron ion Fe~XVI. 
\label{fig:Fes262}}
\end{figure}

On the disk, the larger density of the innermost layers of the solar atmosphere minimizes the departures from ionization equilibrium so that differences are very limited except for the largest stages of ionization (Fe~XV and Fe~XVI), where some limited non-equilibrium induced emission enhancements are visible in the closed field corona, which do not correspond to any visible structure in the Si~XII disk image in Figure~\ref{fig:eqneq1}.
Figure~\ref{fig:Fes262} shows Fe~XVI emission and the relative change due to the non-equilibrium effect. In polar holes departures larger than 10\% can be seen in images from Fe~XIII and higher stages, where again the presence of lower density and plasma speed contributes to the Delay Effect.

\subsection{Narrowband images}\label{subsec:nbi}

We have investigated to what extent non-equilibrium effects propagate into the response of the narrow-band imagers available from the SoHO, STEREO and SDO spacecraft. Since these instruments convolve the EUV spectral range with their wavelength-dependent effective area, the emission they record is the weighted sum of all lines and continuum emission that their coatings and filters are built to transmit. Thus, multiple ions are contributing to the measured intensity, and the non-equilibrium effects are less straightforward to predict. These filters and coatings are built in order to maximize the instrumental response in a single, strong and isolated line, selected to open a window into the temperature range where the emitting ion is formed. Figures~\ref{fig:eqneq1} to \ref{fig:Fes2} indicate that non-equilibrium effects strongly change the intensity of many ions in the EUV spectral range: however the presence of many other lines in the wavelength adds contributions that broaden both the temperature response of the instrument, as well as its sensitivity to plasma departures from equilibrium. For all these imagers, we do not study the channels including the He~II 303.78~\AA\ line, because of the limitations in the CHIANTI model for the emissivities of this ion, and limit our investigation to the coronal channels of each instrument.

We have synthesized the emission of the channels of the most used narrow-band imagers available, calculating the convolution of the imager's response function with the full spectra calculated over the imager's observational interval by SPECTRUM. The narrow-band images available for CR~2063 were taken by SoHO/EIT, while for EUVI and AIA a later time was selected -- 2022-02-25~UT00:00:00 -- for which the GONG magnetic radial field for CR~2254 was used to calculate the coronal model with AWSoM. 

As Solarsoft provided effective areas for the EIT channels in the 165-350~\AA\ wavelength range, synthetic narrowband images were obtained convolving SPECTRUM synthetic spectra with the SolarSoft effective areas for the three coronal channels. We only considered the coronal channels due to the model's inner boundary setup discussed in \citet{Szente:2022}: 171 \AA, 195 \AA\ and 284 \AA. For simplicity we used the 'clr' filter setting only when preparing these figures, as the results are qualitatively similar for each filter. Although the number of channels is limited, two of them (171 and 195) are centered on lines observed by other instruments as well (e.g. EUVI, AIA) so that the present result give a qualitative picture of the expected nonequilibrium effects on these imagers as well.

\begin{figure}[htb!]
\includegraphics[trim={5cm 0.cm 3cm 4cm},clip,width=6cm]{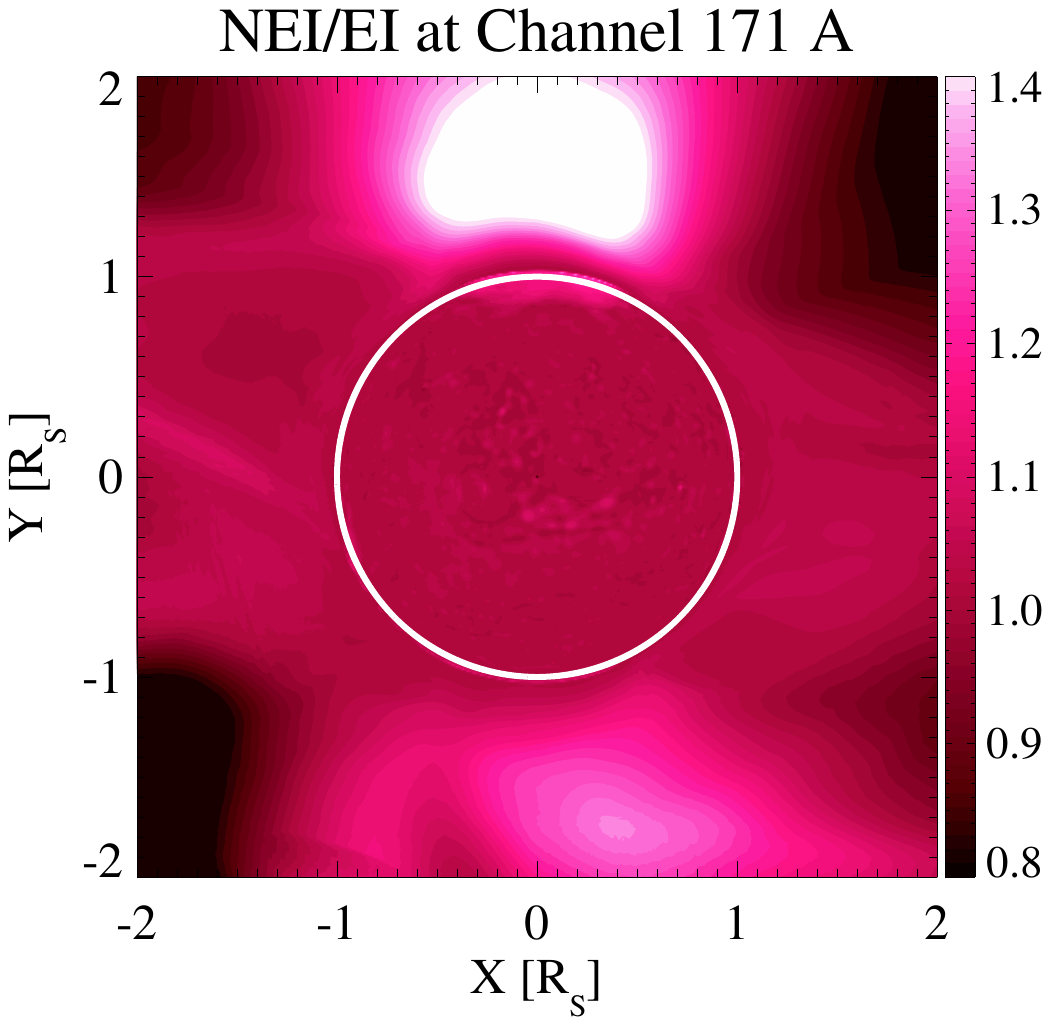}
\includegraphics[trim={5cm 0.cm 3cm 4cm},clip,width=6cm]{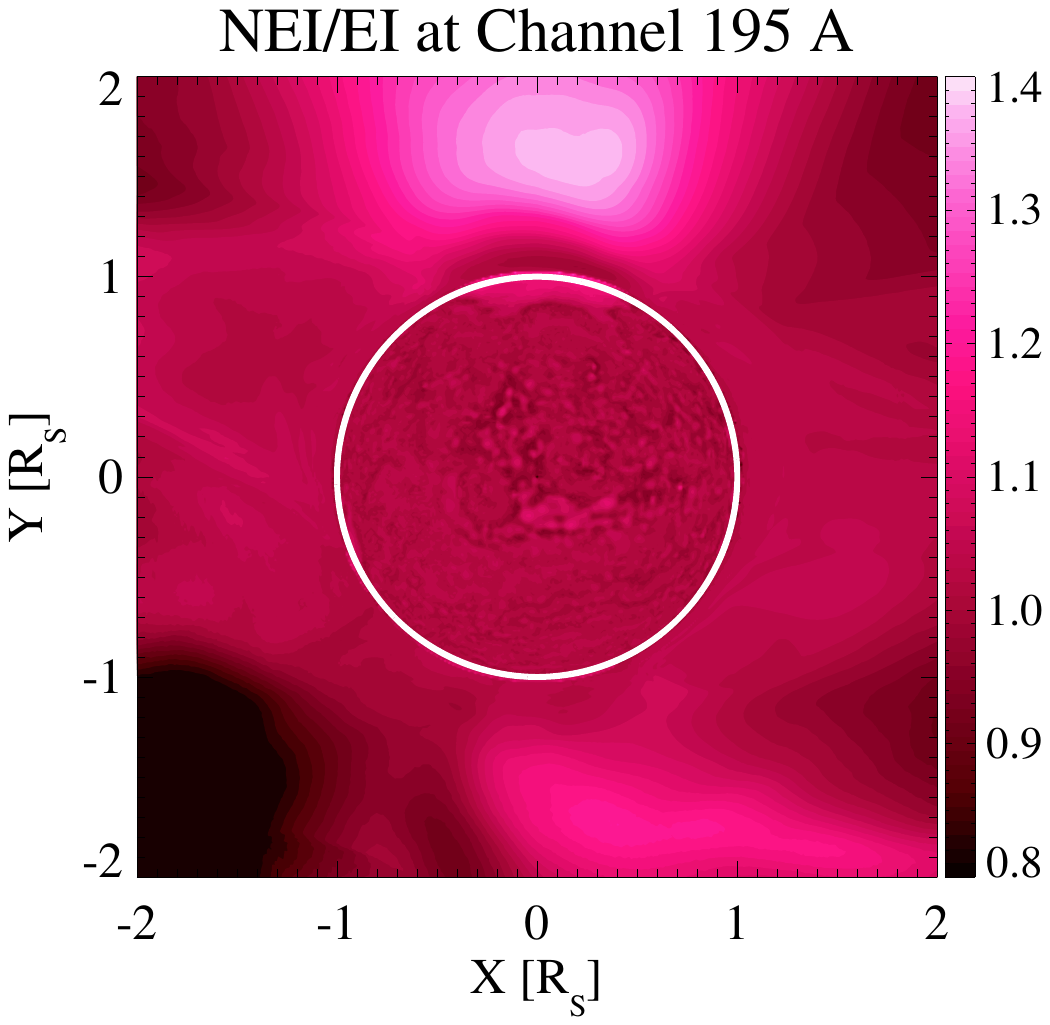}
\includegraphics[trim={5cm 0.cm 3cm 4cm},clip,width=6cm]{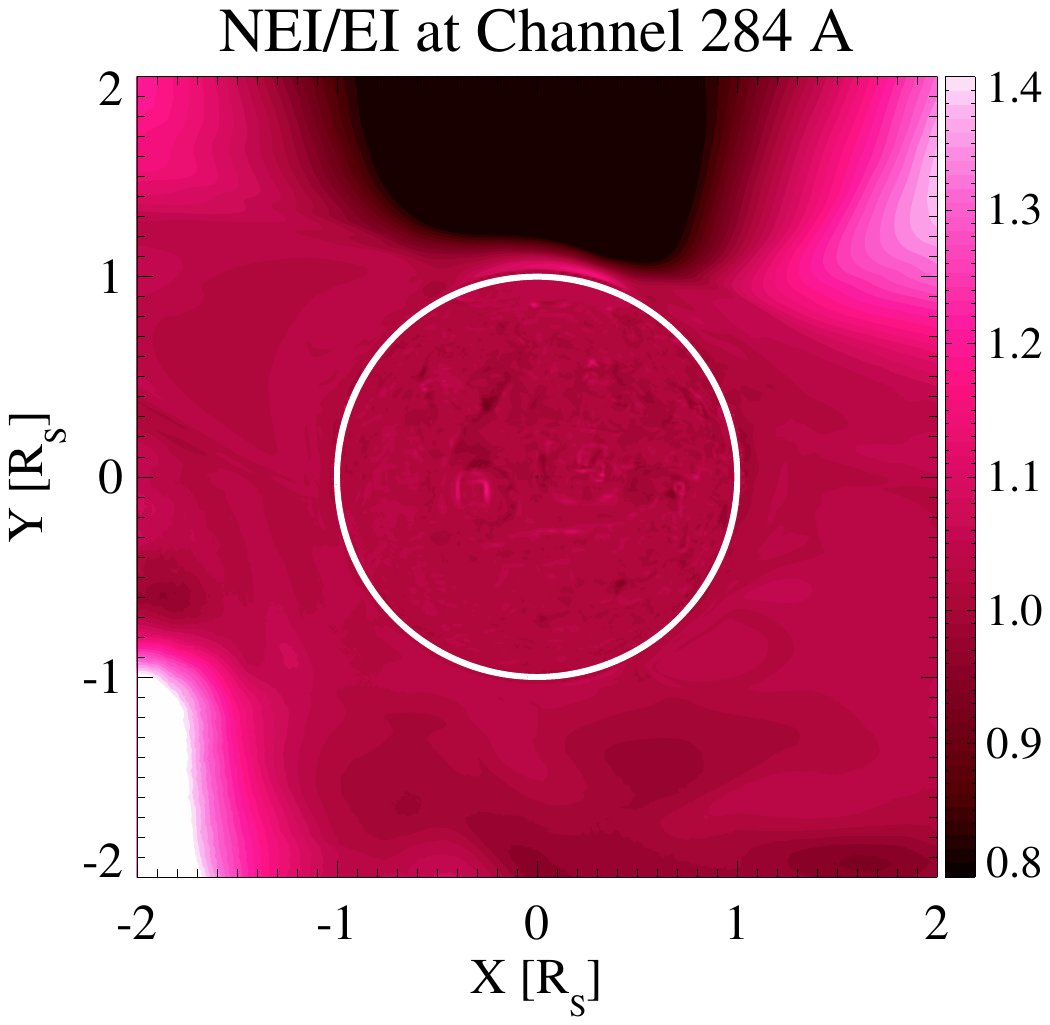}
\includegraphics[trim={5cm 0.cm 3cm 4cm},clip,width=6cm]{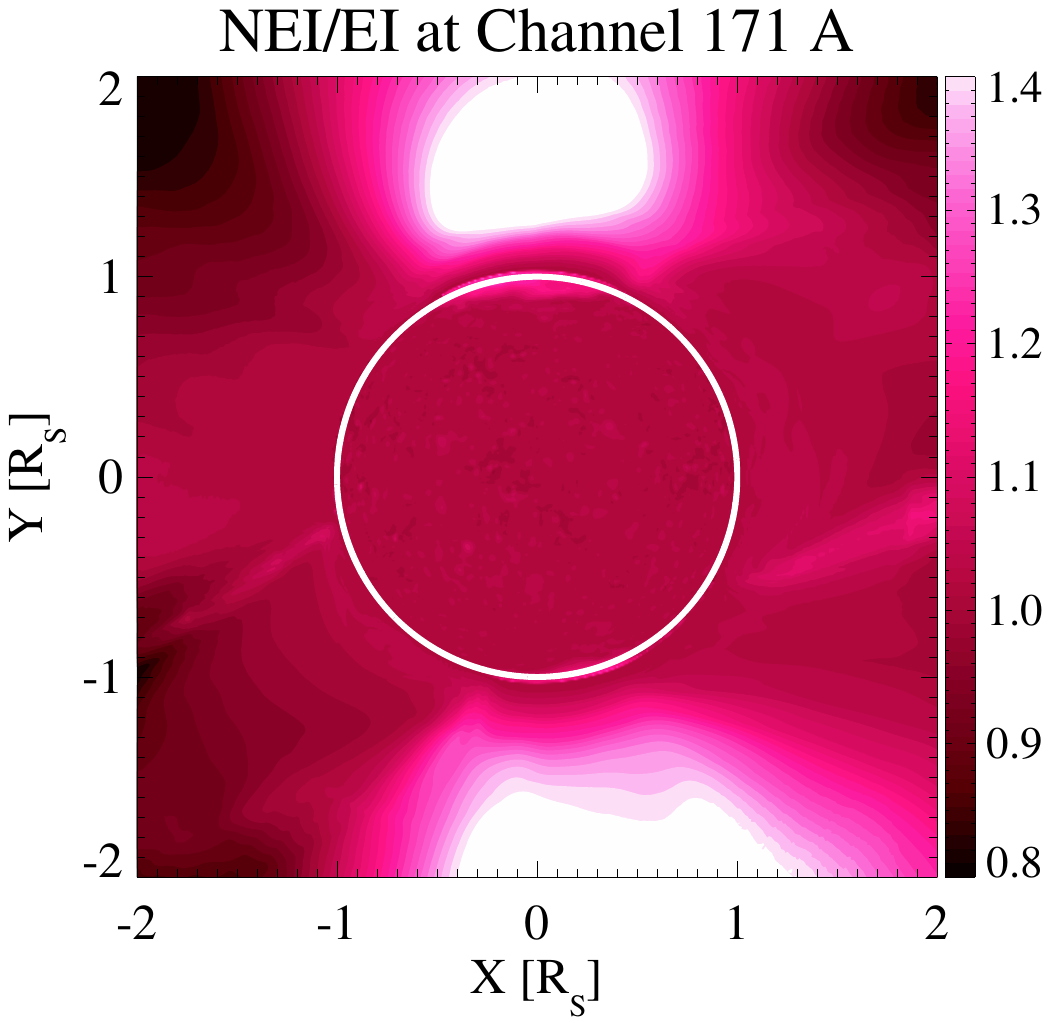}
\includegraphics[trim={5cm 0.cm 3cm 4cm},clip,width=6cm]{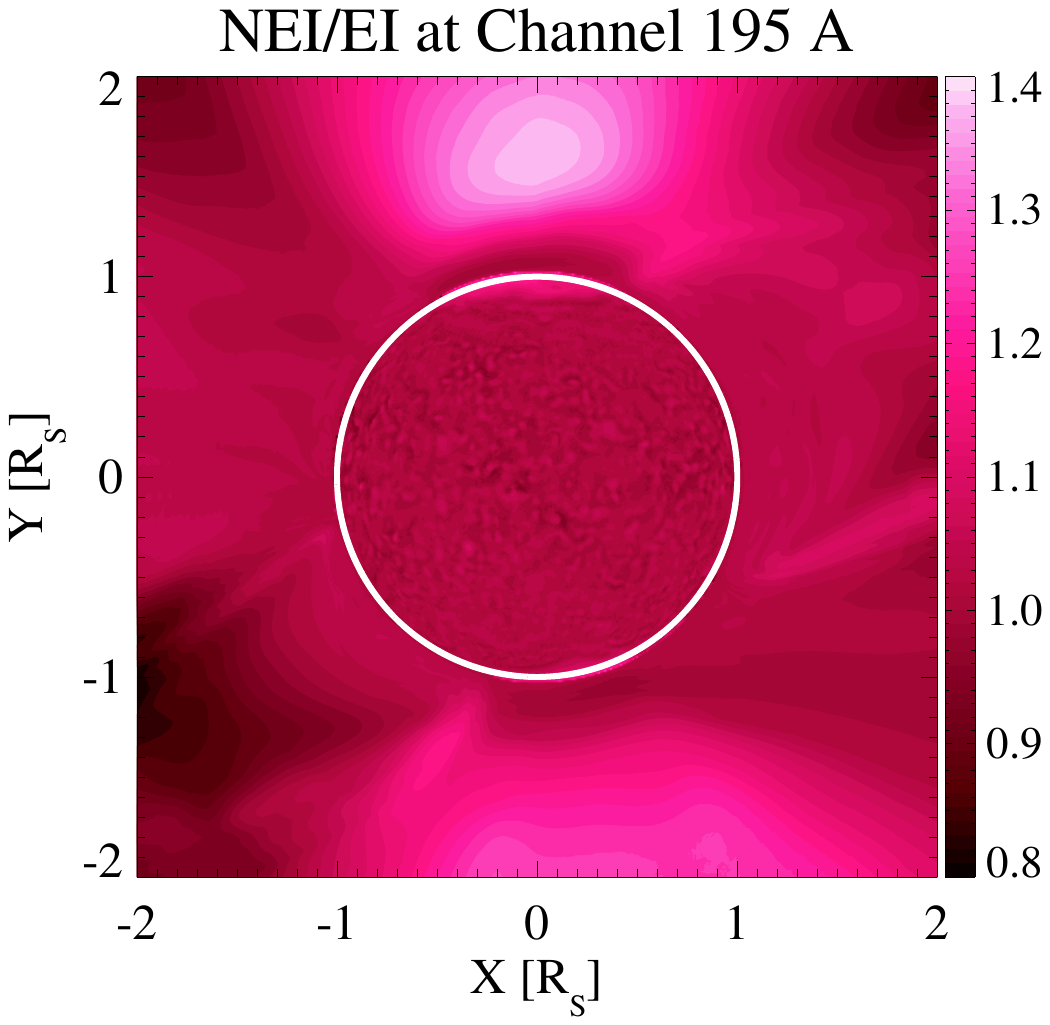}
\includegraphics[trim={5cm 0.cm 3cm 4cm},clip,width=6cm]{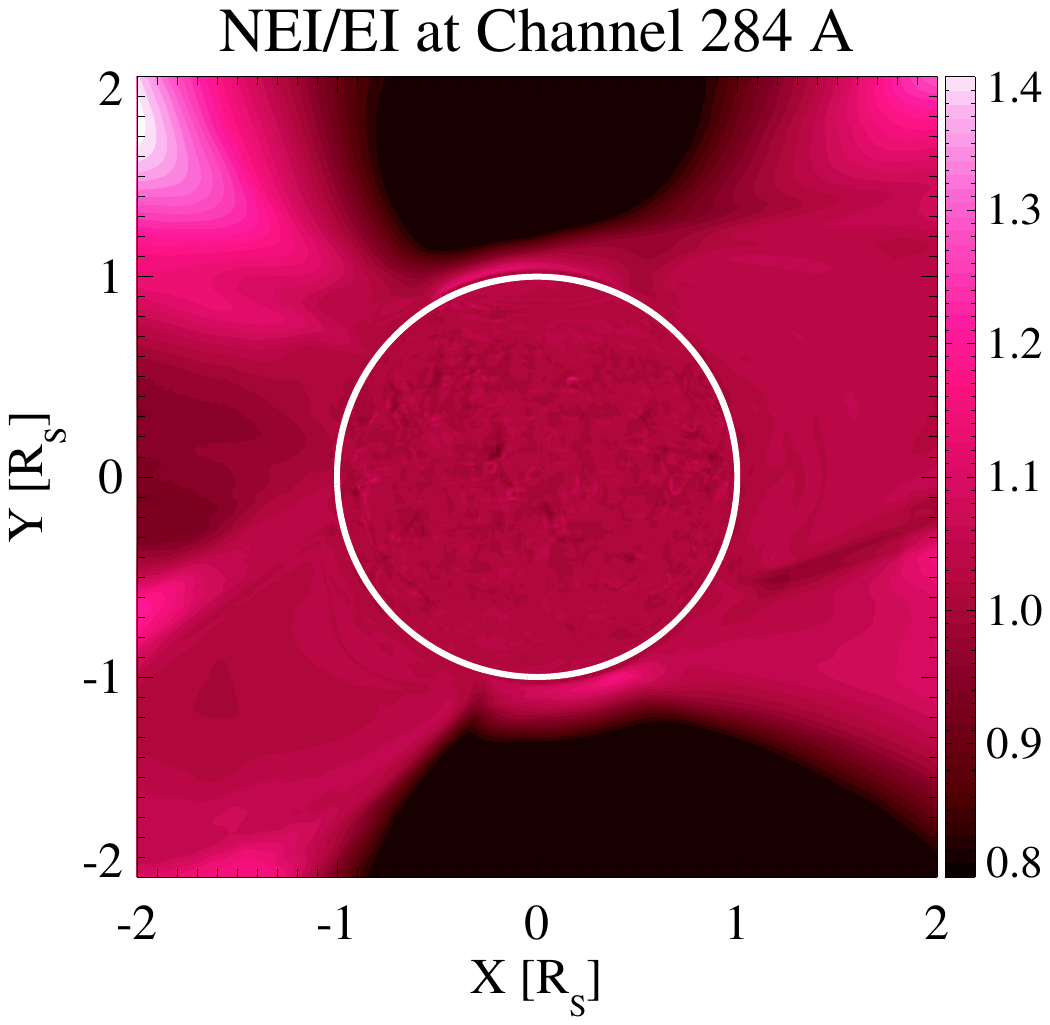}
\caption{Synthetic narrowband images of EIT channels 171~\AA, 195~\AA\ and 284~\AA\ (\emph{left to right}) The two rows correspond to two observational times from \citet{Szente:2019}.
\label{fig:eit}}
\end{figure}

Figure~\ref{fig:eit} shows the ratio of emission from the NEI vs equilibrium ionization solutions for the three EIT coronal channels. We show two observational views corresponding to observational sites discussed in \citet{Szente:2019}. As expected, the 171~\AA\ and 284~\AA\ channels, dominated by Fe~IX-X and Fe~XV respectively, are the most affected, given the largest variations of the ion abundances of these ions; on the contrary, Fe~XII is more stable against non-equilibrium. Also, off-disk polar coronal holes are the most affected regions, due to the combined effect of wind speed and lower electron density. However, departures up to 20\% are locally found also in on-disk regions, while other regions showing significant (and sometimes very large) non-equilibrium effects are streamer legs and the NE-SW regions where the plasma iron charge state distribution is over-ionized. However, given the rapid decrease in electron density, EIT images provide significant count rates only at distances up to $\approx$1.3~solar radii, so that the only differences likely to affect quantitative analysis of EIT narrow band images are those in the two polar coronal holes, and the ones on the disk.

\begin{figure}[htb!]
\includegraphics[trim={4.5cm 0cm 1.2cm 4cm},clip,width=6 cm]{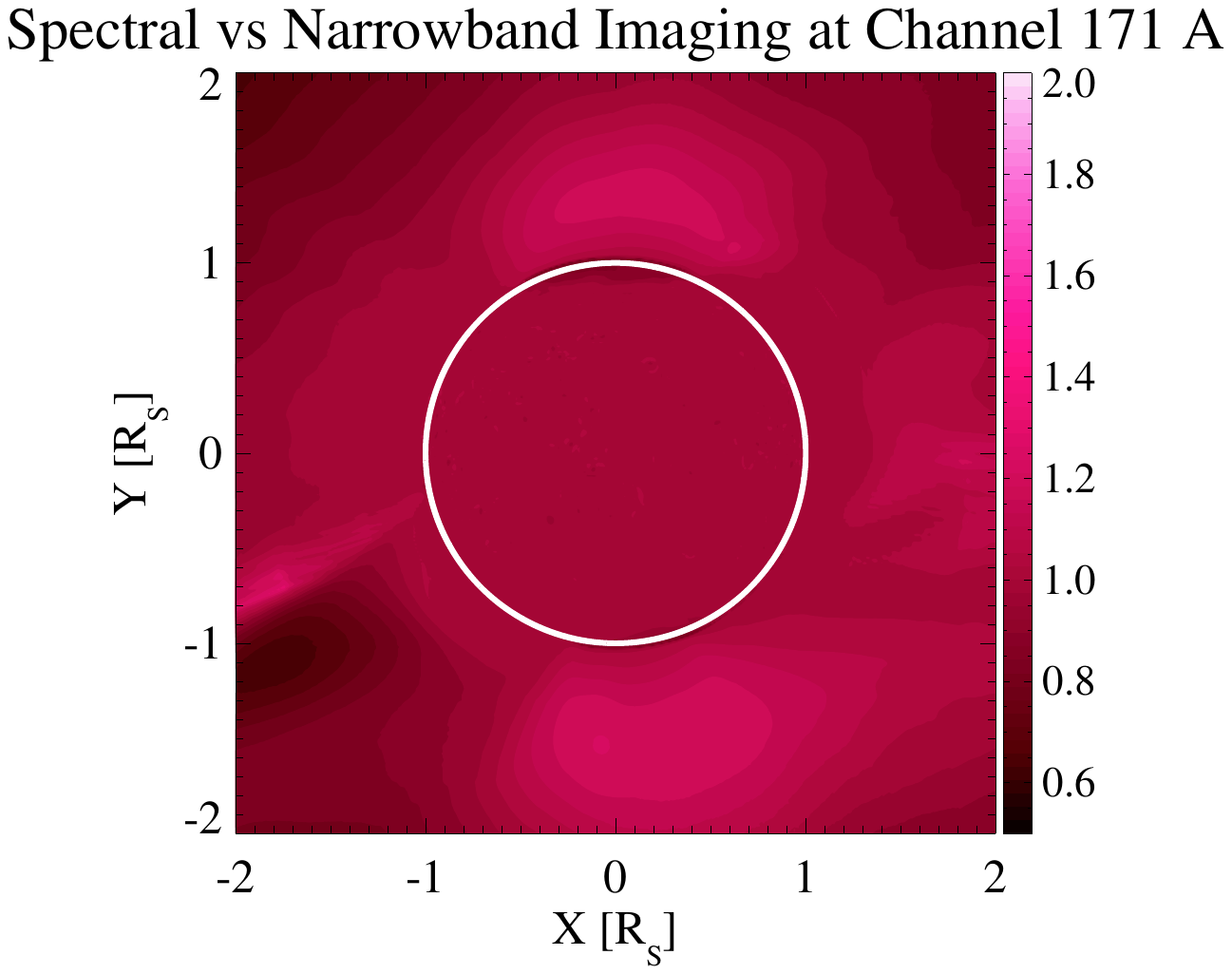}
\includegraphics[trim={4.5cm 0cm 1.2cm 4cm},clip,width=6 cm]{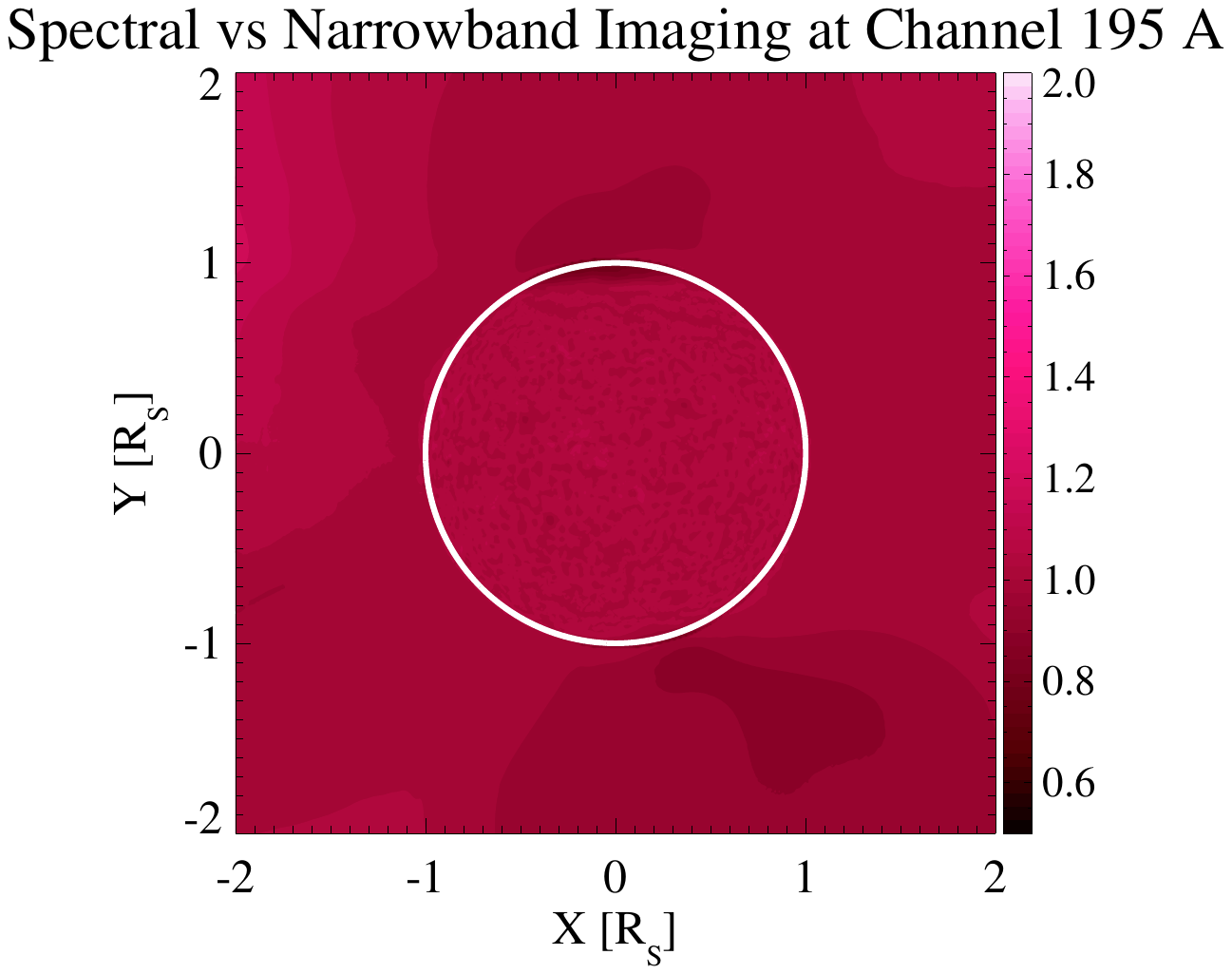}
\includegraphics[trim={4.5cm 0cm 1.2cm 4cm},clip,width=6 cm]{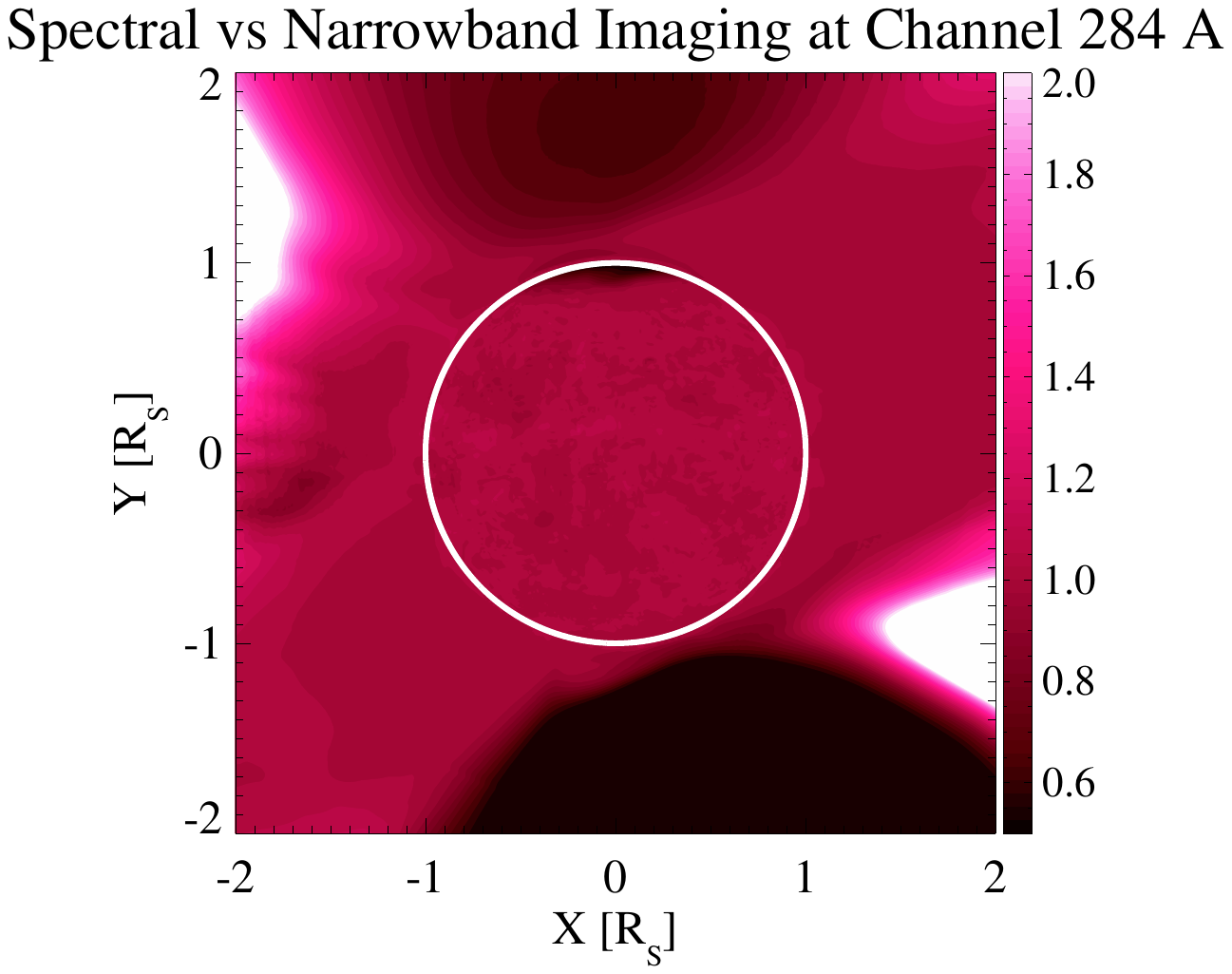}
\caption{Ratio of change in emission due to NEI of synthetic spectral line emission at $171.073$~\AA\ versus synthetic narrowband image at EIT at channel $171$~\AA\ (\emph{left}), line ratios at $195.119$~\AA\ versus EIT channel $195$~\AA, and line ratios at $284.163$~\AA\ versus EIT channel $284$~\AA.  
\label{fig:spnbi}}
\end{figure}

The results shown in Figure~\ref{fig:eit} resemble those obtained with the individual ions dominating each channel, but there are differences in each of them. Such differences are due to the presence of other ions emitting lines in the wavelength range covered by each channel, so the question is, to what extent is the presence of these additional ions mitigating (or worsening) each channel's sensitivity to non-equilibrium effects. To investigate this, we calculated the ratio of emission change of the iron ion spectral emission, corresponding to each channel relative to the narrow-band emission's change (a ratio of ratios). The results are shown in Figure~\ref{fig:spnbi}. We observe an increased sensitivity in spectral imaging at 284\AA. The spectral imaging seems to be more sensitive to non-equilibrium effects in the polar coronal holes, where about 40\% more sensitivity is observed at channels 171~\AA\ and 284~\AA. 195~\AA\ seems to have the closest behavior whether we consider spectral or narrow-band imaging being effected by NEI.  

\begin{figure}[htb!]
\includegraphics[trim={5cm 0cm 3cm 4cm},clip,width=6 cm]{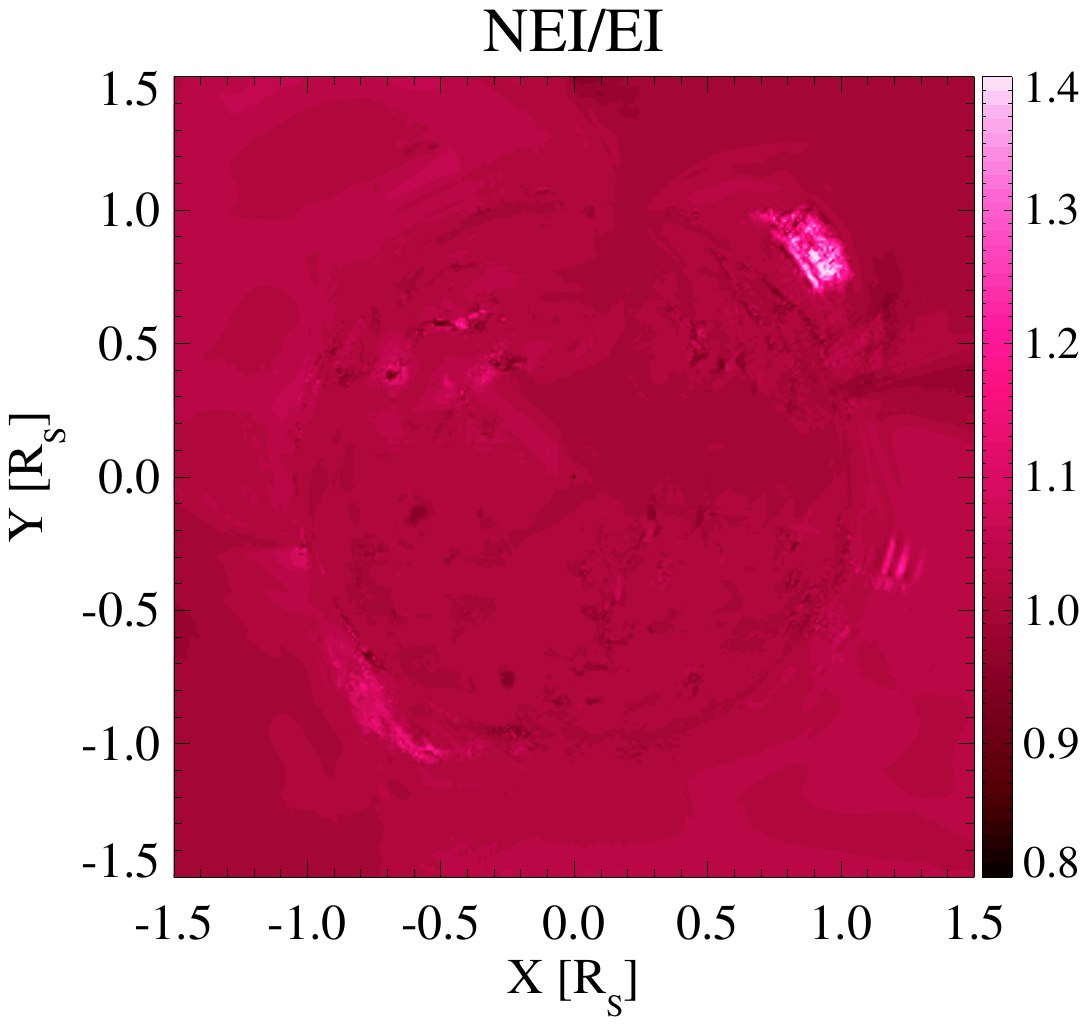}
\includegraphics[trim={5cm 0cm 3cm 4cm},clip,width=6 cm]{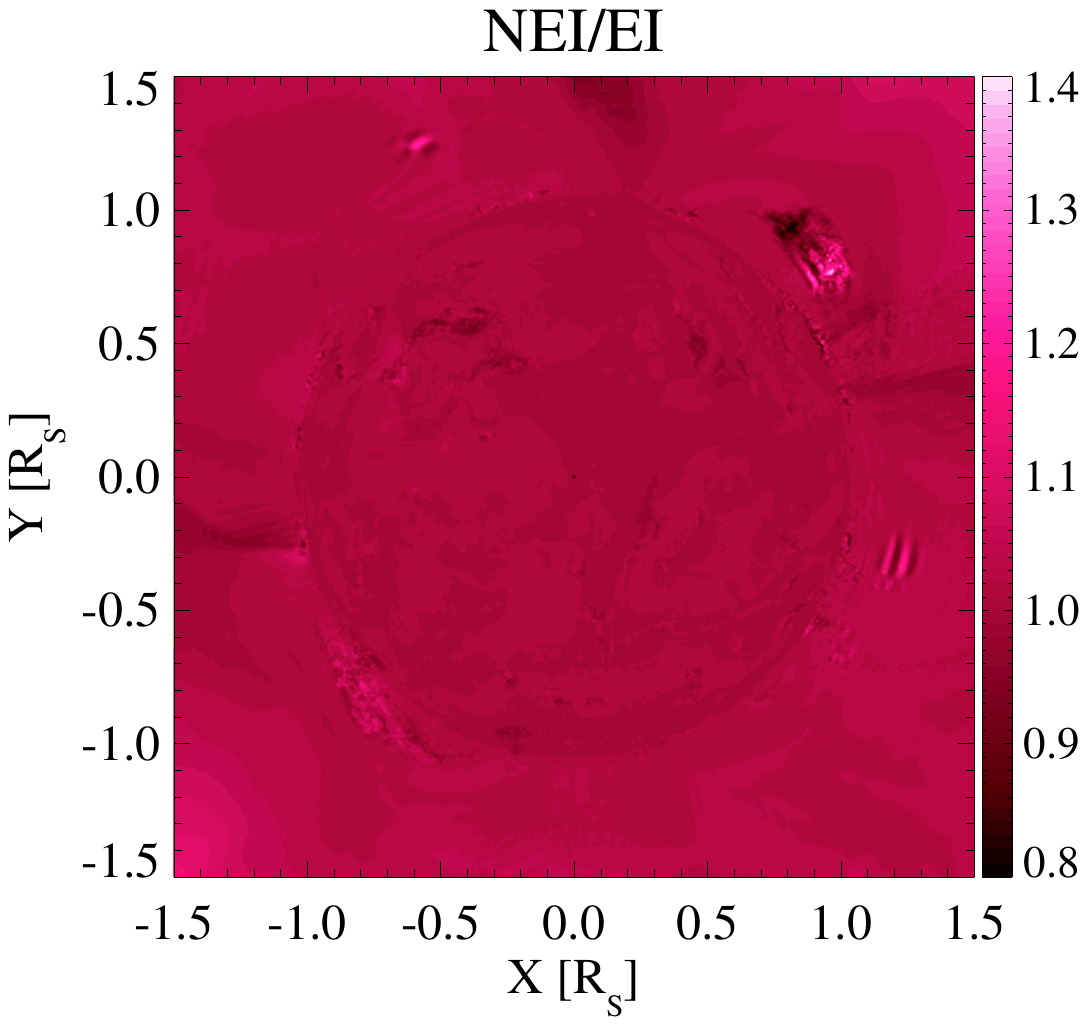}
\includegraphics[trim={5cm 0cm 3cm 4cm},clip,width=6 cm]{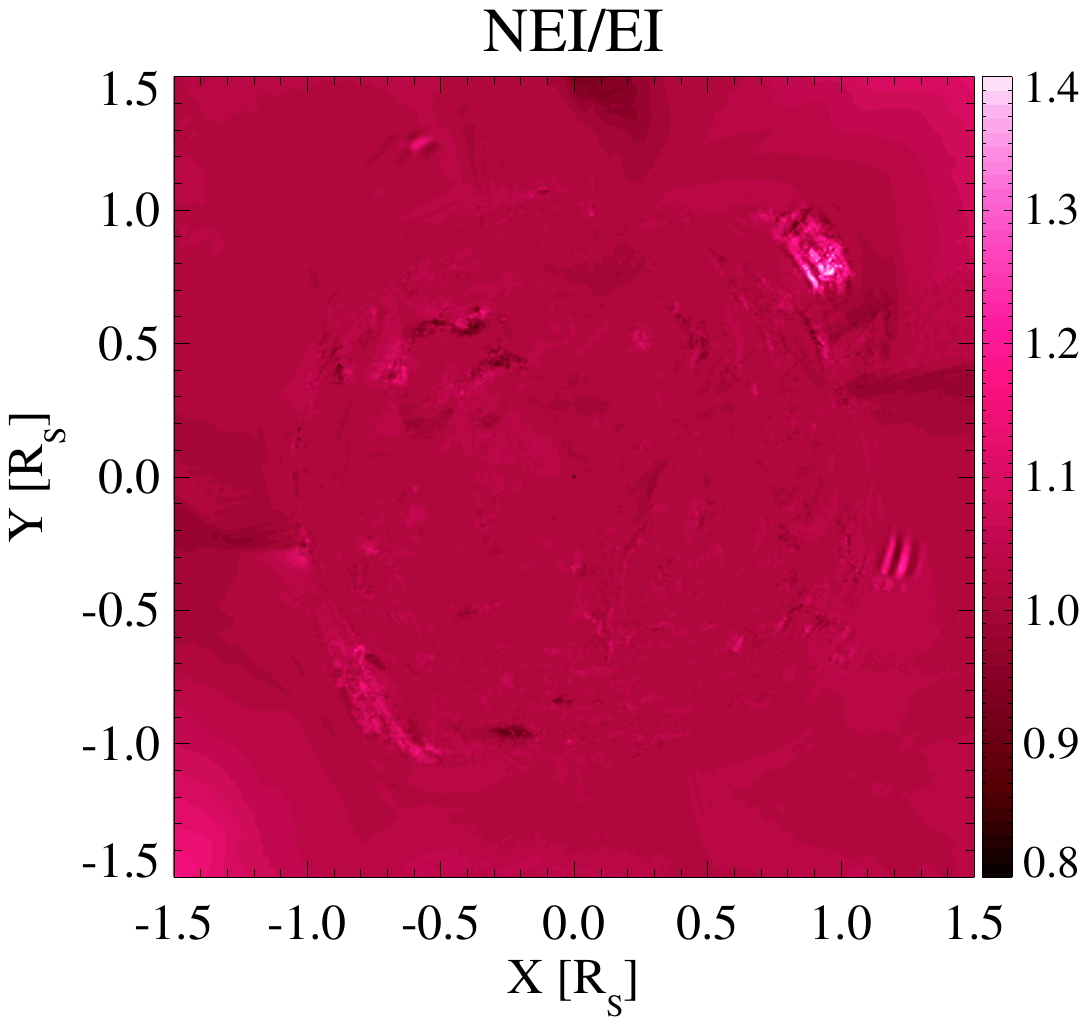}
\includegraphics[trim={5cm 0cm 3cm 4cm},clip,width=6 cm]{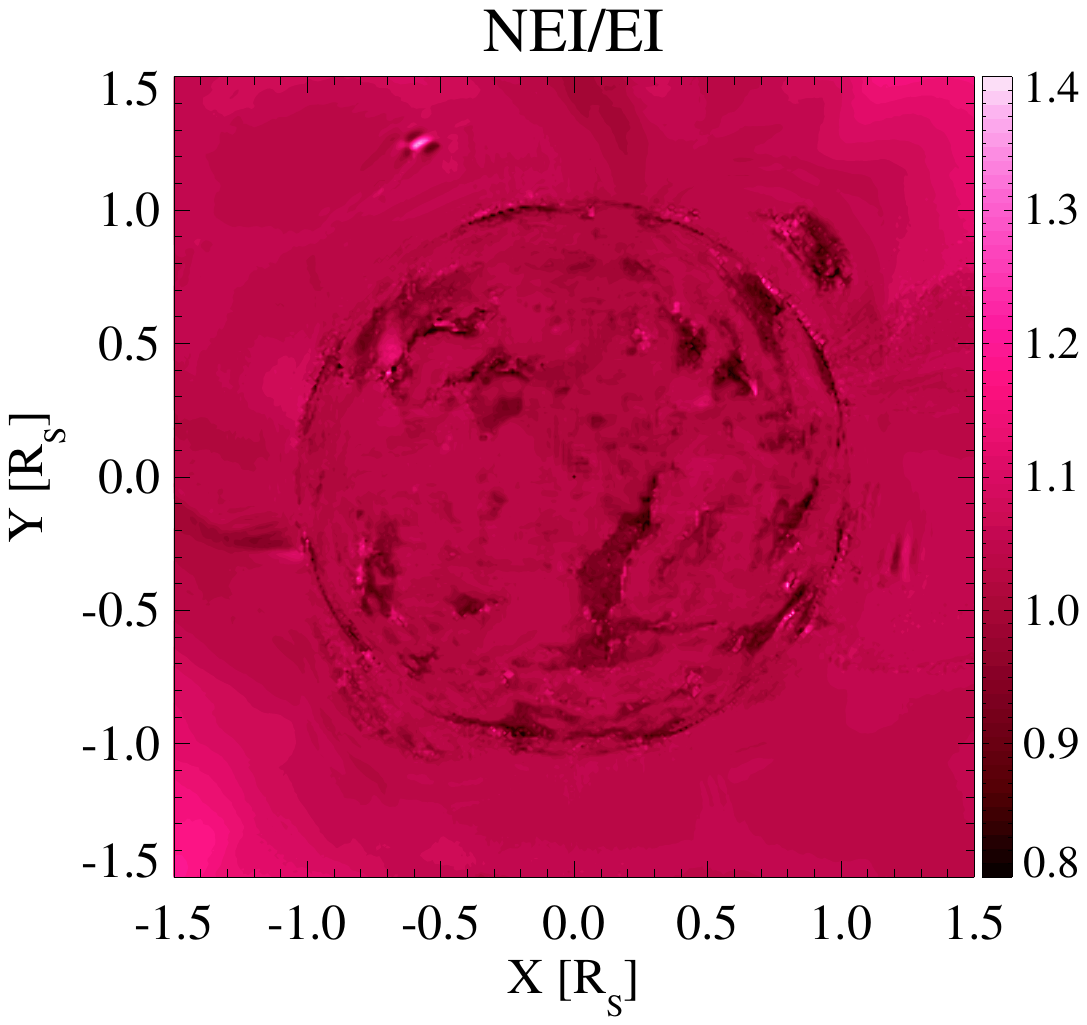}
\includegraphics[trim={5cm 0cm 3cm 4cm},clip,width=6 cm]{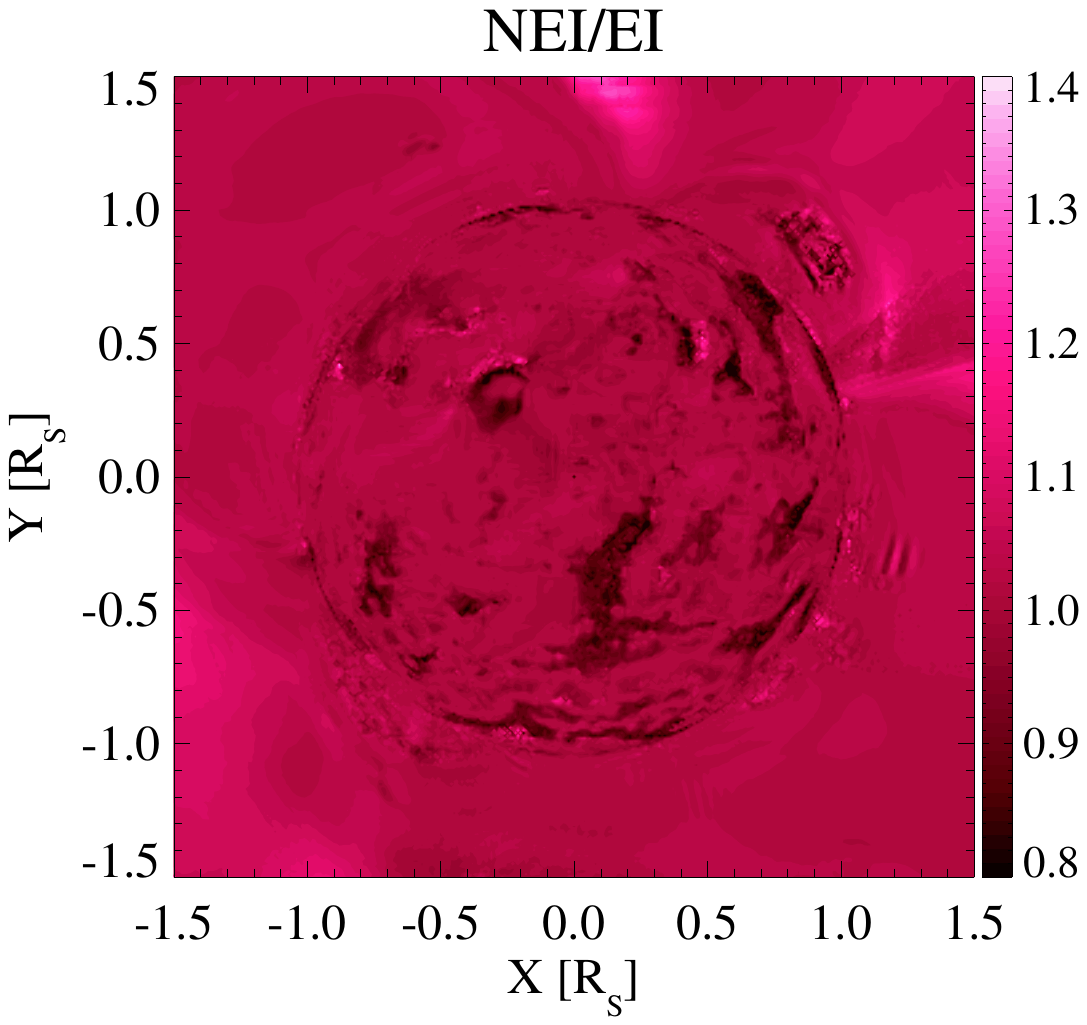}
\includegraphics[trim={5cm 0cm 3cm 4cm},clip,width=6 cm]{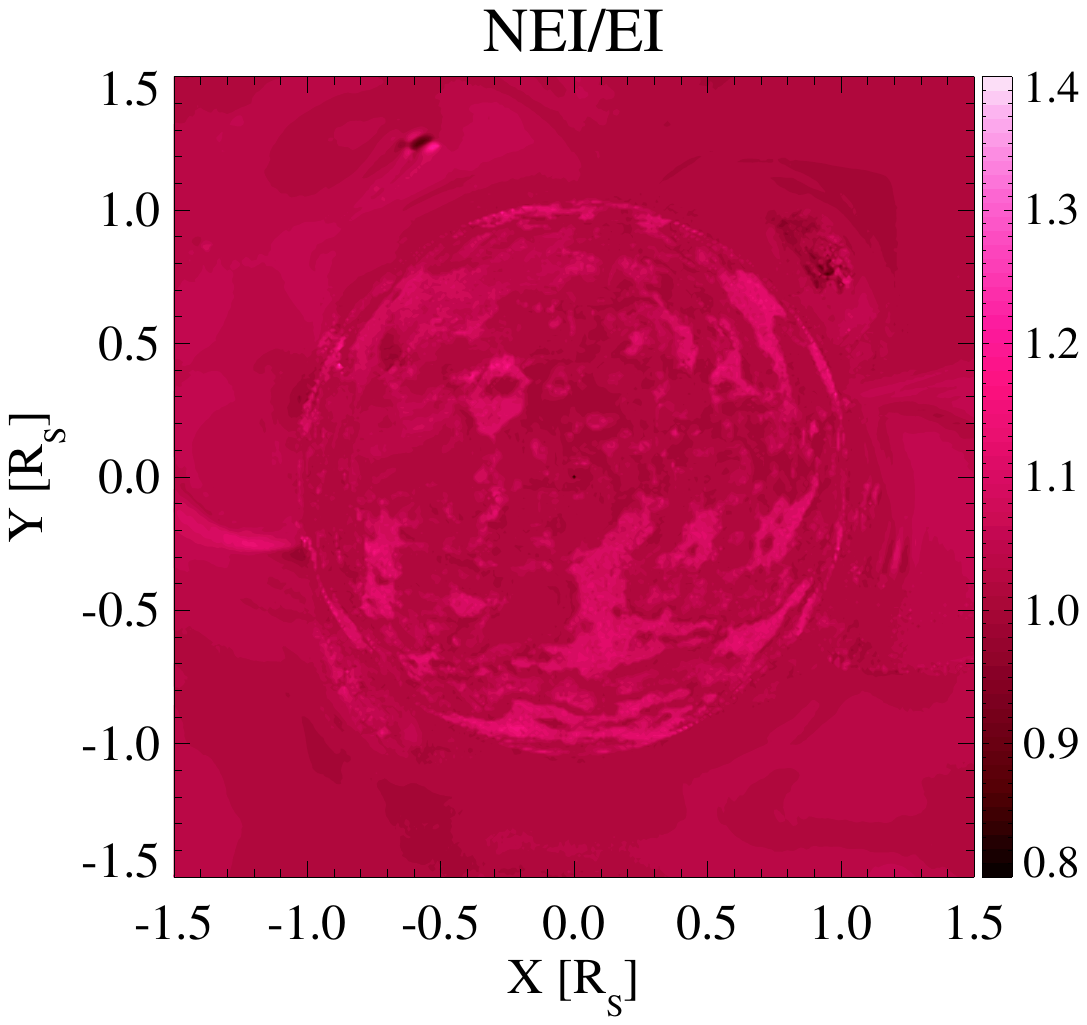}
\caption{NEI effects on synthetic narrowband images of coronal AIA channels 94~\AA, 131~\AA, 171~\AA, 193~\AA, 211~\AA\ and 335~\AA\ for observation time 2022-02-25~UT00:00:00. 
\label{fig:aia}}
\end{figure}

\begin{figure}[htb!]
\includegraphics[trim={5cm 0cm 3cm 4cm},clip,width=6 cm]{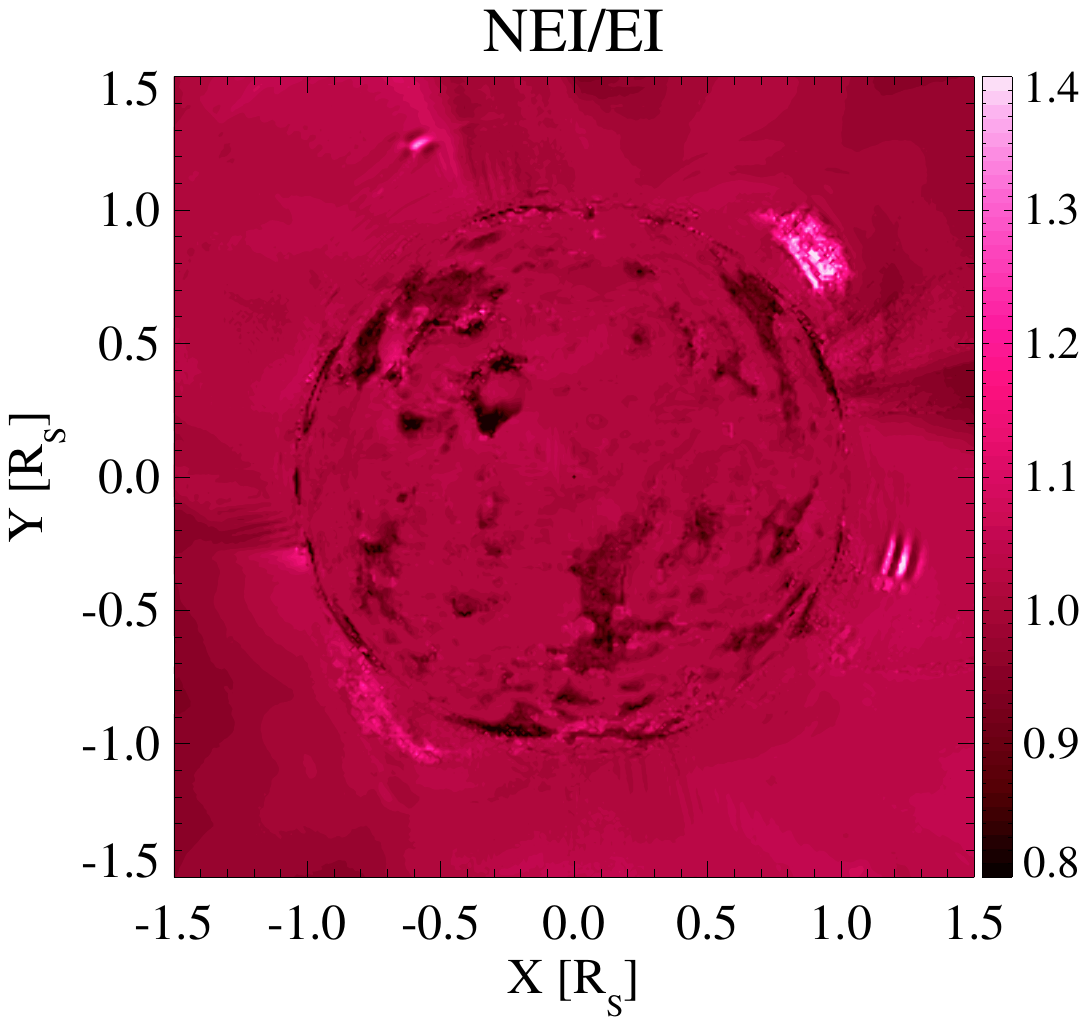}
\includegraphics[trim={5cm 0cm 3cm 4cm},clip,width=6 cm]{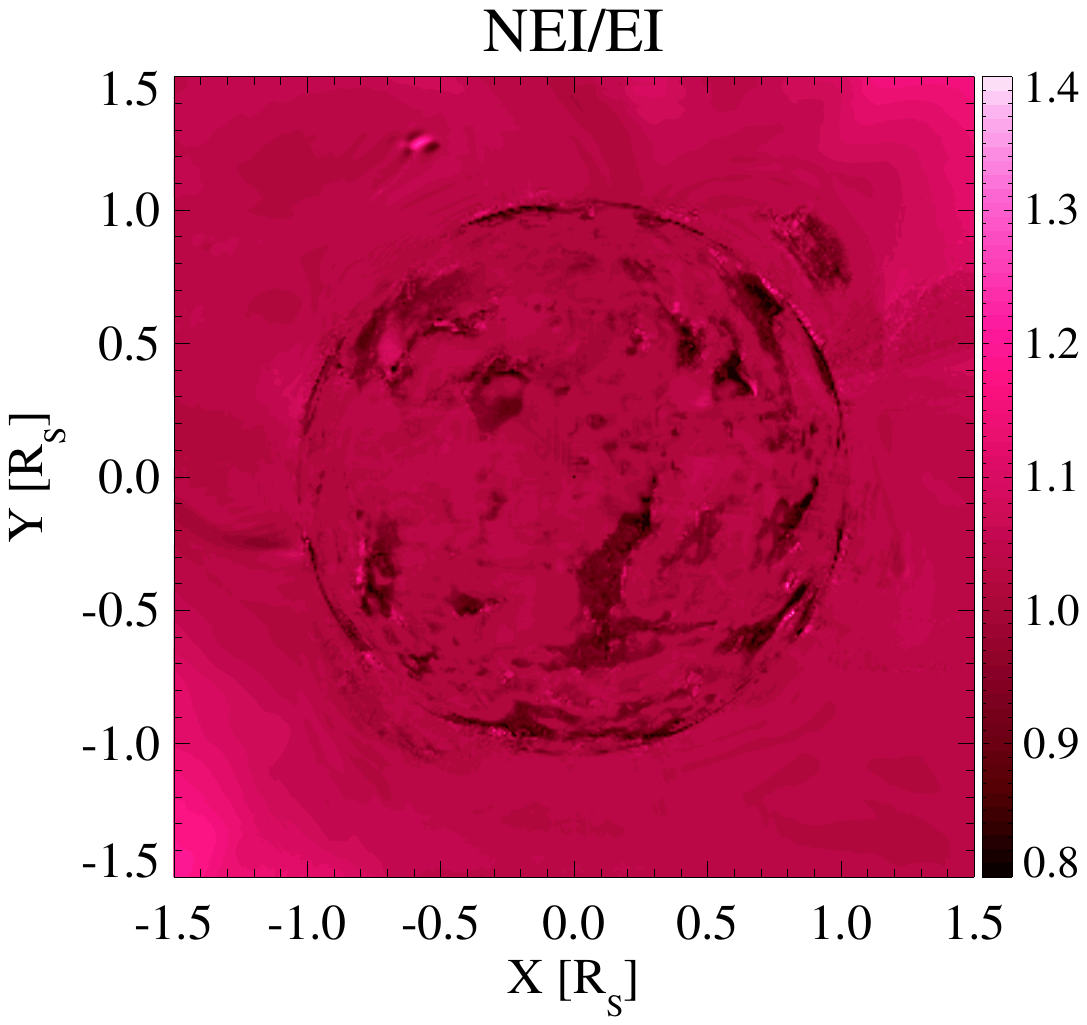}
\includegraphics[trim={5cm 0cm 3cm 4cm},clip,width=6 cm]{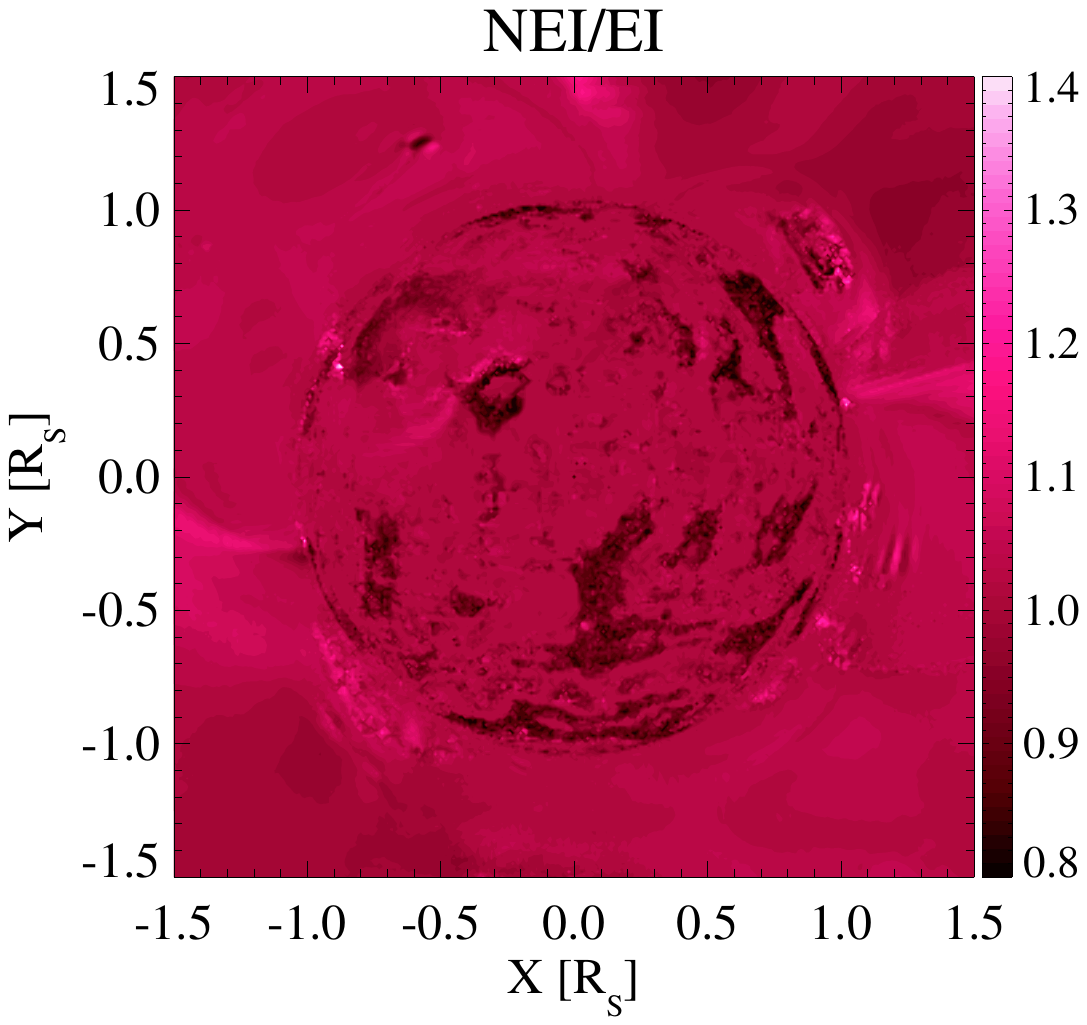}
\caption{NEI effects on synthetic narrowband images of coronal EUVI channels 171~\AA, 195~\AA\ and 284~\AA\ for observation time 2022-02-25~UT00:00:00. 
\label{fig:euvi}}
\end{figure}

The SDO/AIA coronal channels are also affected differently by the departure from ionization equilibrium, see Figure~\ref{fig:aia}. The sensitivity of EUVI narrowband images are shown in Figure~\ref{fig:euvi}.

The response of the SDO/AIA channels (Figure~\ref{fig:aia}) and of the SECCHI/EUVI channels (Figure~\ref{fig:euvi}) is predictably different from the EIT one, because of the differences in the instrumental bandpass, even if some of the channels are centered on the same lines as EIT. Also, the different phase of the solar cycle where the data were taken has consequences. In fact, the EIT data we used were taken during the minimum of solar cycle~24, while both SDO and EUVI are taken close to the maximum of solar cycle 25. The difference in solar activity phase causes the polar coronal holes to disappear, so that on the overall the non-equilibrium worst effects are not present in Figures~\ref{fig:aia} and \ref{fig:euvi}; also, the presence of several active regions on the disk caused a much larger and more structured variability in the solar disk than present at solar minimum. Still, differences at the 20\% level are scattered throughout all the channels in both instruments -- though not dramatic, they do have consequences for the plasma diagnostics carried out with those images.

\subsection{Comparison to observations}\label{subsec:obs}

\begin{figure}[htb!]
\includegraphics[trim={2cm 0.5cm 1cm 4cm},clip,width=6cm]{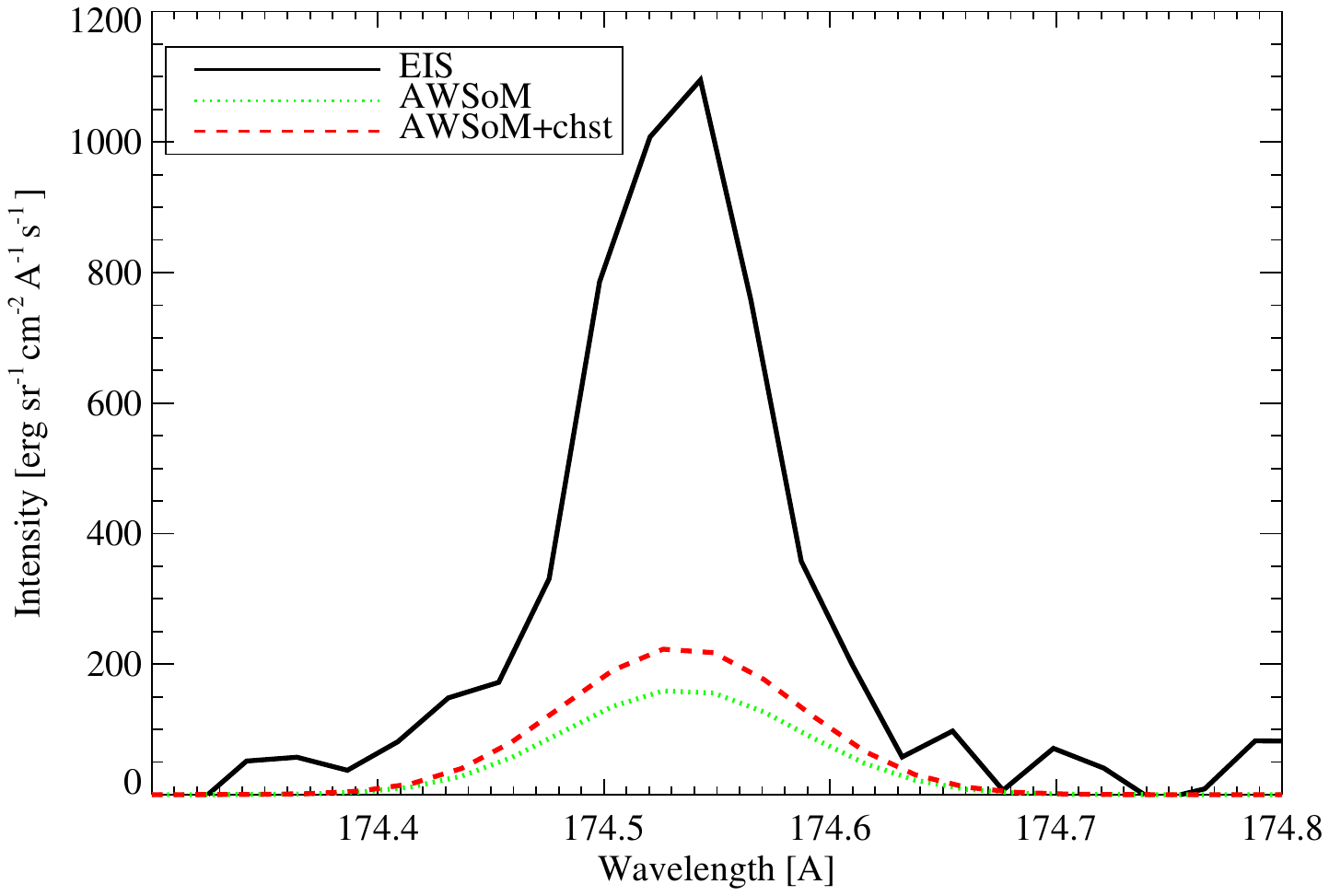}
\includegraphics[trim={2cm 0.5cm 1cm 4cm},clip,width=6cm]{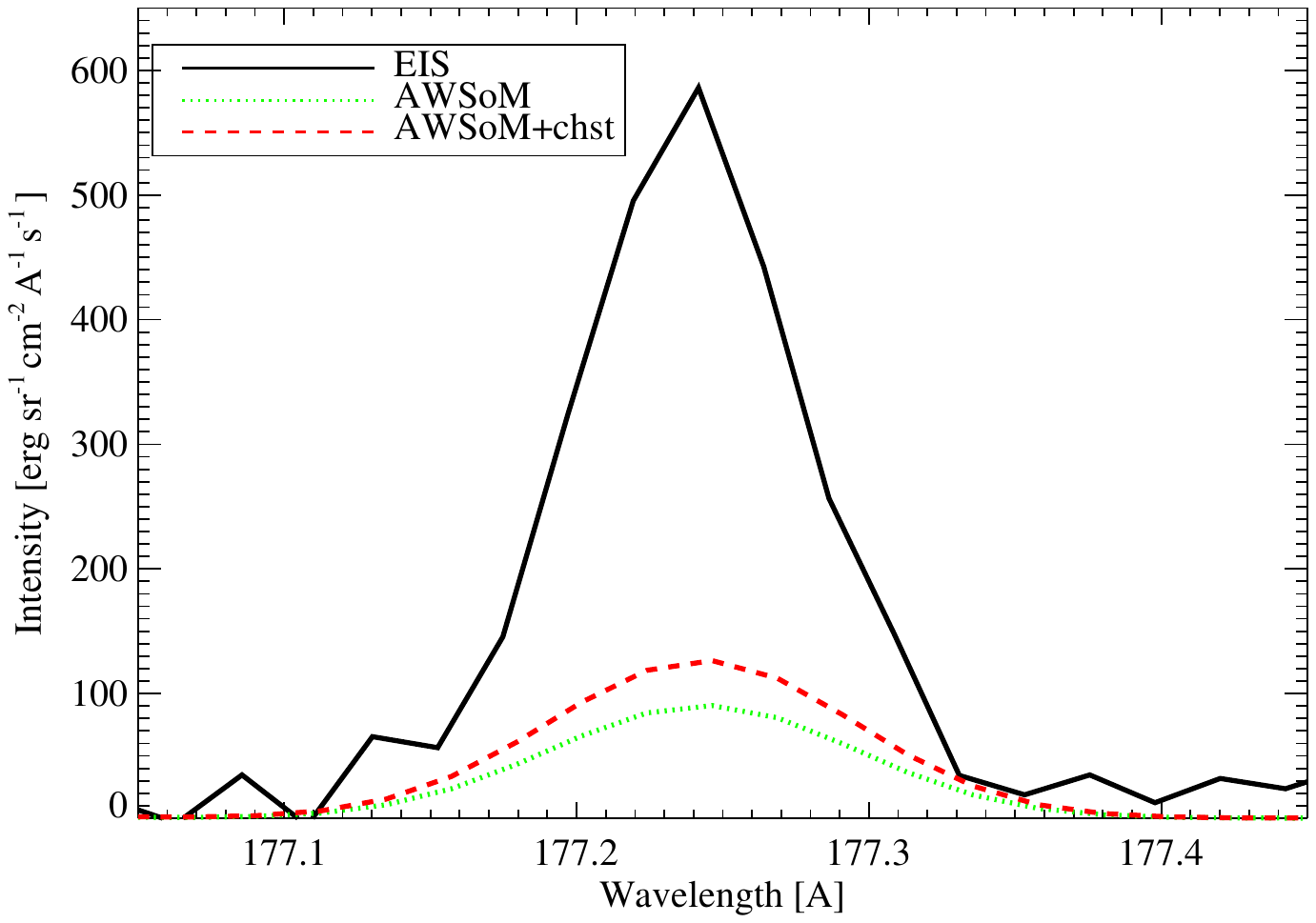}
\includegraphics[trim={2cm 0.5cm 1cm 4cm},clip,width=6cm]{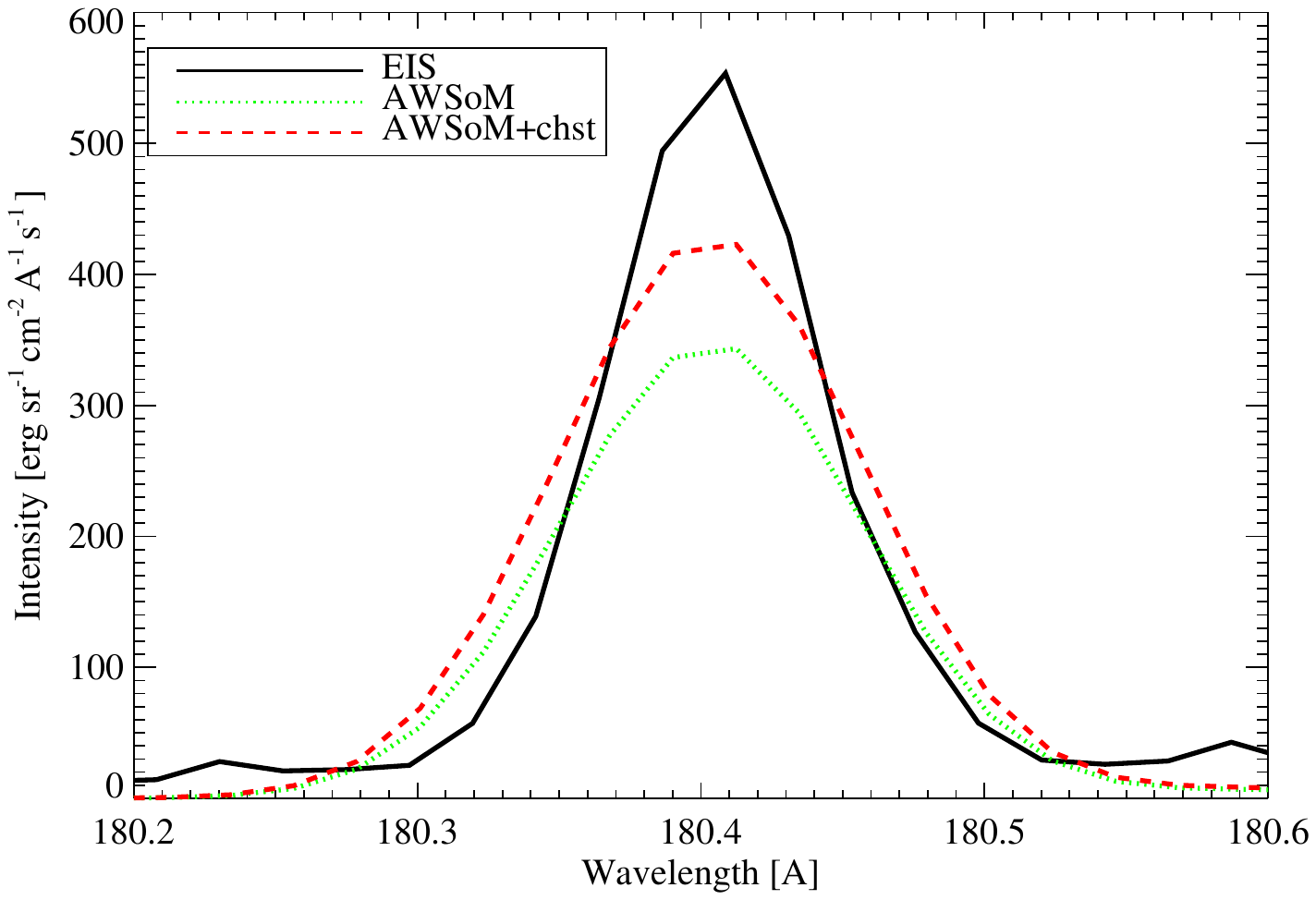}
\includegraphics[trim={2cm 0.5cm 1cm 4cm},clip,width=6cm]{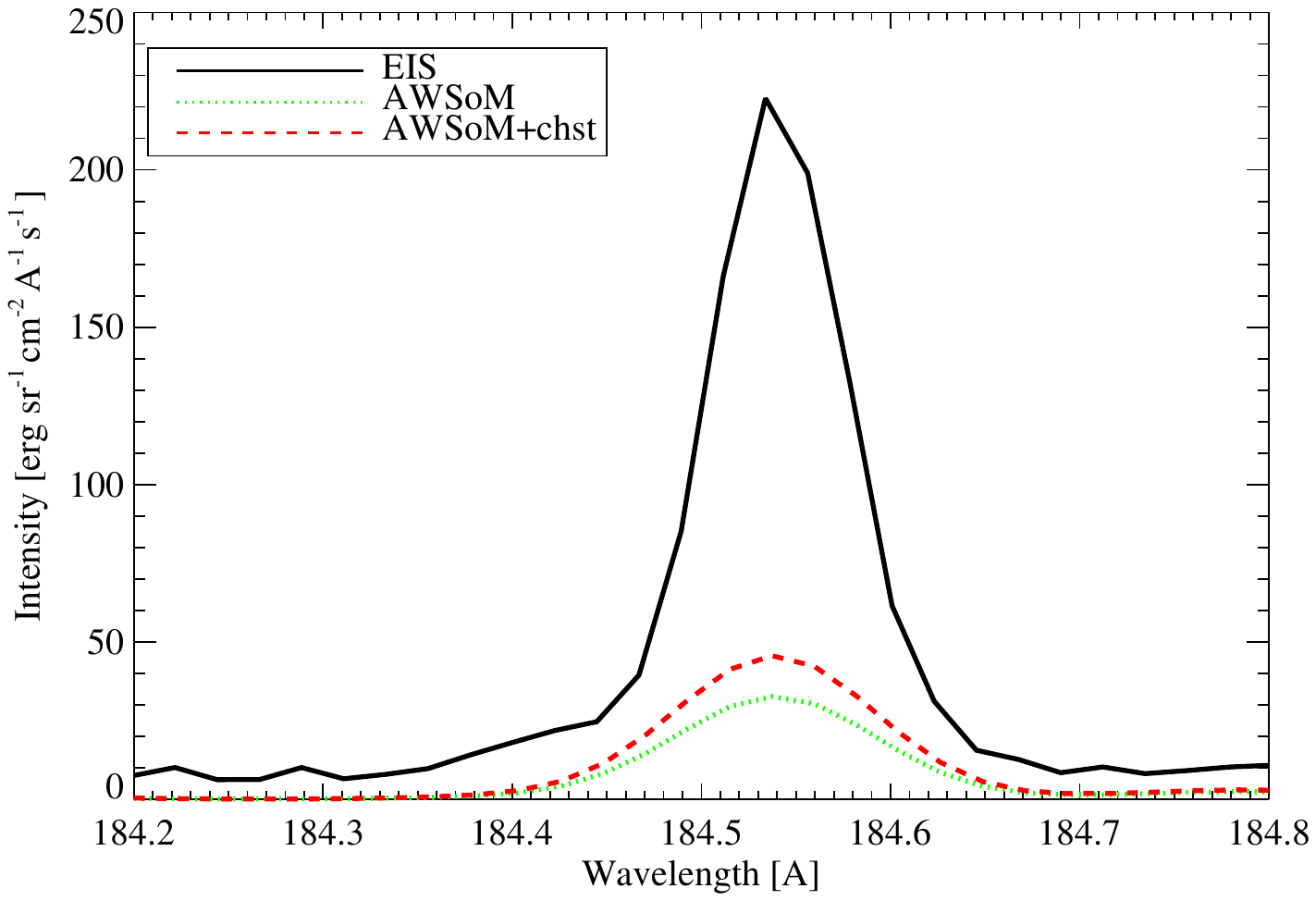}
\includegraphics[trim={2cm 0.5cm 1cm 4cm},clip,width=6cm]{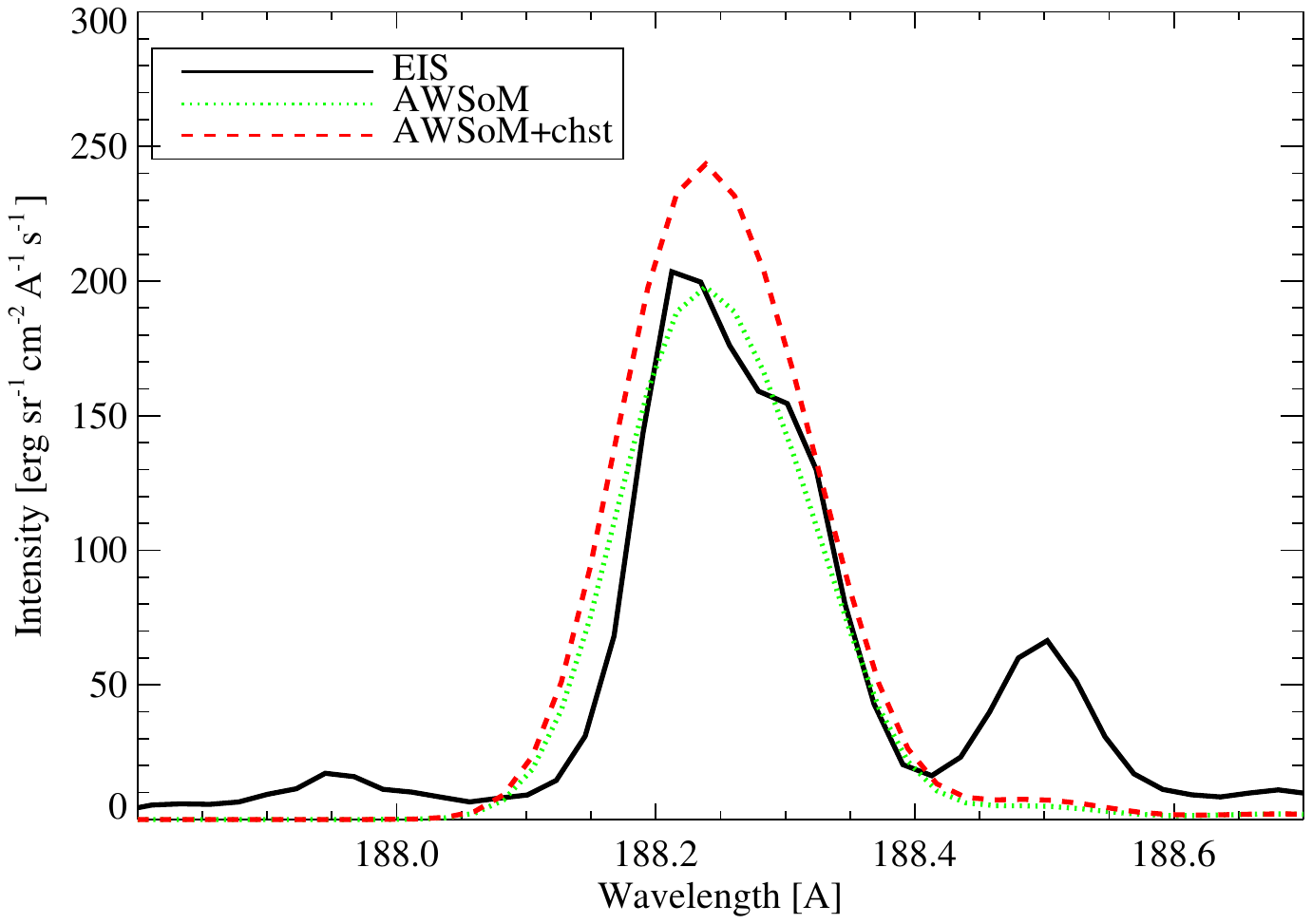}
\includegraphics[trim={2cm 0.5cm 1cm 4cm},clip,width=6cm]{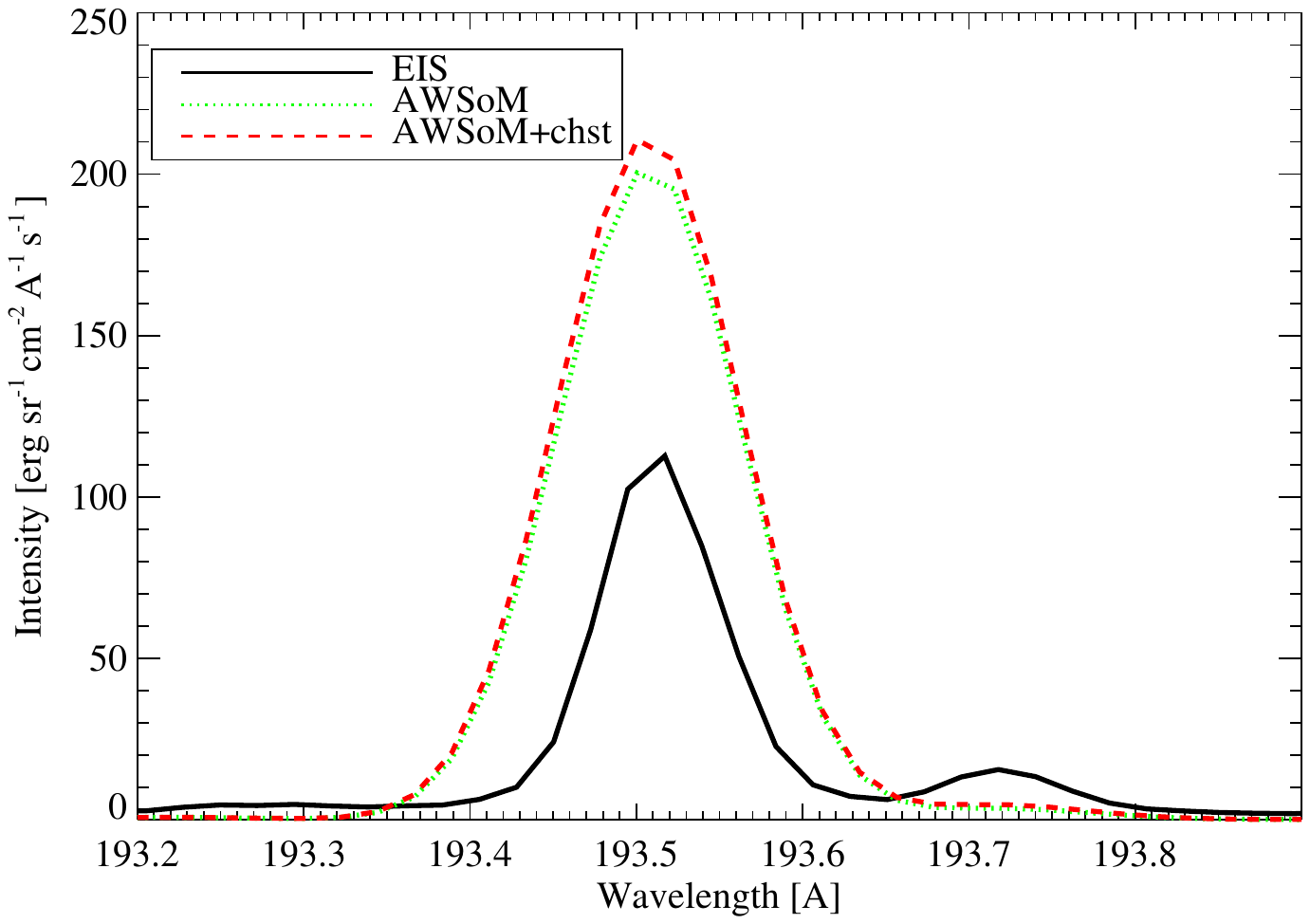}
\includegraphics[trim={2cm 0.5cm 1cm 4cm},clip,width=6cm]{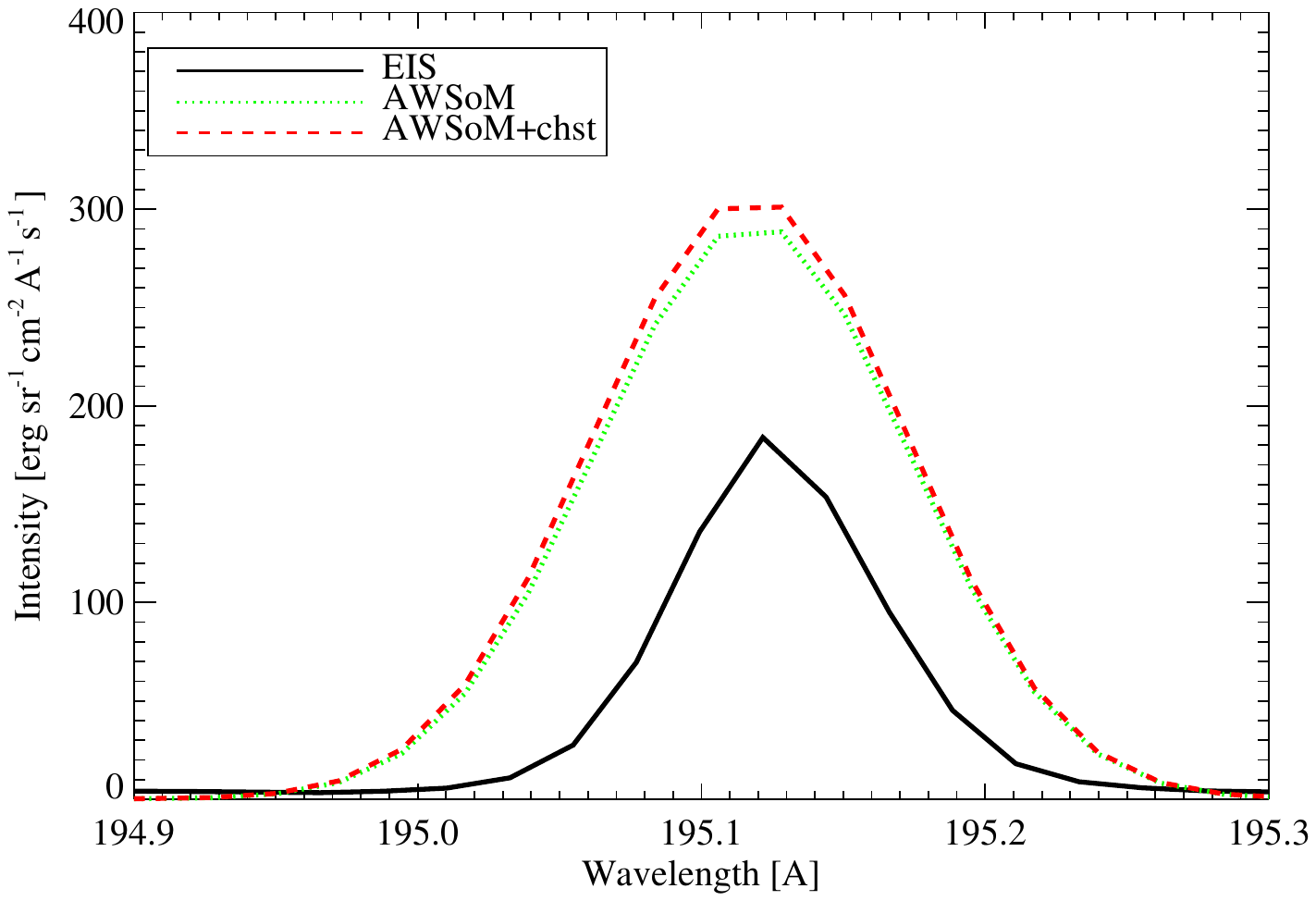}
\includegraphics[trim={2cm 0.5cm 1cm 4cm},clip,width=6cm]{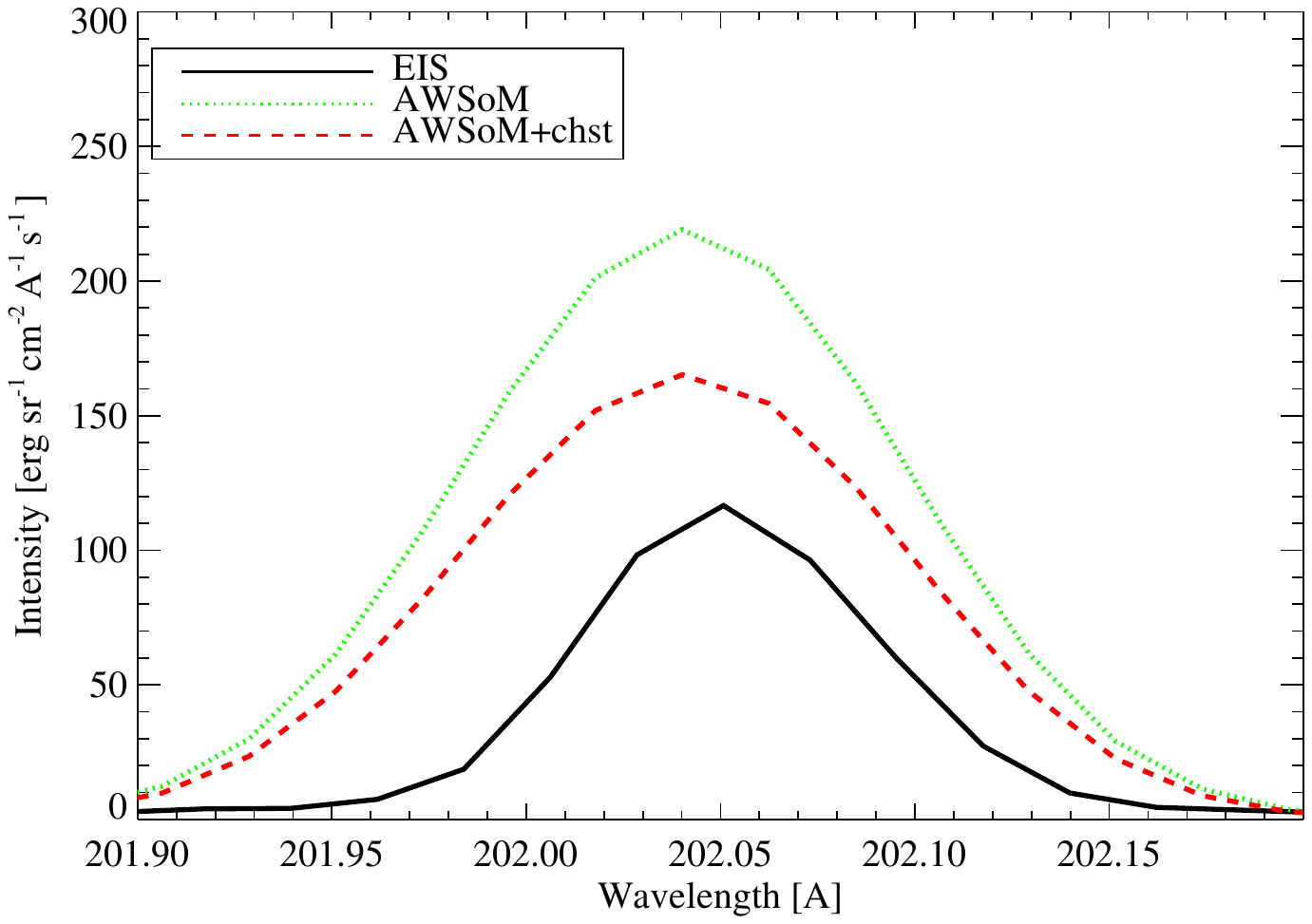}
\includegraphics[trim={2cm 0.5cm 1cm 4cm},clip,width=6cm]{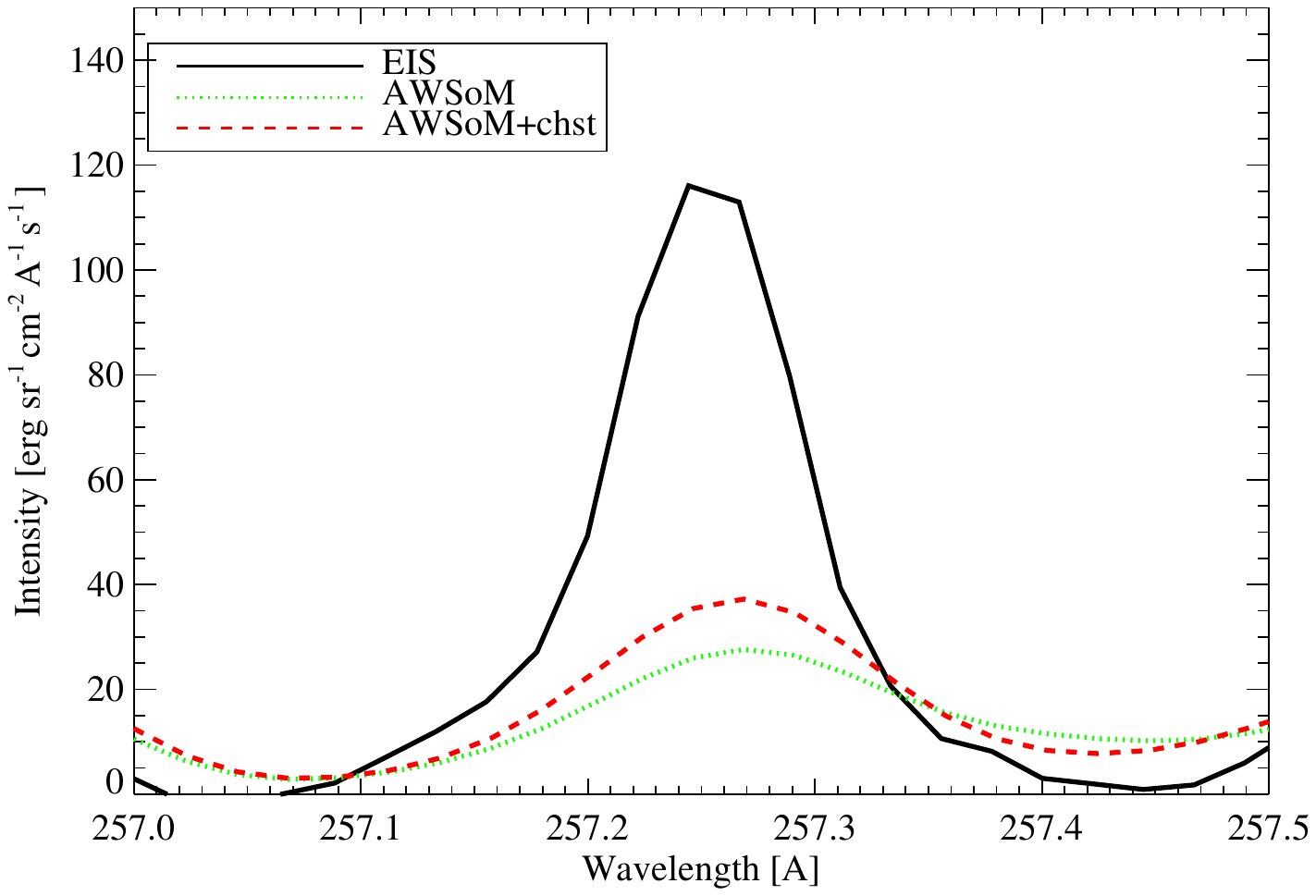}
\caption{Coronal hole observation comparison to data. Black shows Hinode/EIS observations, green the equilibrium solution, red the non-equilibrium solution's emission.
\label{fig:chole}}
\end{figure}

With the addition of available non-equilibrium charge state calculation, we calculated the spectral emission and compared it to observations taken by Hinode/EIS. The observation sites are (also described in \citet{Szente:2019}): an open magnetic field region on the side of a coronal hole at the northern pole measured at 2007-11-12~UT12:32:02 and a closed magnetic field region on the west limb of the disk 2007-11-04~UT19:12:27. Due to the relative darkness of the northern polar coronal hole, we compared only a handful of lines to synthetic spectra calculated with ionization equilibrium and non-equilibrium charge states. Figure~\ref{fig:chole} shows that the difference between observation (calibrated using the Warren method \citep{Warren:2014}) and simulation results are greater than any effects due to NEI. This suggests that there are still issues in the simulation with the plasma solution, both for the electron density and the electron temperature.

\begin{figure}[htb!]
\includegraphics[trim={2cm 0.5cm 1cm 4cm},clip,width=6cm]{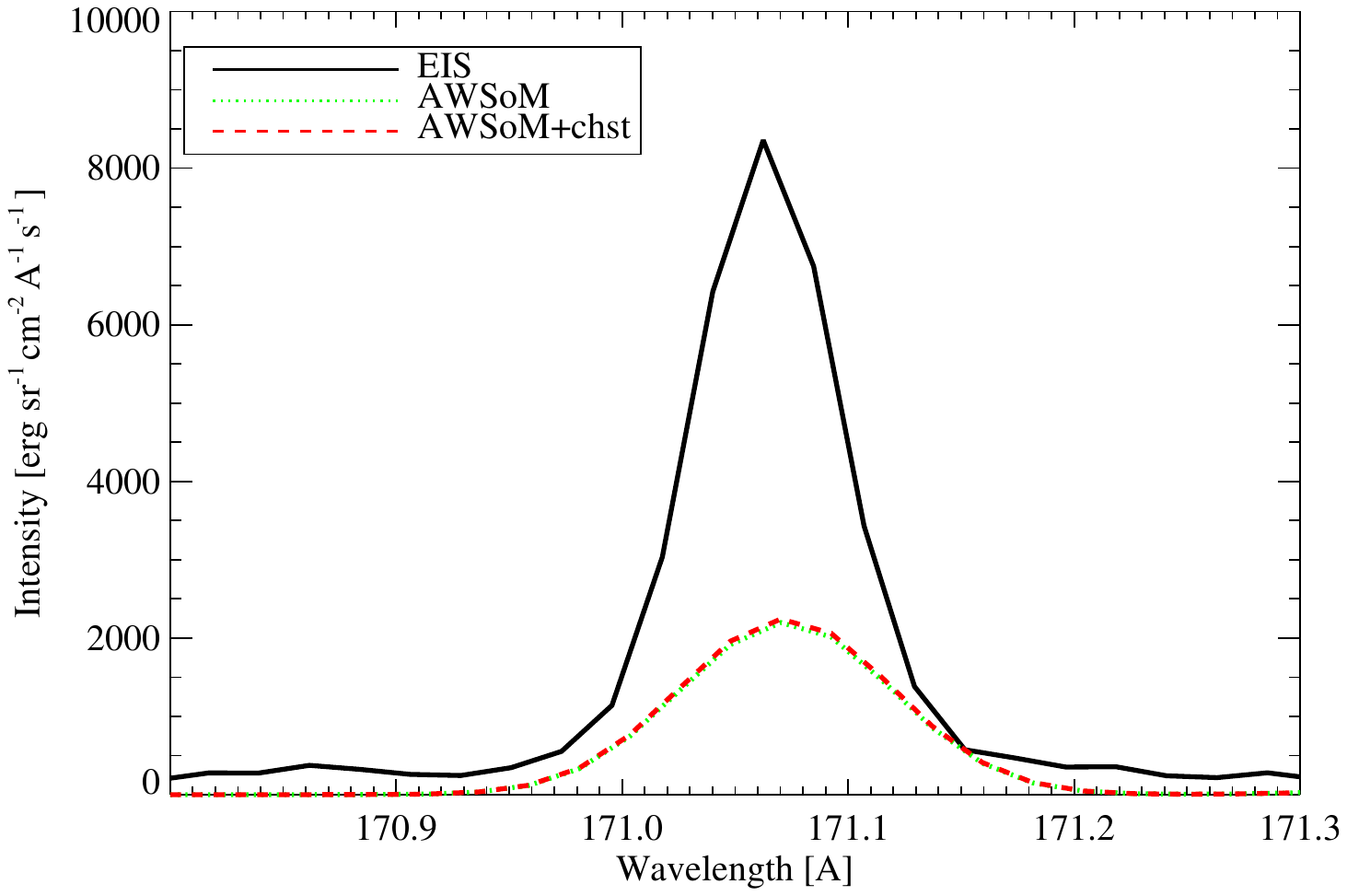}
\includegraphics[trim={2cm 0.5cm 1cm 4cm},clip,width=6cm]{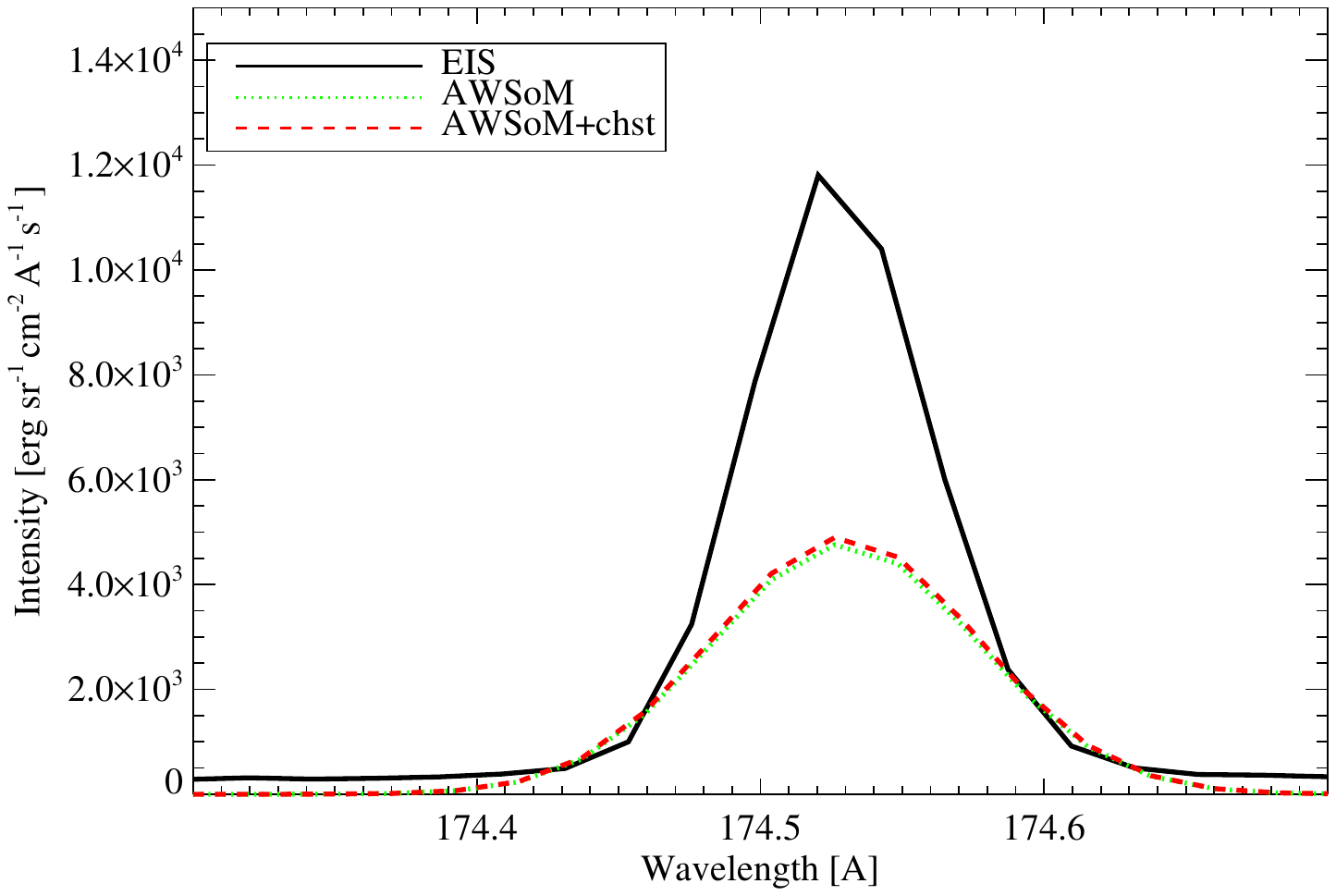}
\includegraphics[trim={2cm 0.5cm 1cm 4cm},clip,width=6cm]{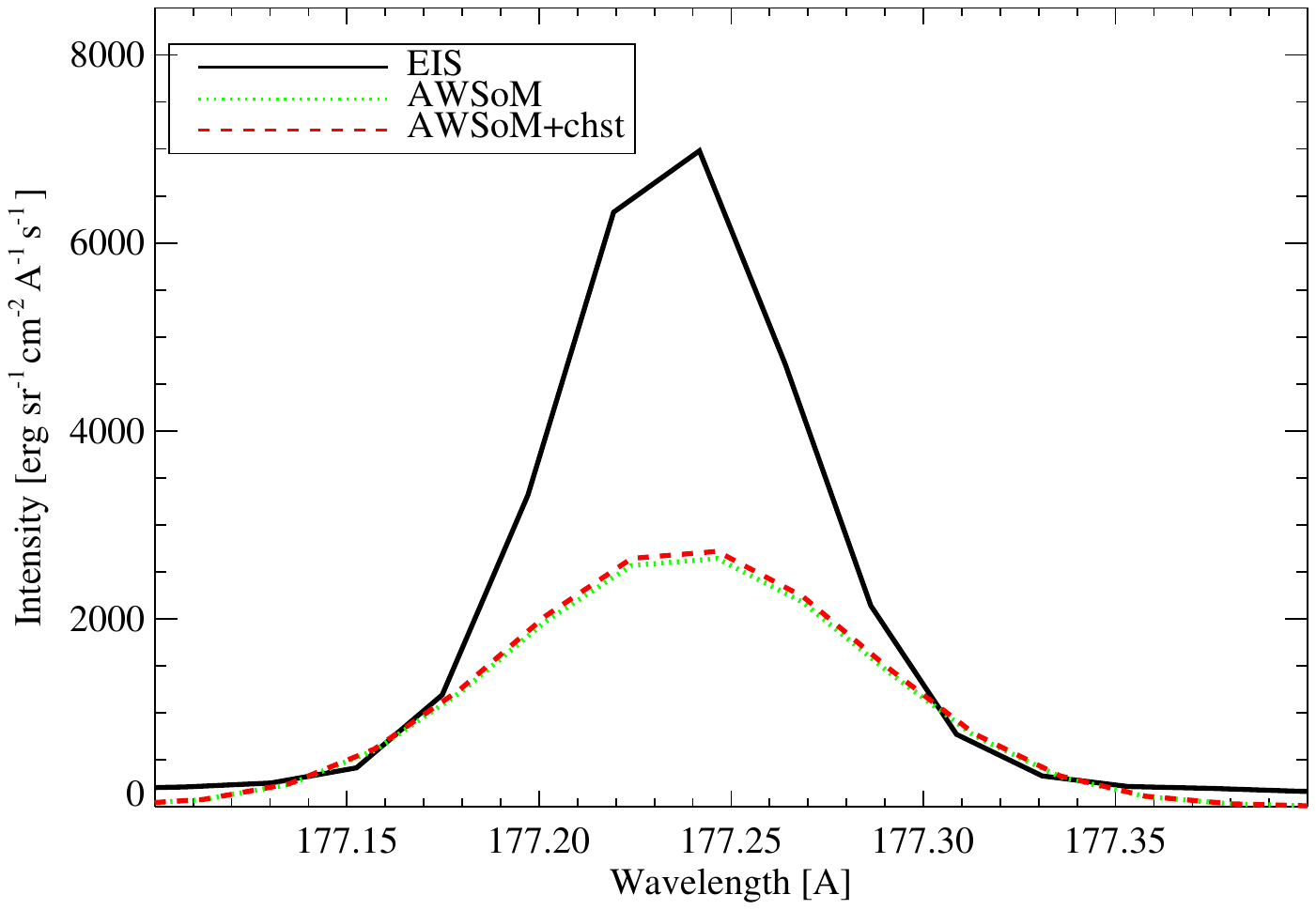}
\includegraphics[trim={2cm 0.5cm 1cm 4cm},clip,width=6cm]{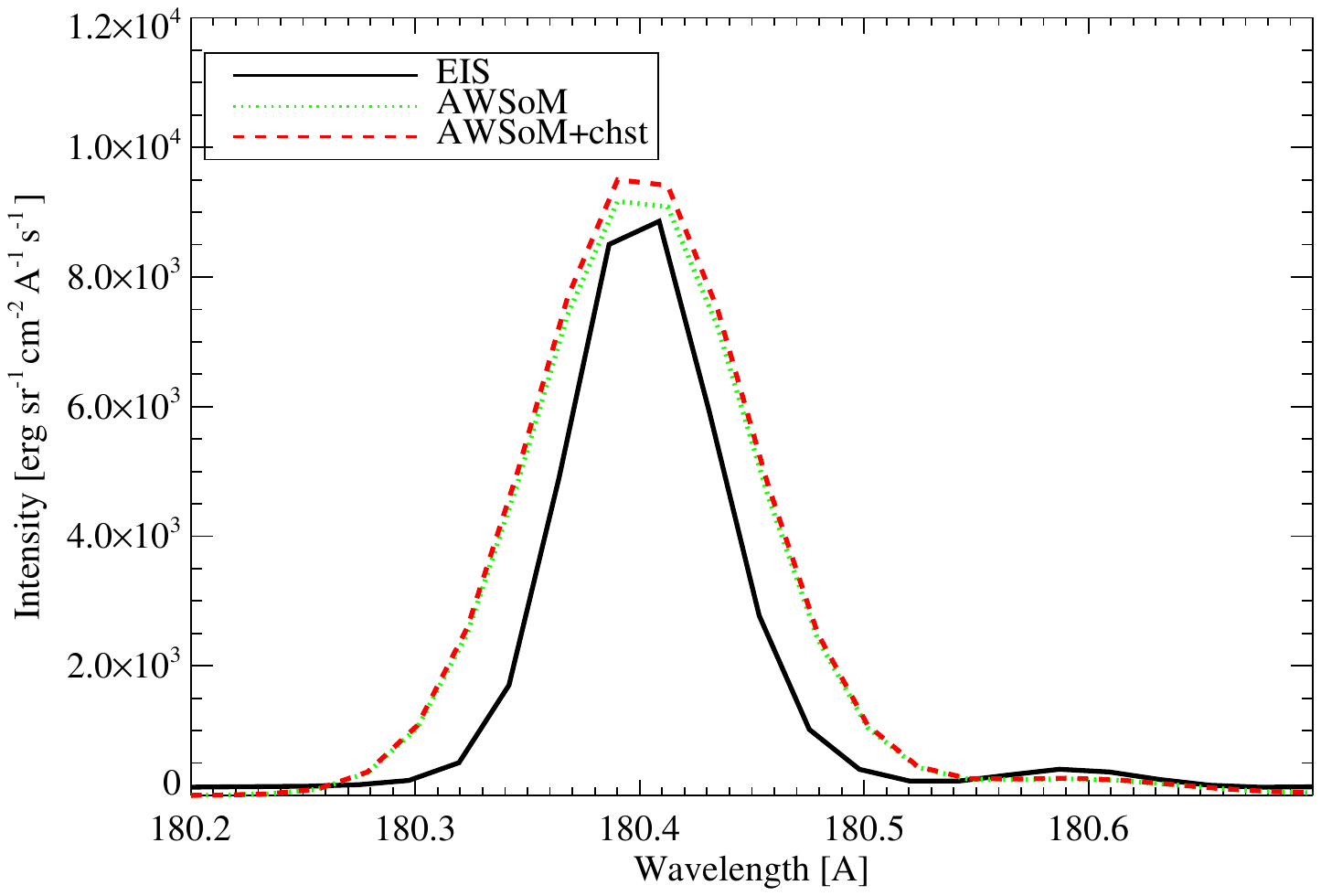}
\includegraphics[trim={2cm 0.5cm 1cm 4cm},clip,width=6cm]{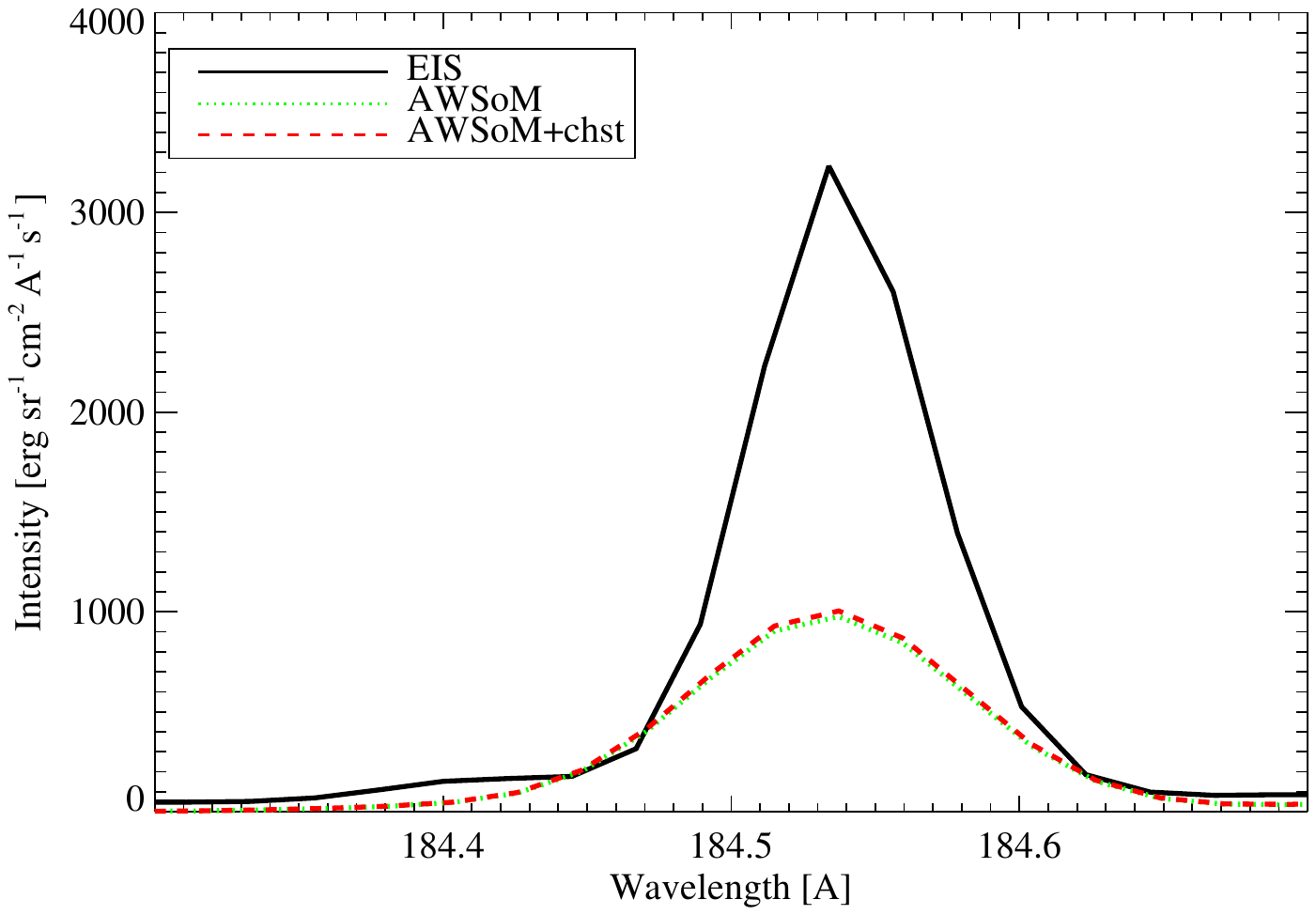}
\includegraphics[trim={2cm 0.5cm 1cm 4cm},clip,width=6cm]{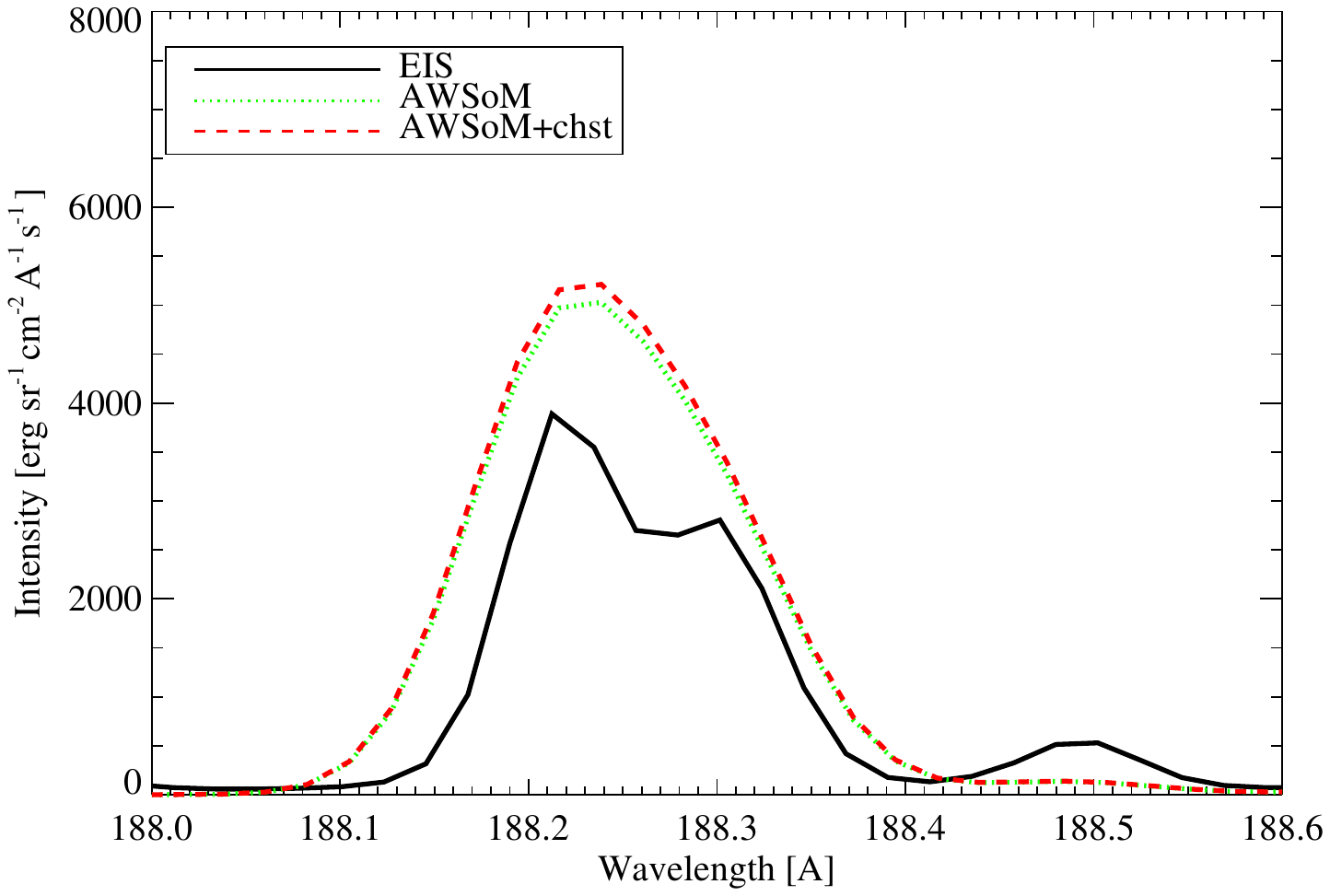}
\includegraphics[trim={2cm 0.5cm 1cm 4cm},clip,width=6cm]{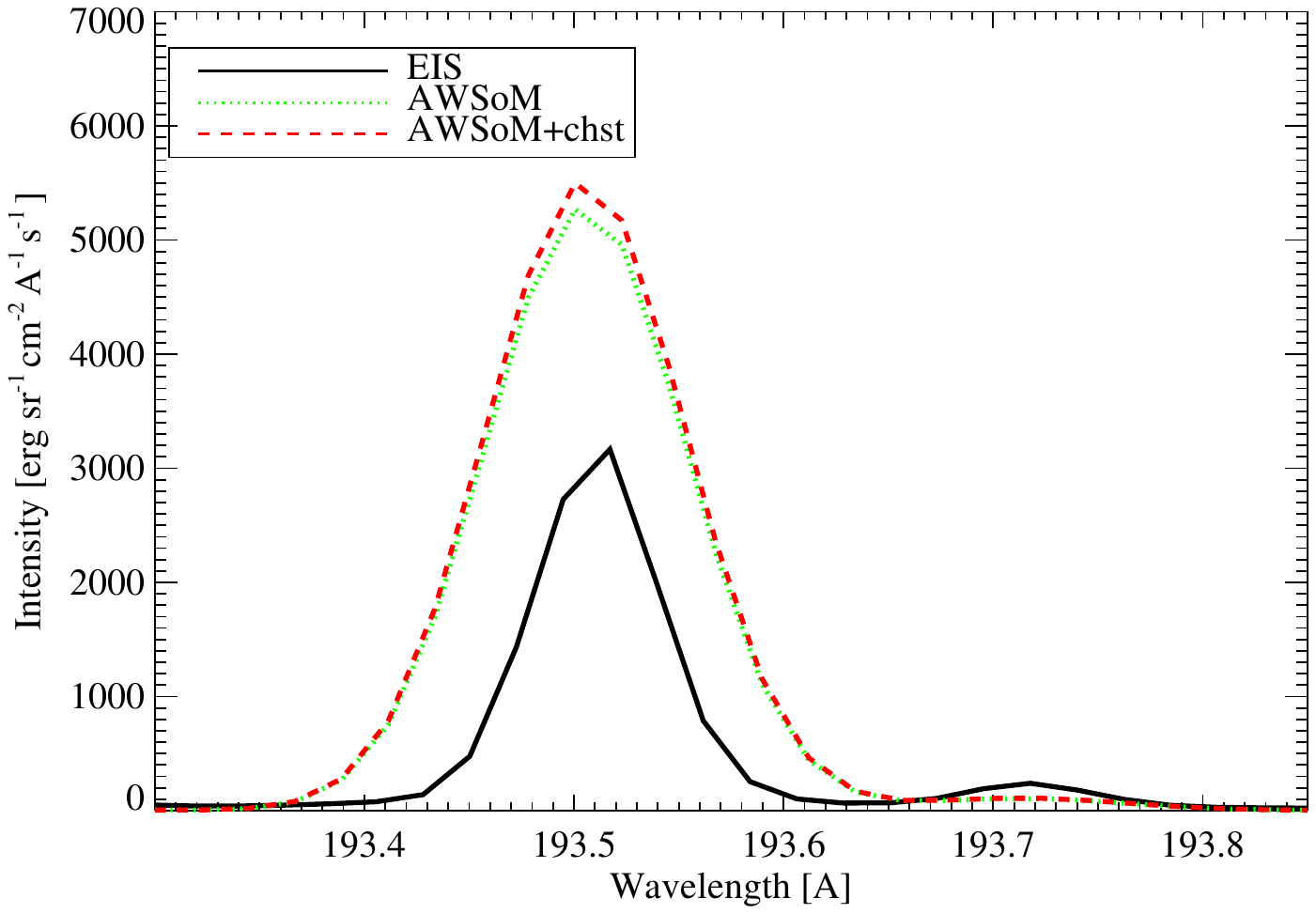}
\includegraphics[trim={2cm 0.5cm 1cm 4cm},clip,width=6cm]{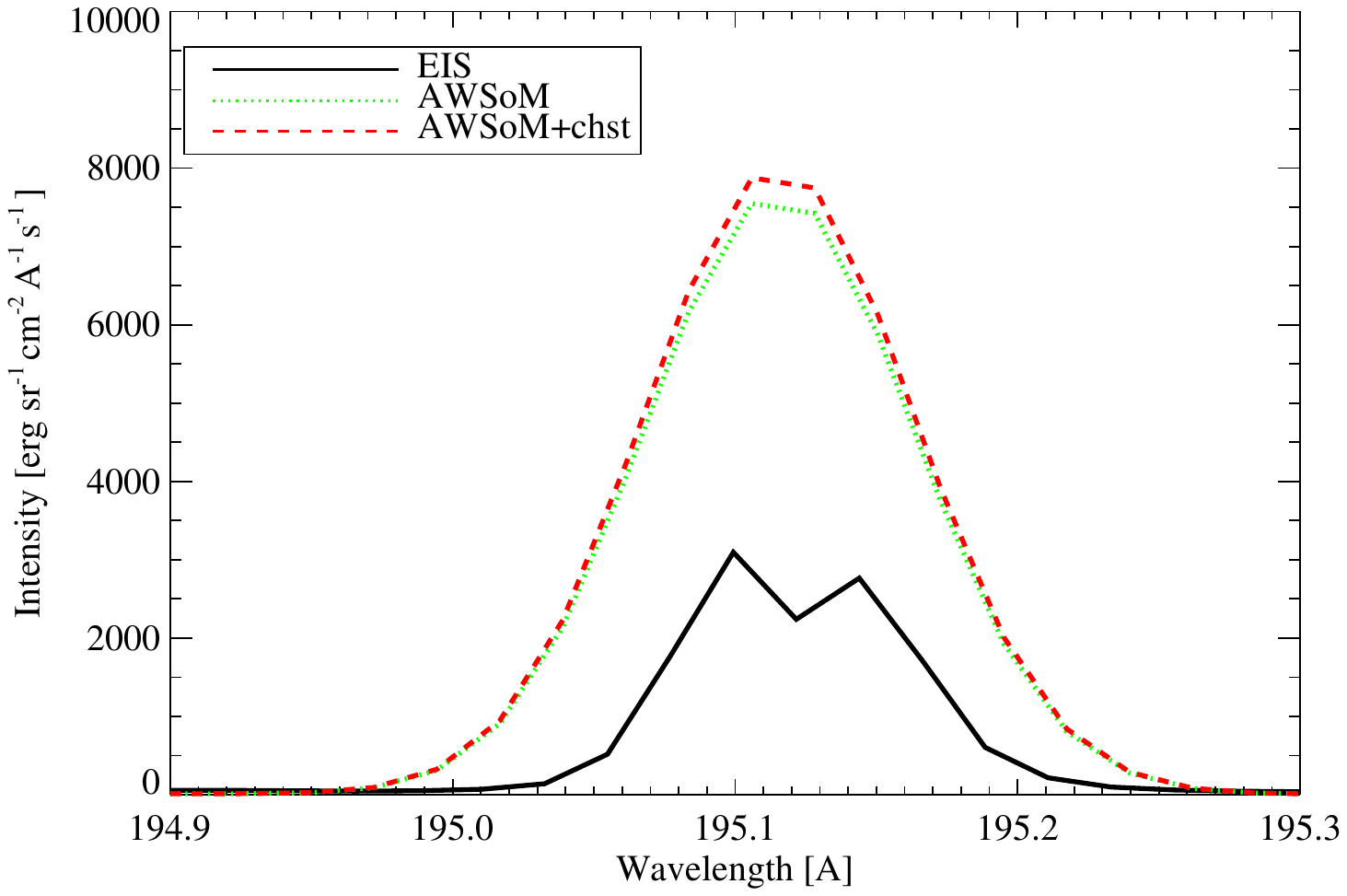}
\includegraphics[trim={2cm 0.5cm 1cm 4cm},clip,width=6cm]{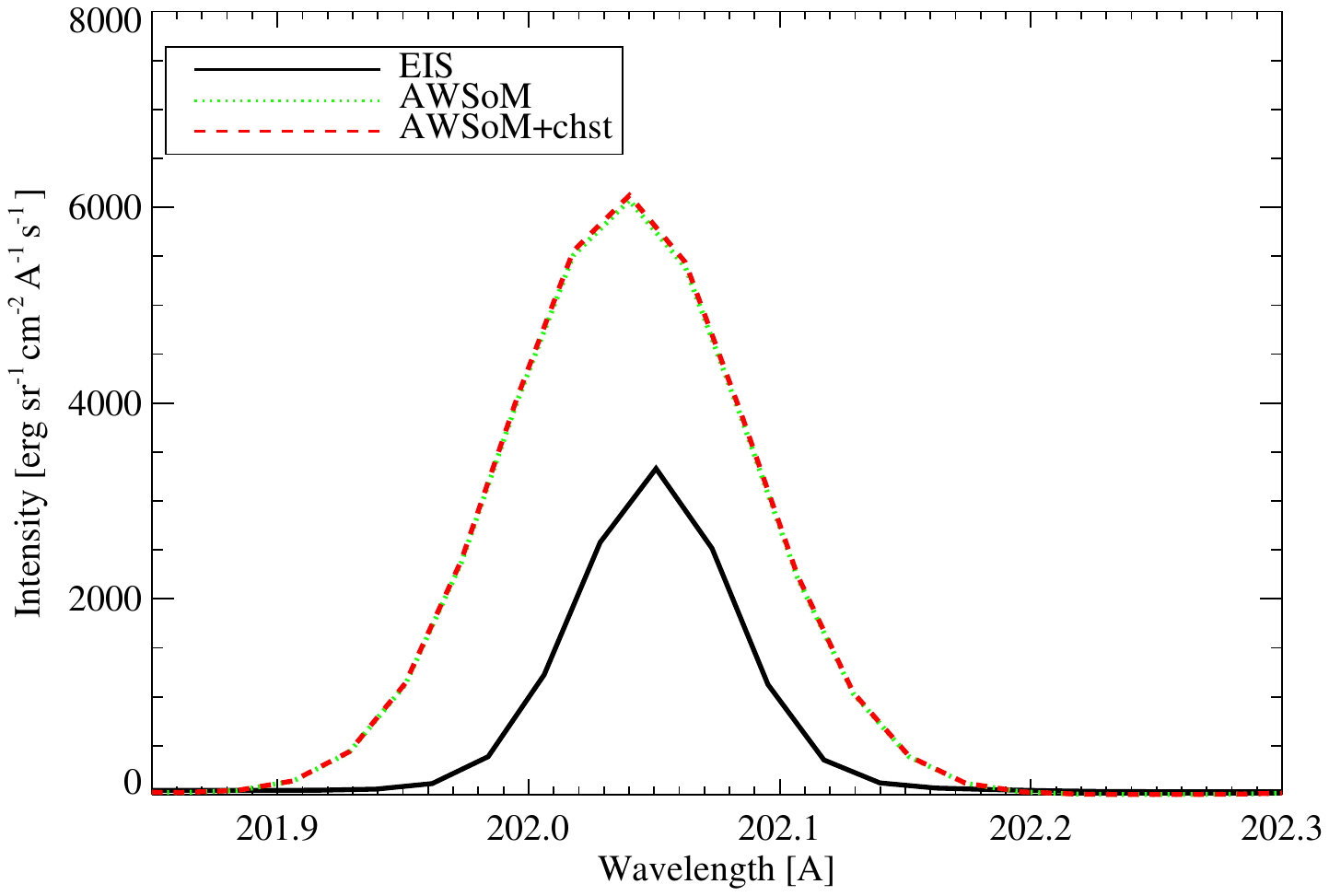}
\includegraphics[trim={2cm 0.5cm 1cm 4cm},clip,width=6cm]{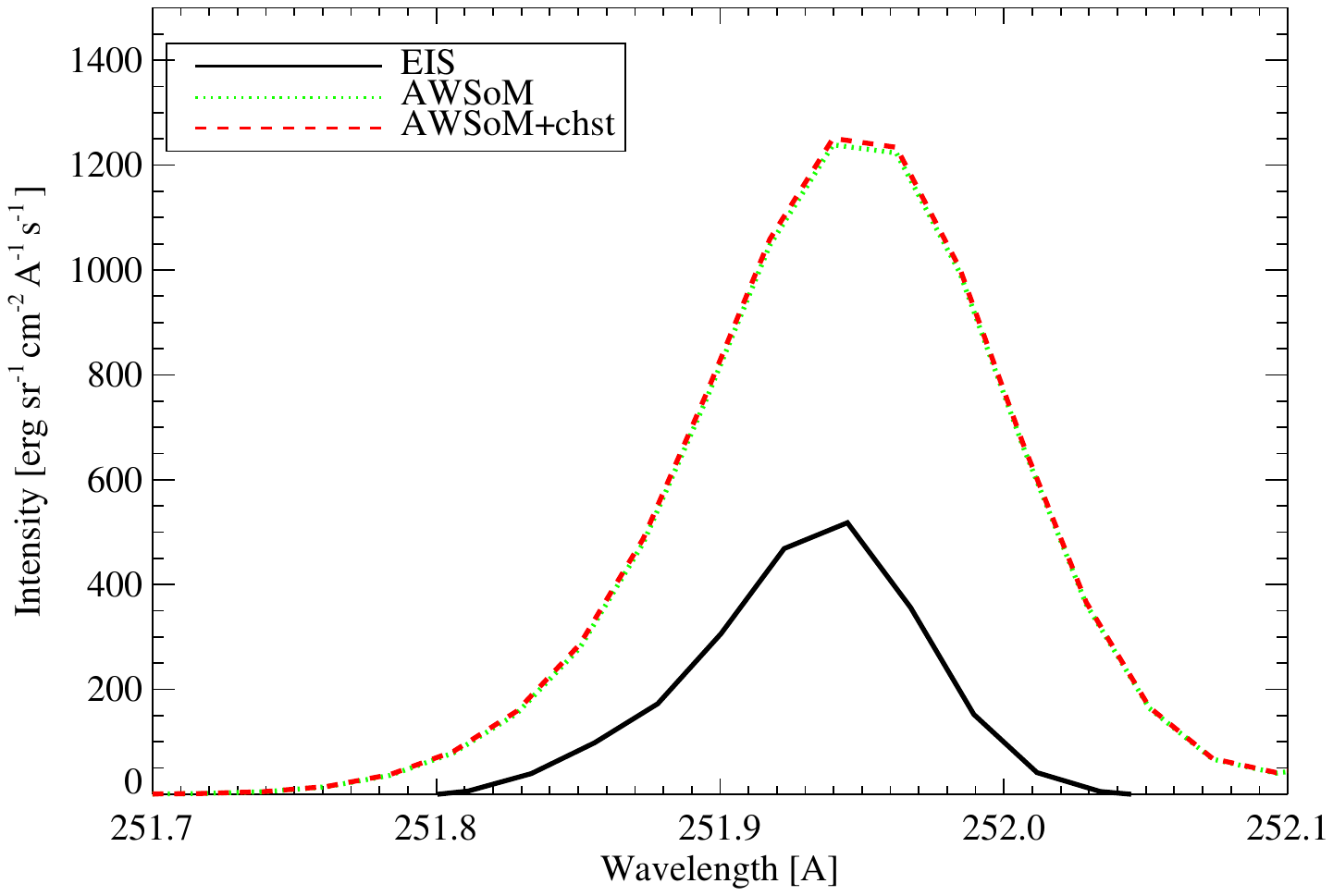}
\includegraphics[trim={2cm 0.5cm 1cm 4cm},clip,width=6cm]{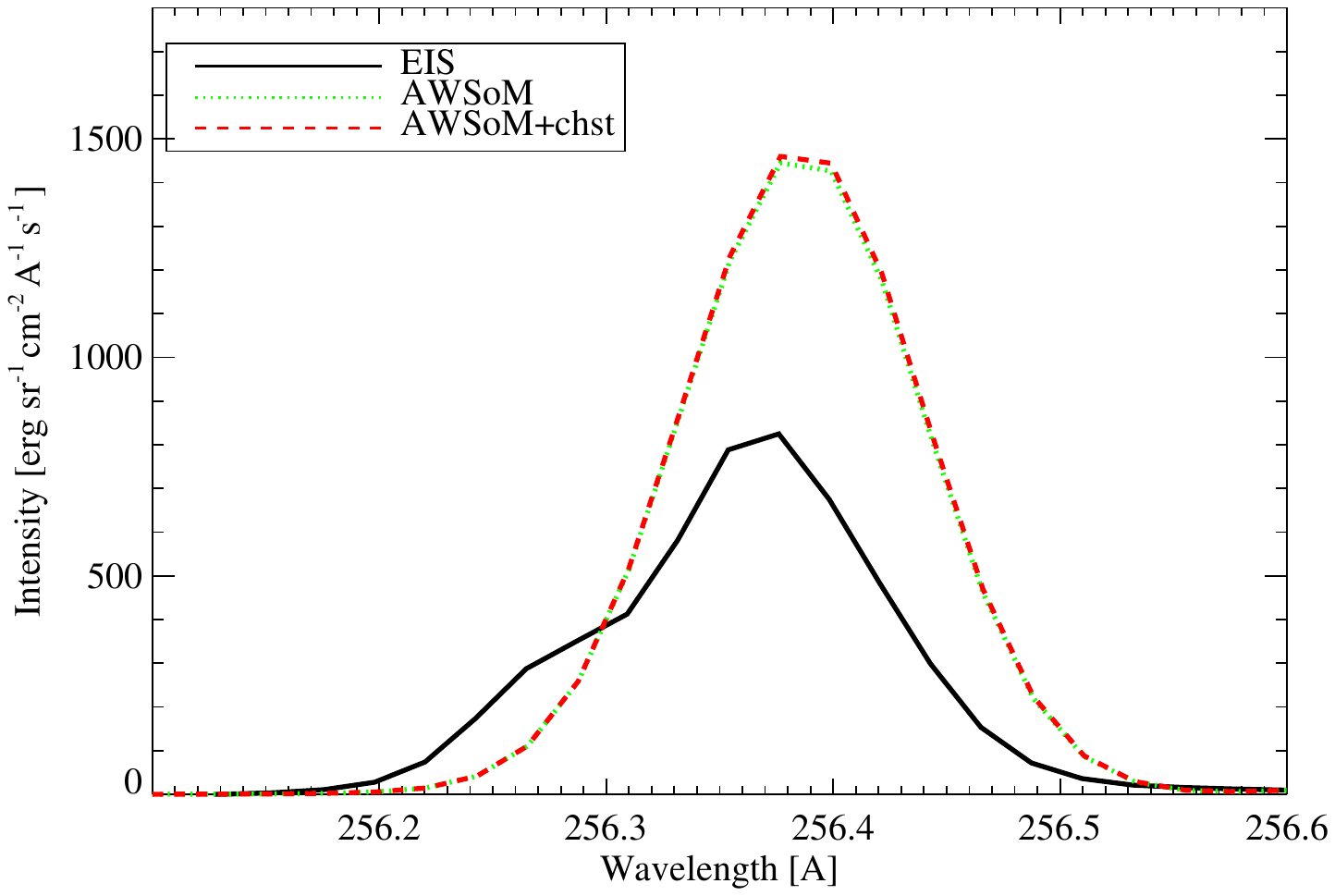}
\includegraphics[trim={2cm 0.5cm 1cm 4cm},clip,width=6cm]{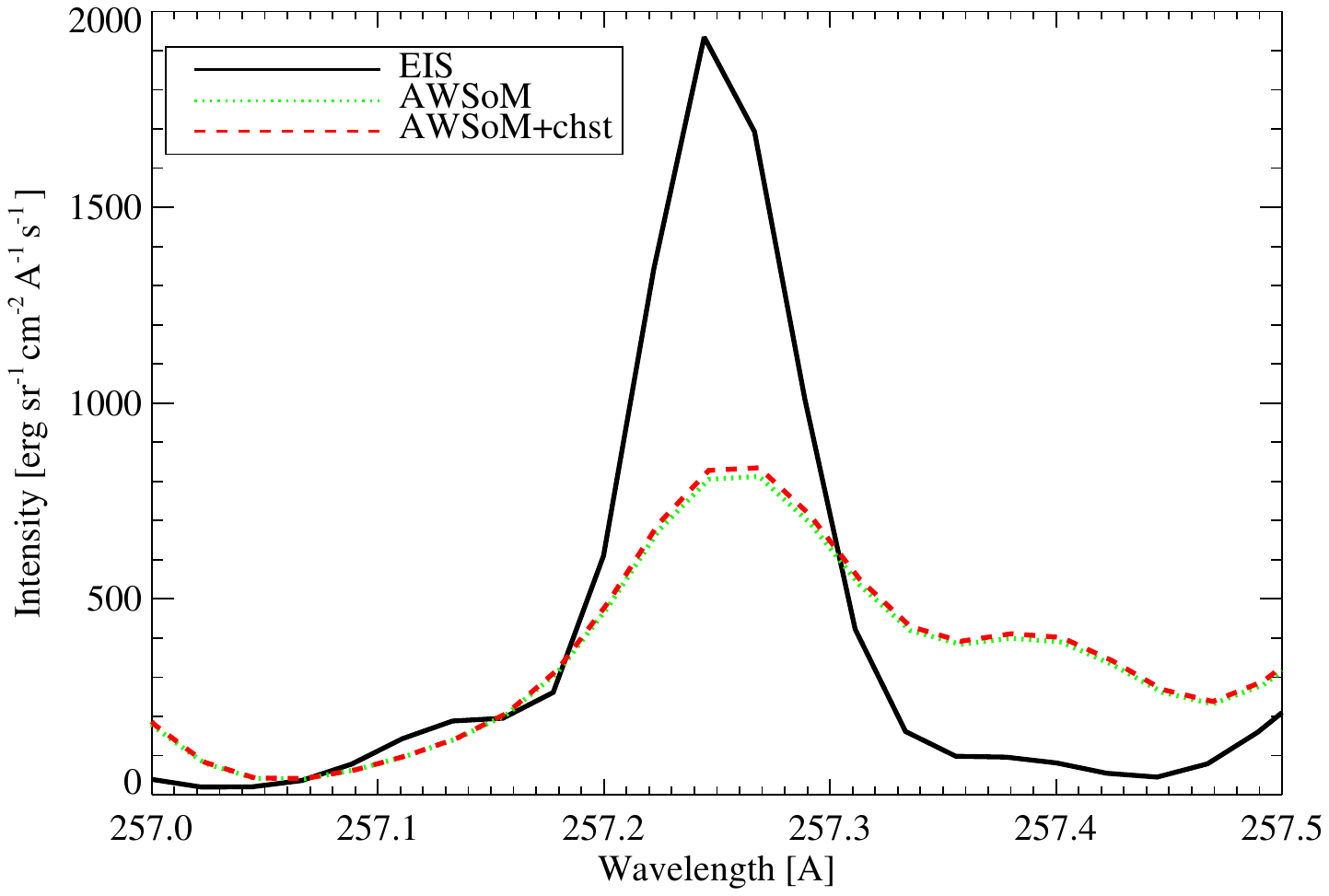}
\includegraphics[trim={2cm 0.5cm 1cm 4cm},clip,width=6cm]{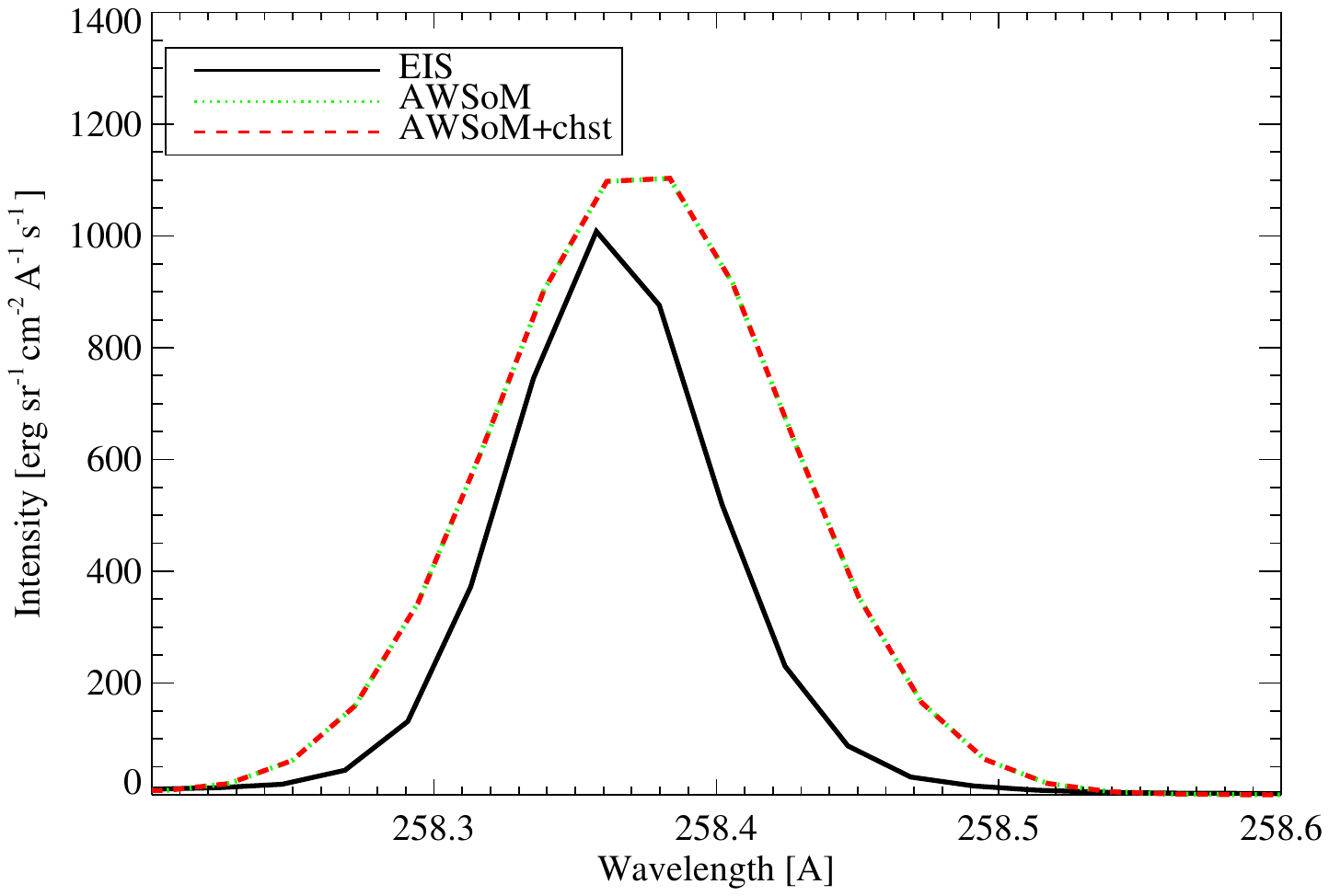}
\includegraphics[trim={2cm 0.5cm 1cm 4cm},clip,width=6cm]{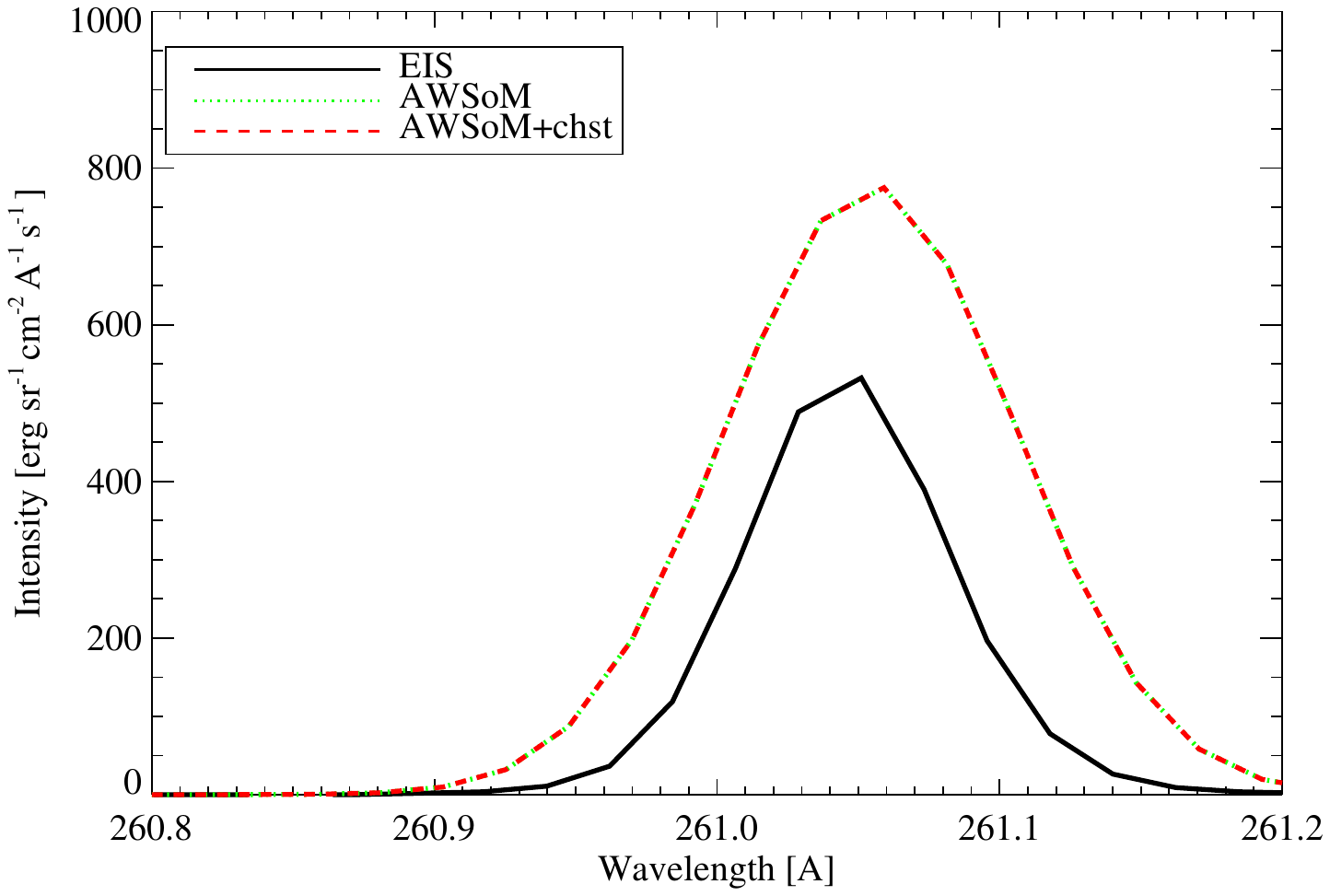}
\includegraphics[trim={2cm 0.5cm 1cm 4cm},clip,width=6cm]{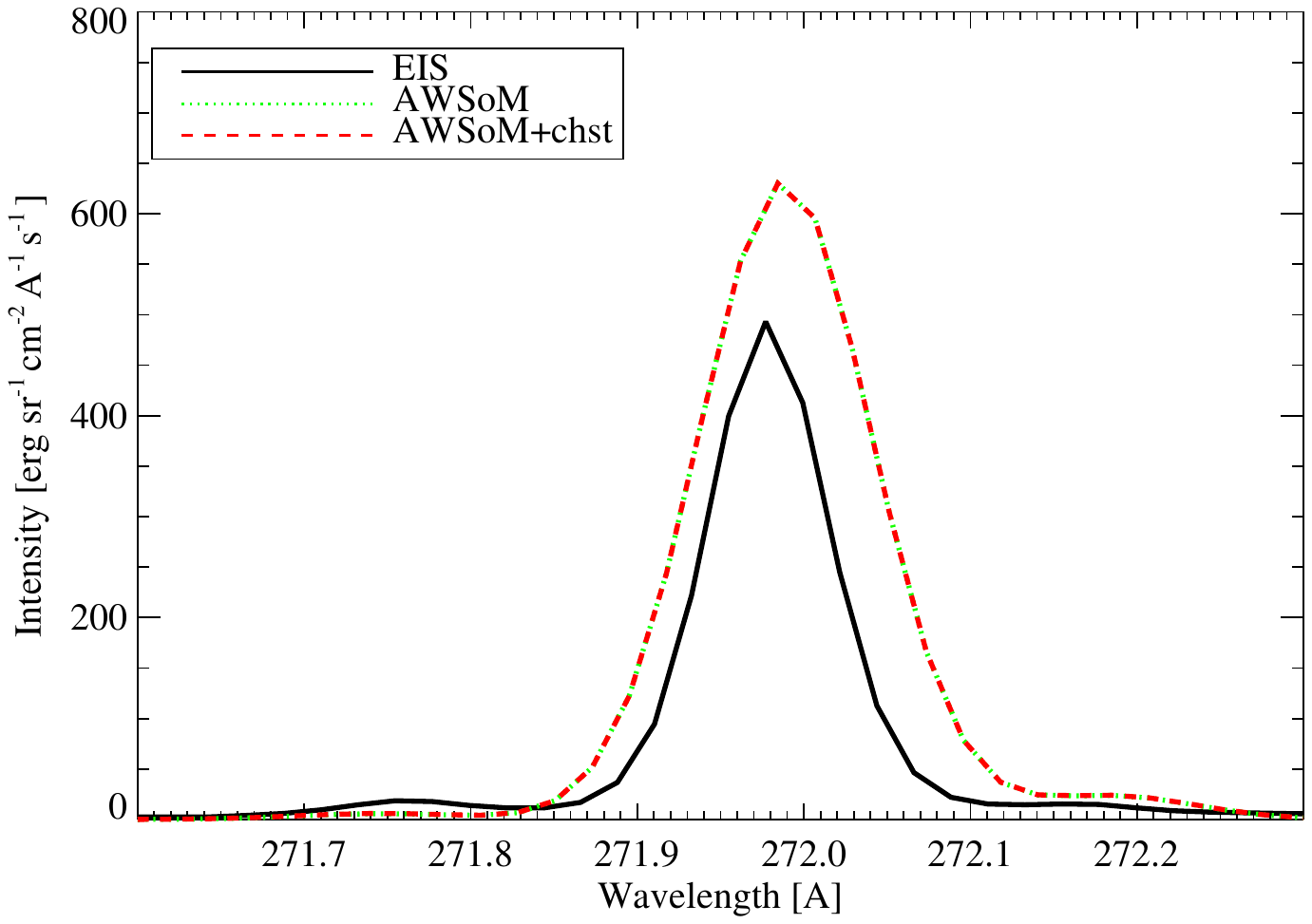}
\caption{West limb observation comparison to data. Black shows Hinode/EIS observations, green the equilibrium solution, red the non-equilibrium solution's emission.
\label{fig:wlimb}}
\end{figure}

Figure~\ref{fig:wlimb} shows the comparison on the West limb, where due to the relative brightness a larger set of lines could be discussed. As expected, the larger electron density and the lack of significant plasma speed in this region minimize the importance of non-equilibrium effects, so that there is no significant difference between the two predicted spectra, while limitations in the model (likely due to resolution) are still present.

The discrepancy between observations and simulations are likely partially due to the inherent limitations of 3D MHD modeling: the transition region extension and Maxwellian electron distributions. Both are resolvable by implementing a different, approach to the modeling of the transition region \citep{Zhou:2021} and using non-Maxwellian distributions of electrons when calculating ionization- and recombination rates \citep{Dere:2023}. These options will be be studied in the future.

\section{Summary}\label{sec:summary}
In this work we combined AWSoM's NEI calculations from \citet{Szente:2022} with the synthetic spectral calculations of SPECTRUM \citet{Szente:2019}, to calculate non-equilibrium line intensities across the entire domain of the AWSoM 3D global model. We find that the resulting spectra are strongly affected by non-equilibrium effects in the fast wind regions and streamer edges and that these effects propagate to narrowband images. The dependence shows a different nature for each line observed and resulted in significant changes in line intensity, which should be accounted for during plasma diagnostics. Non-equilibrium effects on individual line intensities also propagate in the narrow-band images, causing variations in the predicted line intensities somewhat smoothed from those in individual spectral lines, but still significant for plasma diagnostics. However, we also find that these effects depend on the local plasma properties (and consequently, even on the phase of the solar cycle), and that no single correction can be developed to account for non-equilibrium effects in observed spectra and images. Comparing to high-resolution synthetic spectra with Hinode/EIS measurements we saw that the changes due to NEI, while significant, are still unable to account for the differences between EIS spectra and AWSoM/SPECTRUM predictions, suggesting that the model itself still needs improvements.

\begin{acknowledgments}
This work was supported by NASA grants 80NSSC20K0185 and 80NSSC22K0750.\\ 
The authors would like to thank the anonymous referee for their comments, which helped us improve the original manuscript.\\
This work utilizes data obtained by the Global Oscillation Network Group (GONG) program, managed by the National Solar Observatory, which is operated by AURA, Inc. under a cooperative agreement with the National Science Foundation. The data were acquired by instruments operated by the Big Bear Solar Observatory, High Altitude Observatory, Learmonth Solar Observatory, Udaipur Solar Observatory, Instituto de Astrofisica de Canarias, and Cerro Tololo Interamerican Observatory. We acknowledge use of NASA/GSFC's Space Physics Data Facility's OMNIWeb (or CDAWeb or ftp) service, and OMNI data.\\
For SOHO/EIT comparison we used data supplied courtesy of the SOHO/MDI and SOHO/EIT consortia. SOHO is a project of international cooperation between ESA and NASA. SoHO/EIT observational data were provided by The Virtual
Solar Observatory, Solar Data Analysis Center: \url{http://virtualsolar.org}.\\
Hinode is a Japanese mission developed and launched by
ISAS/JAXA, collaborating with NAOJ as a domestic partner,
NASA and UKSA as international partners. Scientific operation
of the Hinode mission is conducted by the Hinode science team
organized at ISAS/JAXA. This team mainly consists of
scientists from institutes in the partner countries. Support for
the post-launch operation is provided by JAXA and NAOJ
(Japan), UKSA (UK), NASA, ESA, and NSC (Norway).\\
CHIANTI is a collaborative project involving George Mason University, the University of Michigan (USA) and the University of Cambridge (UK).\\
Our team also acknowledges high-performance computing support from Pleiades, operated by NASA's Advanced Supercomputing Division.
\end{acknowledgments}

\bibliography{manuscript}{}
\bibliographystyle{aasjournal}

\end{document}